# The Hera Radio Science Experiment at Didymos

Edoardo Gramigna[1], Riccardo Lasagni Manghi[1], Marco Zannoni[1,2], Paolo Tortora[1,2], Ryan S. Park[3], Giacomo Tommei[4], Sébastien Le Maistre[5], Patrick Michel[6], Francesco Castellini[7], Michael Kueppers[8]

[1] Dipartimento di Ingegneria Industriale, Alma Mater Studiorum - Universit,  di Bologna, 47121 - ForlǴ(FC), Italy

[2] Centro Interdipartimentale di Ricerca Industriale Aerospaziale, Alma Mater Studiorum - Universit,  di Bologna, 47121 - ForlǴ(FC), Italy

[3] Jet Propulsion Laboratory, California Institute of Technology, Pasadena, CA, USA

[4] Department of Mathematics, Università di Pisa, Largo Bruno Pontecorvo 5, Pisa, 56127, Italy

[5] Royal Observatory of Belgium, Brussels, Belgium

[6] Université Côte d'Azur, Observatoire de la Côte d'Azur, CNRS, Laboratoire Lagrange, Nice, France

[7] Telespazio Deutschland GmbH

[8] ESA/ESAC, Villanueva de la Cañada (Madrid), Spain

## Abstract

Hera represents the European Space Agency's inaugural planetary defense space mission and plays a pivotal role in the Asteroid Impact and Deflection Assessment international collaboration with NASA DART mission that performed the first asteroid deflection experiment using the kinetic impactor techniques. With the primary objective of conducting a detailed post-impact survey of the Didymos binary asteroid following the DART impact on its small moon called Dimorphos, Hera aims to comprehensively assess and characterize the feasibility of the kinetic impactor technique in asteroid deflection while conducting an in-depth investigation of the asteroid binary, including its physical and compositional properties as well as the effect of the impact on the surface and shape of Dimorphos. In this work, we describe the Hera radio science experiment, which will allow us to precisely estimate critical parameters, including the mass, which is required to determine the momentum enhancement resulting from the DART impact, mass distribution, rotational states, relative orbits, and dynamics of the asteroids Didymos and Dimorphos. Through a multi-arc covariance analysis, we present the achievable accuracy for these parameters, which consider the full expected asteroid phase and are based on ground radiometric, Hera optical images, and Hera to CubeSats InterSatellite Link radiometric measurements. The expected formal uncertainties for Didymos and Dimorphos GM are better than 0.01% and 0.1%, respectively, while their $J_2$ formal uncertainties are better than 0.1% and 10%, respectively. Regarding their rotational state, the absolute spin pole orientations of the bodies can be recovered to better than 1 degree, and Dimorphos' spin rate to better than $10^{-3}$%. Dimorphos reconstructed relative orbit can be estimated at the sub-m level. Preliminary results, using a higher-fidelity dynamical model of the coupled motion between rotational and orbital dynamics, show uncertainties in the main parameters of interest that are comparable to those in standard radio science models. A first-order estimate of the expected uncertainty in the momentum transfer efficiency from DART's impact, obtainable with Hera, yields a value of about 0.25. This represents a significant improvement compared to current estimates. Overall, the retrieved values meet the Hera radio science requirements and goals, and remain valid under the condition that the system is determined to be in an excited but non-chaotic (or tumbling) state. The Hera radio science experiment will play an integral role in the exploration of the Didymos binary asteroid system and will provide unique scientific measurements, which, when combined with other observables such as optical images, altimetry measurements, and satellite-to-satellite tracking of the CubeSats, will support the mission's overarching goals in planetary defense and the deep understanding of binary asteroids.

# 1. Introduction

In the context of planetary defense, space exploration missions to small bodies such as asteroids and comets are of the utmost importance as they provide critical data for understanding the properties, compositions, and orbits of these objects, thereby enabling the development of effective strategies to mitigate potential threats to Earth. By studying the composition, structure, and surface properties of asteroids, missions such as NEAR-Shoemaker [Prockter et al., 2002; Yeomans et al., 2000], Stardust [Brownlee et al., 2003, 2014], Rosetta [Glassmeier et al., 2007; Taylor et al., 2017], Deep Impact [A'hearn et al., 2005; Henderson and Blume, 2015], Hayabusa [Yoshikawa et al., 2015; Fujiwara et al., 2006], Dawn [Russell and Raymond, 2012; Park et al., 2014; Konopliv et al., 2018], Hayabusa2 [Watanabe et al., 2017; Tsuda et al., 2020], OSIRIS-REx [Lauretta et al., 2017; Scheeres et al., 2019], DART [Cheng et al., 2018; Cheng et al., 2023], Lucy [Levison et al., 2021], Psyche [Lord et al., 2017; Zuber et al., 2022], Hera [Michel et al., 2022] and many others contribute to our knowledge of these celestial bodies' characteristics. This information is vital for assessing the potential impact consequences and devising appropriate deflection or disruption techniques. In addition, several international initiatives and organizations focus on planetary defense coordination. The International Asteroid Warning Network (IAWN) and the Space Mission Planning Advisory Group (SMPAG) facilitate information sharing and collaboration among space agencies, observatories, and research institutions. These coordination efforts ensure a unified global response to potential asteroid threats and promote the exchange of data, technology, and expertise. In this context, the NASA-Jet Propulsion Laboratory Solar System Dynamics group, the NEODyS team, and the ESA-NEOCC group actively focus on determining the motion and physical parameters of natural planetary objects, as well as uncertainties in the orbital elements, which are useful in observation planning, hazard assessment, and mission planning [NASA JPL SSD; NEODYS; NEO-CC].

Hera is a European Space Agency (ESA) space mission [Michel et al., 2022], part of the Asteroid Impact and Deflection Assessment (AIDA) international collaboration with the NASA DART mission [Cheng et al., 2015; Michel et al., 2016], with a targeted launch in October 2024 and arrival at the Didymos system in October 2026. This cooperation aims to assess the feasibility and efficiency of the kinetic impactor technique to deflect an asteroid.

On 26 September 2022, DART successfully impacted Dimorphos, the ~151 m diameter secondary body of the Didymos binary asteroid system. The central body is called Didymos, which has a diameter of ~780 m, and the distance between the two asteroids is roughly 1.21 km. A few weeks after the impact, the DART investigation team confirmed that the spacecraft collision with its target asteroid successfully altered the orbit of Dimorphos, decreasing its orbital period by ~33 minutes [Thomas et al., 2023; Cheng et al., 2023]. This outcome is the very first deliberate alteration of the trajectory of a celestial object, representing a significant milestone in human achievement, as well as the inaugural comprehensive exhibition of asteroid deflection capabilities. Furthermore, the Light Italian CubeSat for Imaging of Asteroids (LICIACube) [Dotto et al. 2021] was able to document and witness the collision by taking images of the event and the following few minutes [Dotto et al., 2023], which was not enough to observe the outcome. In this context, Hera will perform a detailed post-impact survey of the Didymos system after the DART impact, which will allow us to fully document the DART impact outcome and to characterize the target's properties that influence its response to impact, offering numerical models the first deflection experiment at asteroid scale for their validation [Richardson et al., 2023].

The main goals of Hera, related to the deflection demonstration, are:

1. Detailed study and characterization of DART's impact outcome on Dimorphos.

2. Analysis of the impact regarding momentum transfer efficiency (i.e., precise measurement of the mass of Dimorphos).
3. Accurate estimation of the physical properties of Didymos and Dimorphos to validate and demonstrate the kinetic impactor technique to deflect potentially hazardous asteroids in the future.
4. Perform in-situ system observations to characterize dynamical effects (e.g., libration imparted by the impact, orbital and spin excitations of Dimorphos).

In addition to the main objectives related to planetary defense, this mission will carry out several analyses and experiments, contributing to asteroid science [Michel et al., 2022].

With Hera data, the mass and mass distribution of Didymos and Dimorphos can be precisely estimated through radio science investigations. In particular, the determination of the gravity field is obtained by precisely reconstructing the trajectory of Hera during selected close encounters with the system. In addition, Hera will carry two CubeSats, Juventas and Milani, which will be released in the vicinity of the Didymos system [Goldberg et al., 2019; Karatekin et al., 2021; Ferrari et al., 2021a, 2021b]. Hera will track the CubeSats utilizing an Inter-Satellite Link system (ISL), the first miniaturized satellite-to-satellite transceiver ever flown in deep space. This represents a crucial addition to the gravity investigation, as the relative Doppler shift between Hera and the CubeSats, computed via the ISL measurements, brings a very high information content on the dynamics of the system, thanks to the closer distance of the CubeSats from the asteroids' center of mass. To this end, the precise trajectory reconstruction of Hera, Juventas, and Milani will allow for estimating the low degree and order gravity fields of Didymos and Dimorphos. Hera's radio science experiment (RSE) will strongly contribute to the planetary defense post-impact survey of DART and to asteroid science, being the first RSE investigation of a binary asteroid system.

This manuscript focuses on the Hera RSE investigation, providing an overview of the observations, measurements, and results obtained from the orbit determination (OD) simulations conducted through a covariance analysis.

The paper is organized as follows: Section 2 presents the Hera mission scenario, including the associated mission phases and orbits. Section 3 introduces the Hera RSE, outlining its objectives, goals, and the measurements used in the investigation. Section 4 describes the covariance analysis procedure, the multi-arc estimation, and the methods employed in our OD simulations. Furthermore, Section 5 defines the dynamical model used in the simulations, while Section 6 introduces the measurements and associated noises incorporated in the OD simulations. Section 7 describes the filter setup and the parameters estimated in the simulations. Section 8 presents the key results of the covariance analysis for Hera's RSE. Subsection 8.1 shows the asteroid phase nominal mission results, including Hera radiometric measurements (Doppler and range), optical images, and ISL Doppler-range measurements. Furthermore, within this Subsection, we discuss the contribution of the experimental phase and perform parametric analyses to examine the influence of various factors, such as measurement uncertainties, duty cycles, spacecraft stochastic accelerations, and mission phases on the estimated parameters. Additionally, in Subsection 8.2, we present the contribution of LIDAR altimetric measurements and crossover estimations within Hera's RSE. In Section 9, we establish the connections between the estimated parameters derived from Hera's RSE and the geophysics of the investigated bodies, including their internal structure, dynamics, and momentum transfer efficiency following the DART impact. Finally, in Section 10, we present preliminary covariance analysis results utilizing higher-fidelity dynamical models to investigate the coupled orbital and rotational motion of the Didymos system. The focus is on both pre- and post-impact non-chaotic scenarios.

## 2. Hera mission scenario

The Hera mission is set to launch in October 2024 on board a Falcon 9 rocket, with a launch window between October 7$^{th}$ and 27$^{th}$. After a deep-space maneuver a few weeks after launch, Hera will perform a Mars flyby in March 2025 at a minimum altitude of 5000-8000 km. The spacecraft will then perform a second deep-space maneuver in January 2026, redirecting itself toward the Didymos system. During the cruise, a flyby of the Martian moon Deimos is also tentatively planned. Upon arrival, the insertion into the Didymos system consists of seven maneuvers starting in October 2026 and ending with the spacecraft entering orbit in late 2026 or early 2027. Backup launch opportunities are available in 2026, which would result in an early 2031 arrival at Didymos [Michel et al., 2022].

Once it arrives at Didymos, Hera will deploy the two CubeSats, Juventas and Milani, to complement the observations of the mother spacecraft. Juventas, which is built by GomSpace, hosts a low-frequency monostatic radar (JuRa), which will perform the first direct measurement of an asteroid interior, a gravimeter for small solar system objects (GRASS), an accelerometer, and the ISL; Milani, which is built by Tyvak, carries a near-infrared imager (ASPECT), a micro thermogravimeter (VISTA), and the ISL [Michel et al., 2022].

The Hera mission scenario is characterized by five main phases, as shown in Figure 1 [Michel et al., 2022]:

- Early Characterization Phase (ECP, 6 weeks): a series of hyperbolic arcs, each lasting 3 to 4 days, at distances around 20–30 km from the Didymos system. This phase is used to start mapping the surface and retrieve the global shapes and mass of the system in preparation for closer flybys. In this phase, Hera trajectory and attitude guidance are based on ground navigation performed by ESOC flight dynamics;

- Payload Deployment Phase (PDP, 2 weeks): CubeSats release and commissioning. Hera will focus on supporting the early operations of Juventas and Milani through satellite-to-satellite tracking;
- Detailed Characterization Phase (DCP, 4 weeks): a series of hyperbolic arcs at distances around 8–20 km from the Didymos system, following the same ECP scheme of consecutive 3- to 4-day arcs. In this phase and the following ones, the close flybys require Hera to employ an autonomous attitude guidance, using centroiding information for Didymos (and potentially also Dimorphos) from onboard processing of the optical images for navigation. Furthermore, Hera starts collecting the ISL measurements between the mothercraft and the CubeSats.
- Close Observation Phase (COP, 6 weeks): same as DCP but closer flybys as low as 4 km. Within this phase, a fully autonomous attitude guidance is also required. Below 8 km, the camera is expected to be mainly pointed to Dimorphos. High-resolution investigations will be performed on the DART crater on Dimorphos. In terms of gravity science, this phase will allow recovering the Didymos extended gravity field[a], thanks to the Juventas' low orbits. Within this phase, support to CubeSats operations for their potential landing and disposal is foreseen, too.
- Experimental Phase (EXP, 6 weeks): flybys with altitudes as low as 1-2 km from Didymos and Dimorphos will be performed, requiring autonomous attitude guidance and trajectory control maneuvers. An experimental feature-tracking algorithm involving onboard processing of Didymos close-up images will be tested. If successful, this algorithm will be used in the autonomous navigation loop. Depending on the actual flybys performed and the possible landing of the spacecraft on the surface of the bodies, a further improvement in the recovered Didymos extended gravity field could be obtained.

[a] Extended gravity field: higher order and degree, ≥ 2, Stokes coefficients used in the spherical harmonics expansion of the body's gravitational potential.

The CubeSats orbits are shown in Figure 2. Milani performs a series of hyperbolic arcs similar to Hera, while Juventas will orbit the system at closer distances using Self Stabilized Terminator Orbits (SSTO) at 3.3 and 2.0 km altitudes.

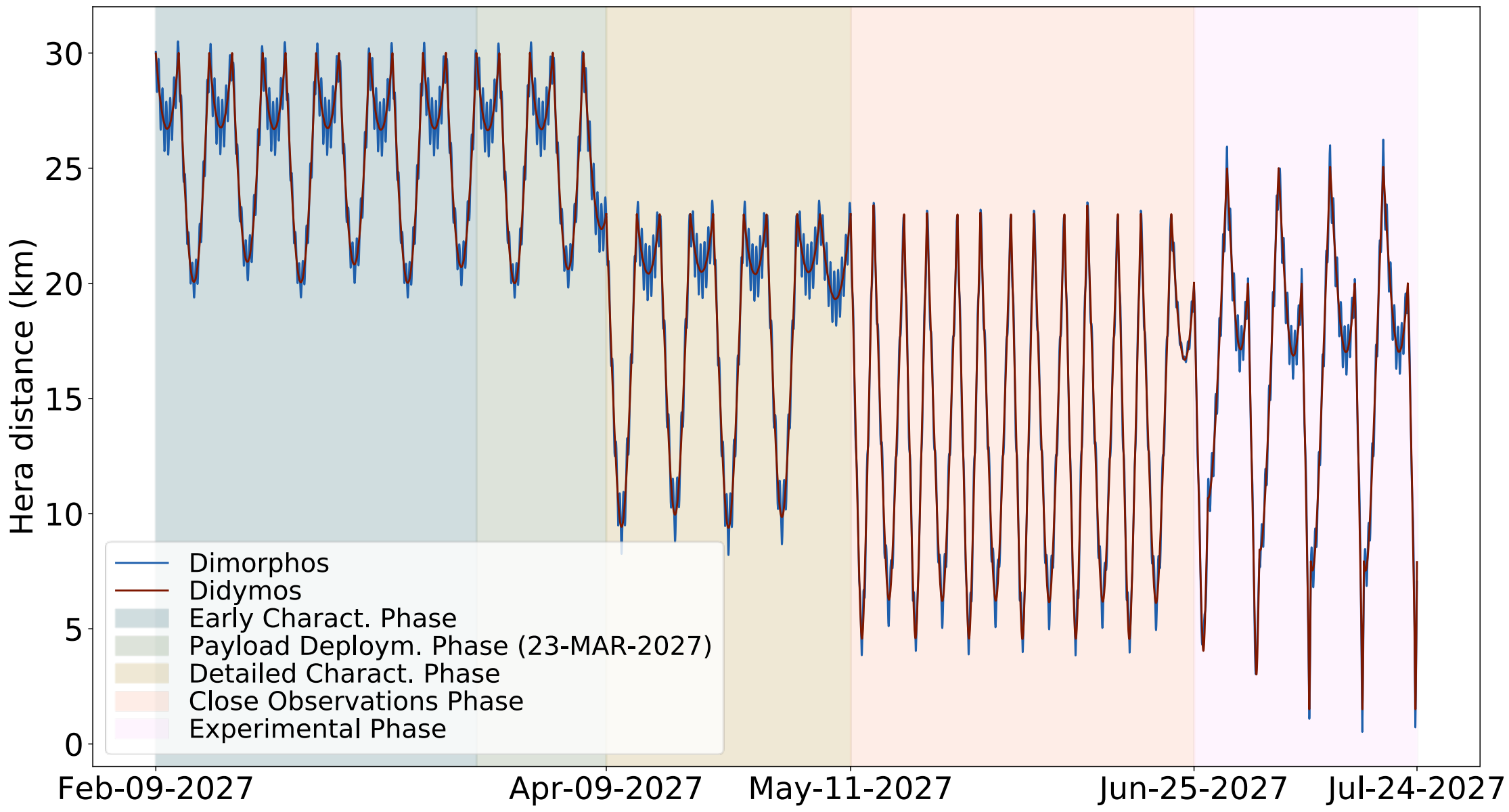

**Figure 1: Hera distances from Didymos and Dimorphos as a function of the mission phase.**

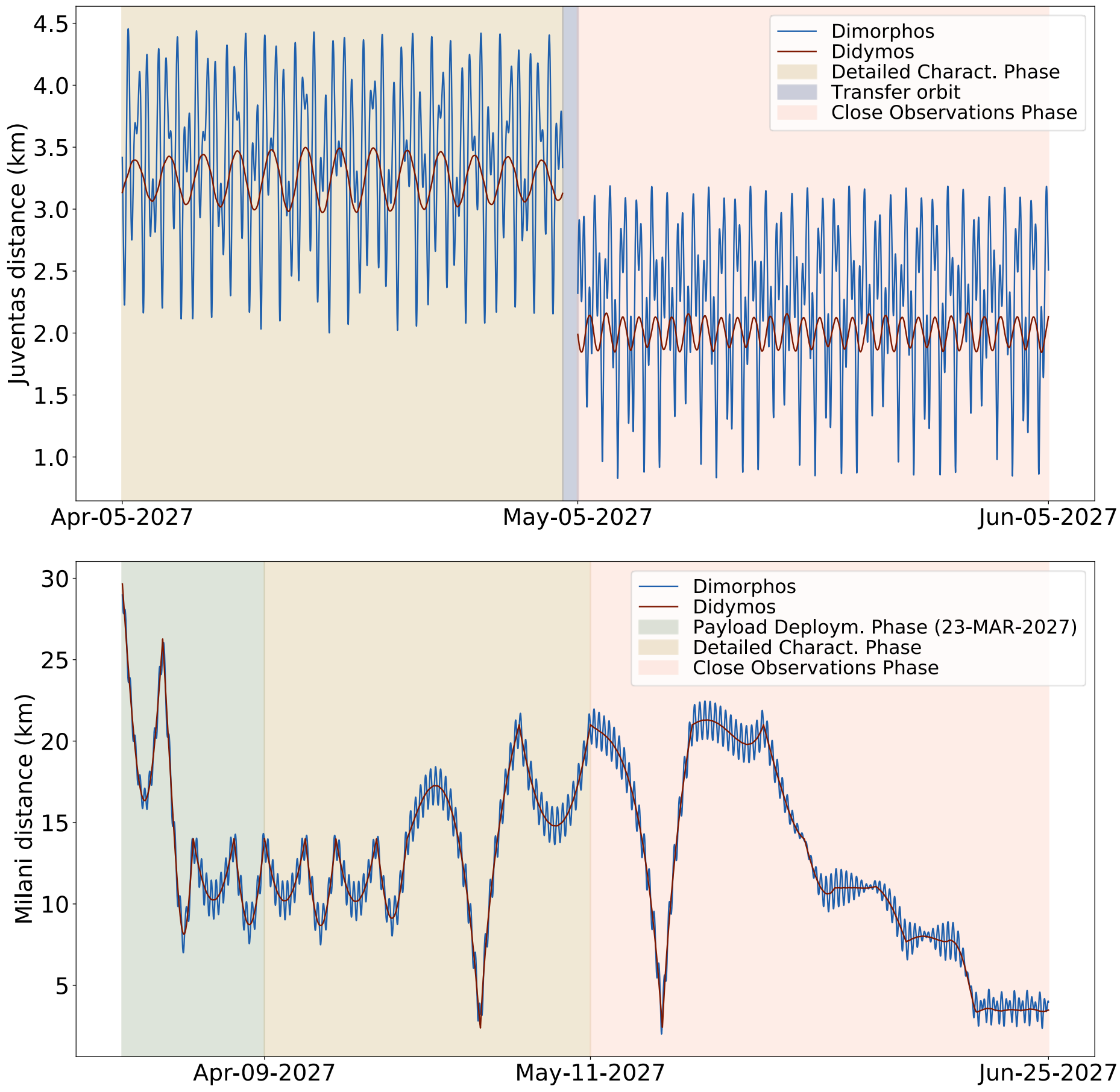

**Figure 2: Juventas (top) and Milani (bottom) distances from Didymos and Dimorphos as a function of the mission phase. Juventas is placed in SSTO orbits at 3.3 km and 2.0 km altitude, while Milani performs hyperbolic arcs like Hera.**

# 3. Hera radio science experiment

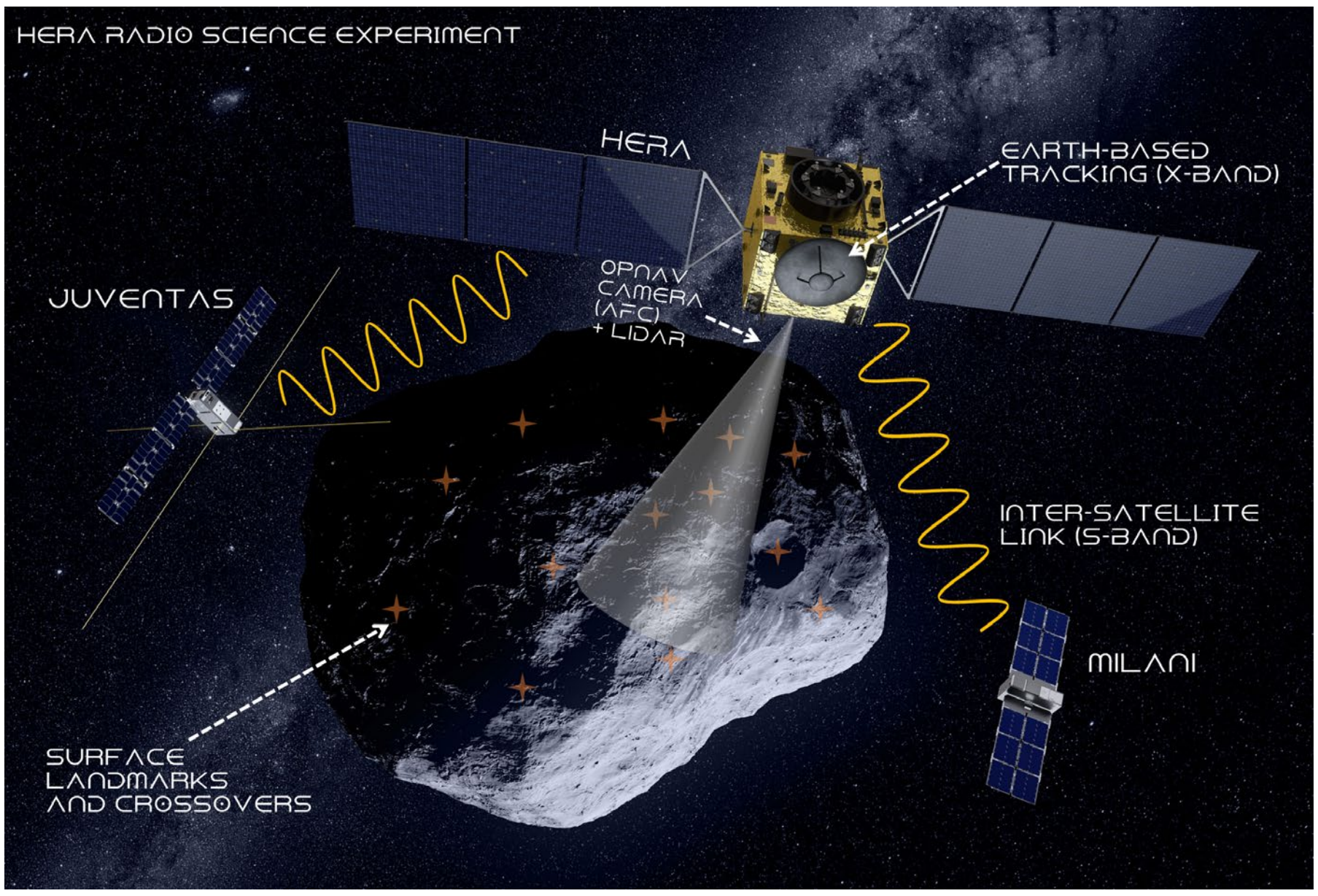

**Figure 3: Hera radio science experiment. The nominal simulations include ground-based radiometric measurements, Hera optical images, Hera to CubeSats Inter-Satellite Link radiometric measurements, and Hera PALT LIDAR measurements to surface landmarks of both Didymos and Dimorphos.**

The Hera RSE is crucial to fulfilling the core mission requirements for planetary defense and is highly relevant to asteroid science. Figure 3 summarizes the experiment, while the main Hera RSE objectives and requirements are:

1) Accurate trajectory reconstruction for Hera, Juventas, and Milani;
2) Estimating the mass and extended gravity fields of Didymos and Dimorphos to constrain their interior structures and global properties (e.g., size, shape, volume, density, porosity). Dimorphos' mass shall be estimated with a relative error of at most 10%, with a goal of 1%, which will allow the momentum enhancement factor estimation (see Section 9) [Meyer et al., 2022; Feldhacker et al., 2017];
3) Characterization of the post-impact mutual and heliocentric orbits, rotational states, and dynamics of the asteroids. Specific requirements include an absolute accuracy of 5 m in the semimajor axis (with a goal of 1 m), eccentricity uncertainty better than 0.001, Dimorphos' spin rate relative error better than 0.1%, and absolute spin pole orientations of Didymos and Dimorphos better than 1 degree (with a goal of 0.1 degrees). Accurate estimation of Dimorphos' libration is crucial in determining the energy dissipation of the system, post-impact dynamics, and further constraining its interior structure.

The Hera RSE uses different types of observables to achieve these goals and requirements. The first class of observables is represented by the Earth-based radiometric measurements. Specifically, two-way Doppler and range measurements will be acquired through Hera's X-Band link Deep Space Transponder (X-DST) developed by Thales Alenia Space Italia (TAS-I)[b], which will also provide the standard capability for simultaneous transmission of housekeeping telemetry data and reception of commands from the ground. This system, which will support Delta-Differential One-way Ranging (Delta-DOR) tracking from Earth ground stations [Book, 2013; Curkendall and Border, 2013], will also support Wide Band Delta Differential One-way Ranging (WBDDOR), allowing Hera to be the first mission to perform an in-flight demonstration of the WBDDOR capabilities [Cardarilli et al., 2019]. In terms of performances, the X-DST end-to-end Allan standard deviation [Riley, 2008] is $8\cdot 10^{-15}$ at 60 s integration time ($2\cdot 10^{-15}$ at 1000 s integration time), which, following Equation (1), corresponds to a Doppler error contribution of 2.4 μm/s at 60 s integration time (0.6 μm/s at 1000 s).

$$y = \frac{\Delta f}{f_0} = -\frac{\dot{\rho}}{c} \,, \qquad (1)$$

where y is the fractional frequency shift, $\dot{\rho}$ is the relative two-way range rate between the source and the observer that produces a Doppler shift equal to the input frequency shift $\Delta f$, $f_0$ is the nominal carrier frequency, and c is the speed of light.

[b] Thales Alenia Space Internal Report n. HERA-TAI-XDST-TR-00208, Issue 1, 20/01/2023.

Ranging data is expected to be affected by errors in the order of 3.4 ns (100 cm) for the high-frequency noise, 3 ns (90 cm) for the calibration error, and 3 ns (90 cm) for aging and drift (other error contributions are leading to higher Doppler and ranging noises, which are discussed in Section 6.1). These specifications align with the RSE requirements, first defined by Zannoni et al. (2018).

The second class of observables includes the ISL S-Band two-way Doppler[c] and range between Hera and the CubeSats. These measurements are exploited using an ISL system developed by Tekever following the design of the PROBA-3 ISL system [Llorente et al., 2013]. Without having a direct Earth-based communication system, the CubeSats rely on Hera to relay their data and commands to and from the operation centers on the ground. In this context, the main goals of the ISL transceiver are: 1) to guarantee the correct communication and data relay (i.e., sending telecommands, receiving housekeeping data, telemetry, and payload data) between Hera and the CubeSats; 2) to form the range and range-rate observables to enable the CubeSats navigation (together with optical data from the CubeSat images of the asteroid system, which are not considered as baseline in this work); 3) to carry out, for the first time in deep space, a RSE involving precise range-rate measurements between the CubeSats and the mothercraft. The ISL subsystem will operate through a spread spectrum signal at S-band frequencies (~2.2 GHz), whose architecture is similar across Hera, Juventas, and Milani. The range is based on the sub-sample accuracy time-of-flight and adopts an Oven Controlled Crystal (Xtal) Oscillator (OCXO) as the reference clock. Furthermore, the use of two patch antennas located on opposing sides of the CubeSats will provide quasi-hemispherical coverage, thus removing the need for a specific spacecraft pointing when the link is active [Gramigna et al., 2022; Goldberg et al., 2019]. The accuracy is expected to be better than 50 cm for ranging and 0.05 mm/sec at 60 sec integration time for the Doppler.

[c] ISL two-way Doppler implementation still under development and work in progress.

A third type of measurement is represented by the optical images collected by the Asteroid Framing Cameras (AFC) on board Hera. The AFCs are two identical cameras developed by JenaOptronik, whose design is based on the Astrohead cameras [JenaOptronik AFC]. The cameras have an array size of 1020x1020 pixels with a field of view of 5.5 x 5.5 deg and an Instantaneous Field of View (IFOV) of 94.1 μrad. The spatial scales are expected to be on the order of 2-3 m/pixel in the ECP, 1-2 m/pixel in the DCP, 0.5-2 m/pixel in COP, and 10 cm/pixel in very close flybys in the EXP [Michel et al., 2022].

The last data type exploited by the RSE is represented by the laser altimetry measurements. Hera is equipped with a Light Detection and Ranging instrument (LIDAR), a Planetary Altimeter (PALT) laser that transmits directly toward the nadir and measures the two-way time of flight of a laser beam at 1.5 μm wavelength, with an aperture of 1 mrad, corresponding to a footprint of 10 m at a distance of 10 km [Dias et al., 2022]. The instrument, which operates only when the Hera-to-surface distance is between 0.5 and 14 km, will target both asteroids depending on the mission phase.

## 4. Covariance analysis

A gravity RSE is a particular application of the orbit determination process, which aims to estimate a set of parameters that affect the spacecraft's motion. While the focus of the orbit determination, as a part of the more general process of navigation, is the trajectory of the spacecraft, the focus of gravity science is the accurate modeling and estimation of the physical parameters of interest, such as the gravity field, the rotational state, and the orbits of celestial bodies.

For this reason, a typical approach adopted in the RSE data analysis is the multi-arc approach, where the observables collected during non-contiguous orbital segments, called arcs, are jointly analyzed to produce a global solution of a set of solve-for parameters. The arc length can be selected to maximize the accuracy and reliability of estimating the global parameters, preferring arcs characterized by a high information content. Consequently, the scientific results will be obtained at the end of the mission by post-processing all available data, as opposed to the output of the operational OD, which must be carried out in real time to navigate the spacecraft safely.

When dealing with future missions, the use of a realistic setup and assumptions within the numerical simulations allows inferring, before the execution of the RSE, the expected formal accuracies on the estimation of scientific parameters of interest, performing a so-called covariance (sensitivity) analysis [Bierman, 1977]. In particular, the same OD procedure for analyzing real data is adopted, and the real measurements are replaced by simulated ones. Furthermore, the effects of the main design parameters on the experiment's performance can be analyzed by controlling the dynamical model used to generate the simulated measurements, performing parametric studies in this way.

The output of a covariance analysis [Lombardo et al., 2022; Lasagni Manghi et al., 2019] (i.e., computing the covariance matrix of the estimated parameters) is instrumental in understanding the expected performance of an estimation filter setup, as it conveys how well the estimation of the state vector $x_0$ can be made by processing M measurements collected in N arcs [Park et al., 2010, 2012]. However, the real uncertainties associated with the estimated parameters are usually larger than the formal values provided by the OD since the classical procedure does not consider possible estimation biases due to errors in the dynamical model, linearization errors, non-white/non-Gaussian measurement noise, and other un-modeled effects. For this reason, the numerical simulations employ a series of conservative assumptions and safety factors to obtain more realistic uncertainties.

The sensitivity analysis performed in this paper adopts a multi-arc approach and is based on a linearized weighted least-squares principle, where the problem is posed in a single batch. In particular, to retain numerical precision (i.e., when the direct computation of the information matrix inversion is performed), we adopt a Square Root Information Filter (SRIF) algorithm [Montenbruck et al., 2001; Battin, 1999; Bierman, 1977]. The OD simulations are performed using NASA-JPL's orbit determination program MONTE (Mission Analysis, Operations, and Navigation Toolkit Environment), currently used for the operations of all NASA's space missions managed by JPL [Evans et al., 2018], as well as for radio science data analysis and processing (see e.g. Iess et al., (2014a, 2014b); Tortora et al. (2016); Zannoni et al. (2020); Casajus et al. (2021, 2022); Zannoni and Tortora (2013); Buccino et al. (2022); Gramigna et al. (2023); Park et al. (2014, 2016, 2020)). MONTE's mathematical formulation and measurement models are detailed in Moyer (1971, 2005).

## 4.1. Multi-arc estimation

The determination of the gravity field typically uses data from different flybys, or arcs, separated in time by days, weeks, or even months. The spacecraft dynamic during the entire time period is usually not well-known, given the complexity of the non-gravitational interactions and the size of the parameter space. As a result, the errors in the dynamical model can jeopardize the predictions by amounts exceeding the measurement accuracy [Milani and Gronchi, 2010].

To overcome the non-deterministic nature of the orbit determination problem, especially at small bodies, one solution is to adopt the multi-arc approach. This method allows the decomposition of the entire time span of the observations in short, non-overlapping, and non-contiguous intervals, each one characterized by its own set of observables and initial conditions. This approach results in an over-parametrization, with the additional initial conditions absorbing the dynamical model uncertainties [Milani and Gronchi, 2010]. In the multi-arc approach, the solve-for parameters that form the state vector can be classified as:

- Global: parameters that do not vary in time and affect all the arcs simultaneously. Global parameters may include, among others, the GM of a celestial body, the spherical harmonics coefficients of its gravitational potential, and its ephemeris;

- Local: parameters that only affect and depend on a single specific arc. Theoretically, any parameter can be a local parameter. In the multi-arc approach, the S/C trajectory along each arc and the parameters related to its dynamics (e.g., Solar Radiation Pressure (SRP) scale factors, stochastic accelerations, and S/C attitude) are considered local parameters.

In the multi-arc method, the state vector (i.e., the vector of all estimated parameters) $\bar{x} = [\bar{g}, \bar{h}_1, \bar{h}_2, \ldots, \bar{h}_N]$ is split into a vector of global parameters $\bar{g}$ and N vectors of local parameters $\bar{h}_i$, where N is the number of local arcs which partition the entire time span [Milani and Gronchi, 2010]. For the Hera simulations, the arc durations are 2-4 days, depending on the mission phase.

The epoch state vector $\bar{x}_0$ (with dimensions n × 1) includes all the estimated parameters of our simulations (namely spacecraft state, the Didymos barycenter and Dimorphos states, SRP scale factors, stochastic accelerations, pointing errors, asteroid frames, gravity coefficients, and more, see Section 7 for more details) and it can be defined as:

$$\bar{x}_0 = \begin{pmatrix} \bar{x}_{G0} \\ \bar{x}_{10} \\ \bar{x}_{20} \\ \ldots \\ \bar{x}_{N0} \end{pmatrix}, \tag{2}$$

where $\bar{x}_{G0}$ is the state sub-vector including the global solve-for parameters and the $\bar{x}_{i0}$ sub-vectors contain the local solve-for parameters of the i-th arc. The subscript 0 refers to the reference time $t=t_0$ of the sub-vectors: the global reference time is the initial filter epoch (i.e., the filter solution epoch), while the local reference time of each arc is set to the time at the middle of the arc, which corresponds to the closest approach of Hera to the target.

Similarly, the measurement vector $\bar{z}$ (m × 1) can be partitioned in N arcs, namely:

$$\bar{z} = \begin{pmatrix} \bar{z}_1 \\ \bar{z}_2 \\ \ldots \\ \bar{z}_N \end{pmatrix}, \tag{3}$$

where $\bar{z}_i$ includes all the measurements of the i-th arc.

Given the measurement and state vectors, the matrix of measurement partial derivatives H, referenced to the filter state at time $t_0$, can be written as:

$$H_{t_0} = \frac{\partial \bar{z}}{\partial \bar{x}_0} = \begin{pmatrix} \frac{\partial \bar{z}_1}{\partial \bar{x}_0} \\ \frac{\partial \bar{z}_2}{\partial \bar{x}_0} \\ \cdots \\ \frac{\partial \bar{z}_N}{\partial \bar{x}_0} \end{pmatrix} = \begin{pmatrix} \frac{\partial \bar{z}_1}{\partial \bar{x}_{G0}} & \frac{\partial \bar{z}_1}{\partial \bar{x}_{10}} & \emptyset & \cdots & \emptyset \\ \frac{\partial \bar{z}_2}{\partial \bar{x}_{G0}} & \emptyset & \ddots & \cdots & \vdots \\ \vdots & & & & \emptyset \\ \frac{\partial \bar{z}_N}{\partial \bar{x}_{G0}} & & \emptyset & \ddots & \frac{\partial \bar{z}_N}{\partial \bar{x}_{N0}} \end{pmatrix}, \quad (4)$$

Since the matrix H (m × n) is computed in post-processing, it includes all the available measurements and their partial derivatives in a single batch. The matrix H can also be rewritten in a more compact form:

$$H_{t_0} = \begin{pmatrix} H_1^G & H_1^L & \emptyset & \cdots & \emptyset \\ H_2^G & \emptyset & H_2^L & \cdots & \vdots \\ \vdots & \vdots & \vdots & \ddots & \emptyset \\ H_N^G & \emptyset & \emptyset & & H_N^L \end{pmatrix}. \quad (5)$$

This form highlights the separation between the measurements' partial derivatives with respect to the global parameters (first column, superscript G) and the measurements' partial derivatives with respect to the local state vector for each arc (columns 2 to N, superscript L). By definition, each arc is independent of the others, so the only terms different from 0 are the partial derivatives of the local measurement vector with respect to the parameters belonging to the same arc and the partial derivatives with respect to the global parameters.

Furthermore, the measurement weight matrix W (m × m) can be constructed as a diagonal matrix, and it is assumed to be the inverse of the covariance matrix of the measurements noise, namely:

$$W = \begin{bmatrix} \frac{1}{\overline{\delta_1^2}} & \emptyset & \emptyset & \emptyset \\ \emptyset & \frac{1}{\overline{\delta_2^2}} & \emptyset & \emptyset \\ \emptyset & \emptyset & \ddots & \emptyset \\ \emptyset & \emptyset & \emptyset & \frac{1}{\overline{\delta_N^2}} \end{bmatrix}, \tag{6}$$

where $\overline{\delta_i^2}$ (with i = 1,..,N) includes the weights for the i-th arc observations.

Or in a compact form:

$$W = \begin{bmatrix} W_1 & \emptyset & \emptyset & \emptyset \\ \emptyset & W_2 & \emptyset & \emptyset \\ \emptyset & \emptyset & \ddots & \emptyset \\ \emptyset & \emptyset & \emptyset & W_N \end{bmatrix}, \tag{7}$$

where Wi are the weights associated with the observations of the i-th arc.

Given the matrix of measurement partials H and the weight matrix W, the covariance matrix P at time $t_0$ (i.e., the filter solution epoch) can be evaluated, namely:

$$P_{t_0} = \left[H_{t_0}^T W \, H_{t_0} + P_0^{-1}\right]^{-1} = \Lambda_{t_0}^{-1}, \tag{8}$$

where $P_0$ (n × n) represents the *a priori* covariance matrix, which reflects the confidence in the *a priori* knowledge of each global and local estimated parameter, while $\Lambda_{t0}$ is the information matrix. The matrix $P_{t0}$ (again, n × n) is the covariance matrix. It represents the main output of a covariance analysis as it provides the expected accuracy of all the global and local parameters estimated in the filtering process, given at the reference epoch $t_0$. As mentioned above, to improve the numerical performance, the algorithm operates on the square root information matrix R rather than the covariance matrix P directly, defined by:

$$R^T(t)R(t) = \Lambda(t), \tag{9}$$

where the superscript T represents a transpose. Then, following Equation (8), the covariance matrix P can be evaluated from the information matrix $\Lambda$. For more details on the SRIF implementation see Bierman (1977) and Park et al. (2006).

The diagonal elements of the covariance matrix $P_{ii} = v_{ii}$ represent the estimated parameters' variance. In contrast, the off-diagonal ones $P_{ij} = v_{ij}$ indicate the correlation between two estimated variables. Hence, the uncertainty estimates (i.e., standard deviations) of $\bar{x}_0$ can be obtained by computing $\sigma_i = \sqrt{v_{ii}}$ , for i = 1,…,n.

The covariance is usually given at the filter solution reference epoch, but it can be propagated or mapped at a generic time $t_j$ using deterministic covariance updates, namely:

$$P_{t_j} = \Phi(t_j, t_0)\, P_{t_0} \left[\Phi(t_j, t_0)\right]^T, \tag{10}$$

where $\Phi(t_j, t_0)$ is the state transition matrix (STM), which maps the state from $t_0$ to $t_j$, i.e.:

$$\Phi(t_j, t_0) = \frac{\delta\bar{x}(t_j)}{\delta\bar{x}(t_0)}. \tag{11}$$

As a final remark, since we are considering $\bar{x}_0$ characterized by Gaussian random noise (with mean $\underline{x_0}$ and covariance P) its probability density function is Gaussian, and it remains Gaussian also when mapped to different times, since it is invariant under linear operations.

## 5. Dynamical model

Within the OD process, all the relevant dynamics must be modeled and updated, namely: the heliocentric orbit of the Didymos system, the relative orbits of Didymos and Dimorphos around their common barycenter, and the relative orbits of Hera, Juventas, and Milani with respect to the Didymos system. Didymos system heliocentric orbit and relevant orbital parameters are provided in Table 1.

**Table 1: Didymos system and Dimorphos orbital parameters, from DRA v5.01, dimorphos_s527.bsp, sb-65803-205.bsp [NASA JPL SSD], and Richardson et al., 2023.**

| Didymos system | |
|---|---|
| GM system ($km^3/s^2$) | $4.04166824 \cdot 10^{-08}$ |
| GM Didymos ($km^3/s^2$) | $4.0071462 \cdot 10^{-08}$ |
| GM Dimorphos ($km^3/s^2$) | $3.4522065 \cdot 10^{-10}$ |
| Semimajor axis | 1.643 AU |
| Orbital period | 2.109 year |
| Eccentricity | 0.384 |
| Inclination (EMO2000) | 3.414 deg |
| **Dimorphos** | |
| Semimajor axis | 1.204 km |
| Orbital period | 11.9215 h (pre-impact); 11.3685 h (post-impact) |
| Eccentricity | <0.03 (post-impact) |
| Inclination (EMO2000) | 169.3 deg |

The primary gravitational forces acting on Didymos barycenter come from the point-mass gravitational perturbations of the Sun, the Moon, the eight planets, and Pluto, whose positions and velocities are taken from DE430 [Folkner et al., 2014]. Furthermore, the force model considers small gravitational perturbations, namely point-mass perturbations from the 16 most massive main-belt asteroids [Farnocchia, 2017]. In addition, relativistic perturbations are applied for the Sun, the Moon, and the planets based on the Einstein-Infeld-Hoffman formulation [Moyer, 2005]. A simplified model for the Yarkovsky effect [Farinella et al., 1998; Bottke et al., 2002] has also been included, which accounts for the main transverse acceleration and neglects the out-of-plane and radial accelerations. This effect is a non-gravitational acceleration due to the anisotropic emission of thermal radiation, which is associated with solar heating on a rotating body. In the implemented model, the transverse acceleration is expressed in the form $a_T = A_2 r^{-d}$, where $A_2$ is a parameter, r is the heliocentric distance in au, and d is strictly in the range 0.5–3.5 (for most NEAs the value is in the range 2-3, see Farnocchia et al. (2013)). Following Chesley et al. (2021), we use d = 2 (i.e., the value that matches the level of absorbed solar radiation from optical and radar observations) and $A_2 = -9.41 \cdot 10^{-15}$ AU/d$^2$ (from JPL solution *sb-65083-205.bsp* [NASA JPL SSD], which corresponds to $-1.89 \cdot 10^{-16}$ km/s$^2$ at 1 AU Sun distance*)*.

Regarding Dimorphos, the same base force model is applied, including the point-mass and gravity spherical harmonics accelerations from Didymos and the indirect acceleration due to its own spherical harmonics (also called indirect-oblateness). Within our work, we opt for the numerical integration of Didymos system barycenter and Dimorphos translational motion exclusively, as opposed to integrating Didymos and Dimorphos. This approach is adopted to mitigate additional numerical errors. The precise position and velocity of Didymos are uniquely determined by the orbit of Dimorphos and the masses of both bodies. Thus, Didymos' self-oblateness is inherently considered when activating Didymos' spherical harmonics in Dimorphos' integration. As a result, Didymos barycenter heliocentric orbit and Dimorphos' orbit with respect to Didymos barycenter were computed by numerically integrating the equations of motion using the aforementioned force models. The initial conditions for Didymos barycenter are taken from *sb-65083-205.bsp*, while the ones for Dimorphos are obtained from *dimorphos_s527.bsp* [NASA JPL SSD].

Furthermore, to better understand the dynamical environment within the Didymos system, the order-of-magnitude of the main gravitational and non-gravitational accelerations acting on Hera were computed at different distances from Didymos, see Figure 4.

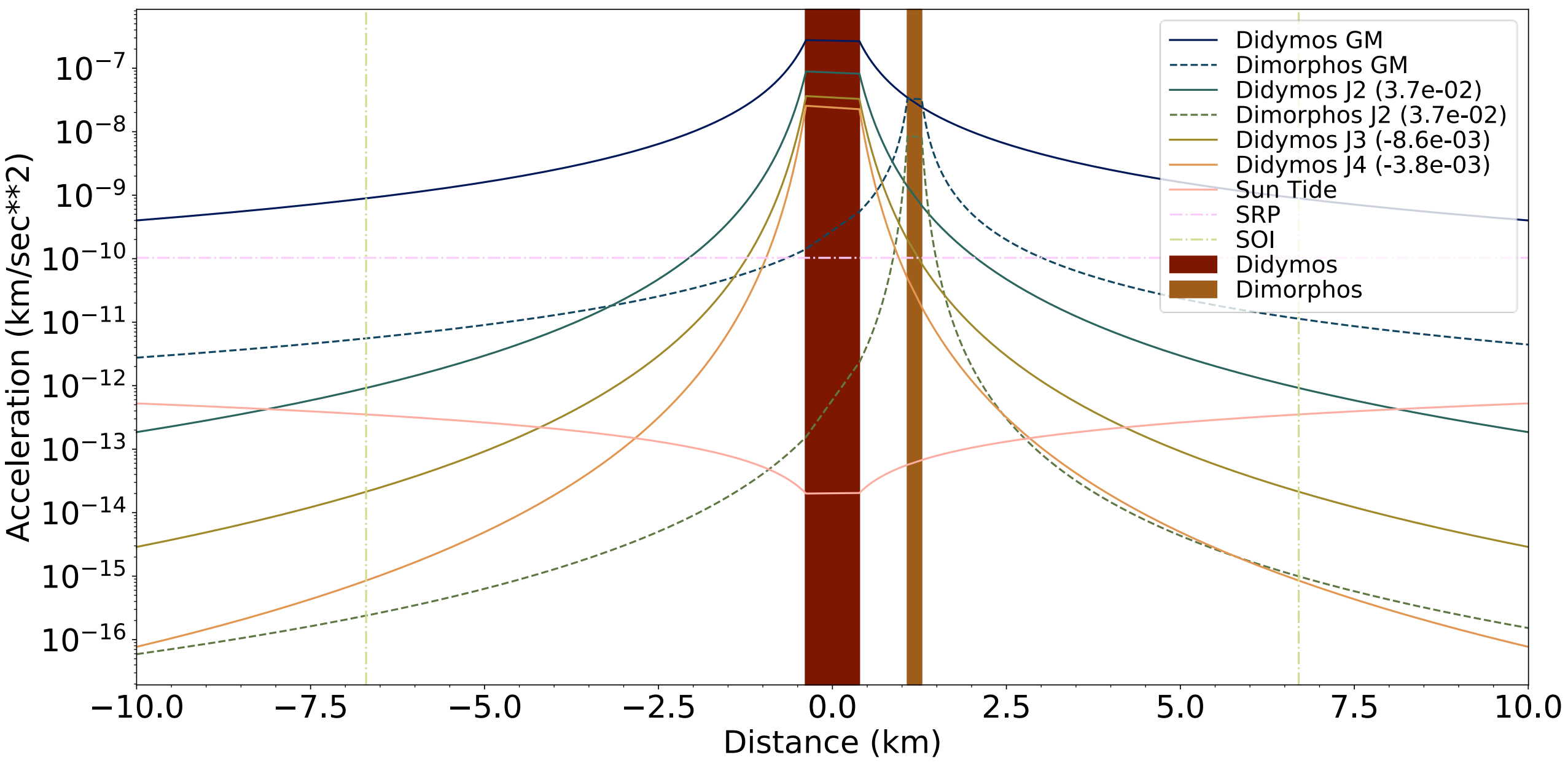

**Figure 4: Accelerations acting on Hera as a function of the distance from Didymos (x-axis).**

The force model for the spacecraft considers the point-mass gravitational perturbations of the Sun, the Moon, the eight planets, and Pluto, along with the point-mass gravity and spherical harmonics accelerations from Didymos and Dimorphos. Regarding the non-gravitational accelerations, the Solar Radiation Pressure acceleration (SRP) acting on the spacecraft is included using the spacecraft's realistic shape models. The total SRP on the spacecraft is computed as the sum of the forces acting on their components, each with different geometry, surface properties, and orientation with respect to the Sun. In addition, stochastic accelerations are considered for all the spacecraft to account mainly for the Thermal Recoil Pressure (TRP) and other non-gravitational accelerations that are not modeled in this work (e.g., albedo and thermal emissions from the asteroids). The TRP is considered to be in the order of 5-10% of the SRP (see Zannoni et al. (2021), Kato and Van der Ha (2012)). For the stochastic accelerations, we assumed a batch time of 18 hours[d] (equivalent to ¼ of each arc, on average) and a magnitude of $5\cdot10^{-12}$ km/s$^2$ (i.e., 5% of Hera SRP; Subsection 8.1.3 shows the influence of different stochastics in the formal uncertainties results) One of the primary sources of error in the OD, especially at low-gravity bodies, are the modeling errors of the non-gravitational accelerations. To account for these effects, a scale factor for the SRP and the three Cartesian components of the stochastic accelerations were estimated as a part of the OD process. The spacecraft orbits with respect to the system center of mass are then computed by numerically integrating the equations of motion adopting the described force models. Hera and CubeSats' initial states and attitudes are taken from the latest kernels released by ESA [Hera ESA Kernels]. Hera s/c dimensions are assumed to be 1.3x1.5x1.8 m, while Juventas and Milani are 6U CubeSats.

The following accelerations were not included in the integration because they were considered negligible at this stage: the non-gravitational accelerations on Dimorphos due to the solar radiation pressure, albedo, and thermal emissions, because of its low area-to-mass ratio.

The gravity spherical harmonics of the two asteroids are retrieved from the latest polyhedral shapes, which are reported in Figure 5, and following the standard and classical work by Werner and Scheeres (1996) (see also the methods described in Zannoni et al. (2018), and Wieczorek et al. (1998)).

[d] A series of estimates for each stochastic parameter is performed, one for each of its batches. Every 18 hours a set of 3 stochastic accelerations (X-Y-Z) are estimated. Each batch is independent from the others.

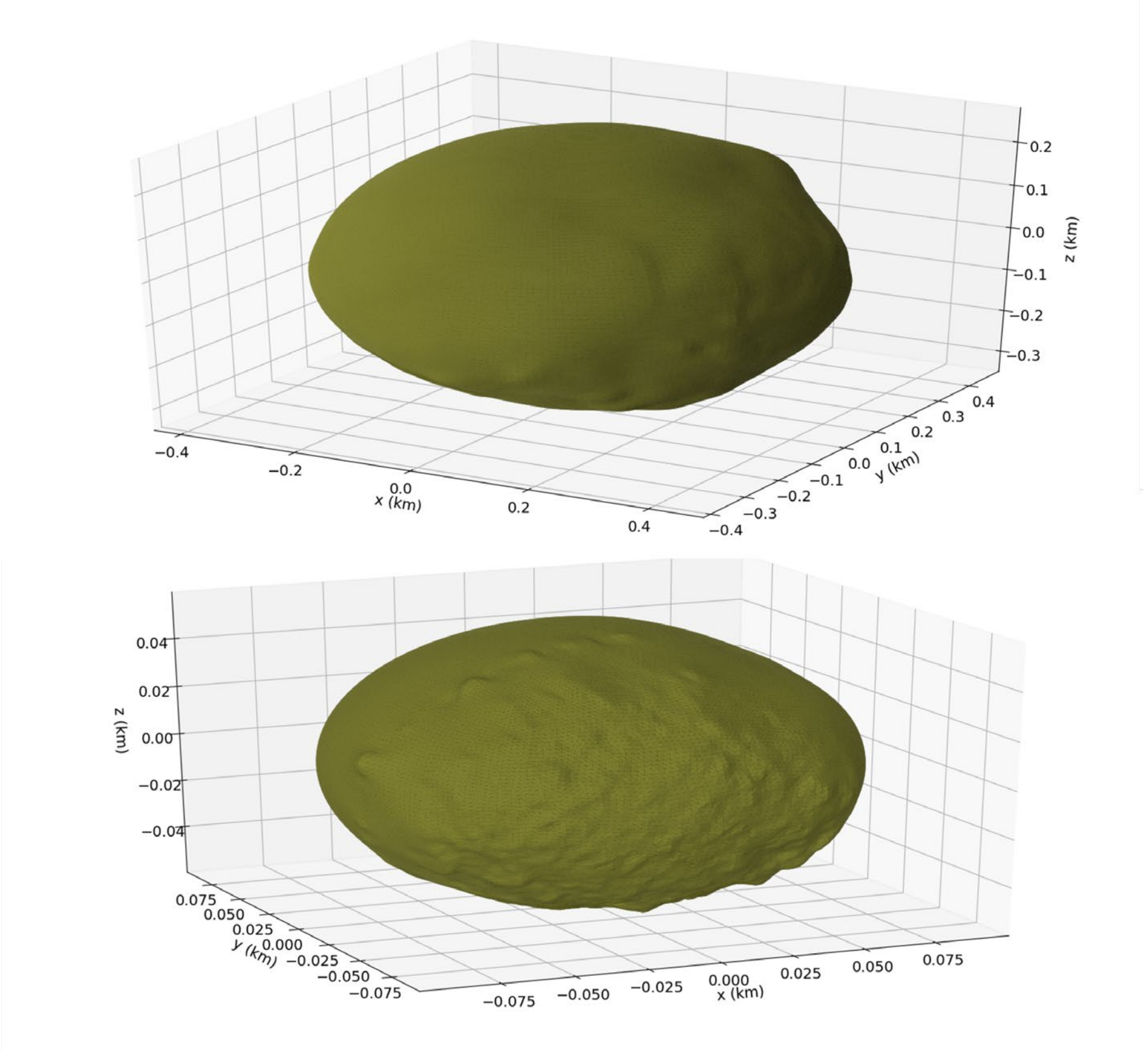

**Figure 5: Didymos (top) and Dimorphos (bottom) polyhedral shape models.**

Figure 6 depicts the gravity field coefficients, computed using a full 20-degree and 10-degree model for Didymos and Dimorphos, respectively (the same model was also adopted in the simulations). For reference, the corresponding un-normalized degree-2 gravity coefficients of Didymos and Dimorphos are collected in Table 6. The reference radii for the primary and the secondary are 0.43 km and 0.104 km, respectively. To this end, the gravitational masses (GM) of Didymos and Dimorphos were computed from their diameter ratio and the system's total mass [Fang and Margot, 2012], assuming the same density. The resulting primary and secondary GM are $4.0071 \cdot 10^{-8}$ $km^3/s^2$ and $3.4522 \cdot 10^{-10}$ $km^3/s^2$, respectively, with a mass ratio of about 0.0086, see Table 1.

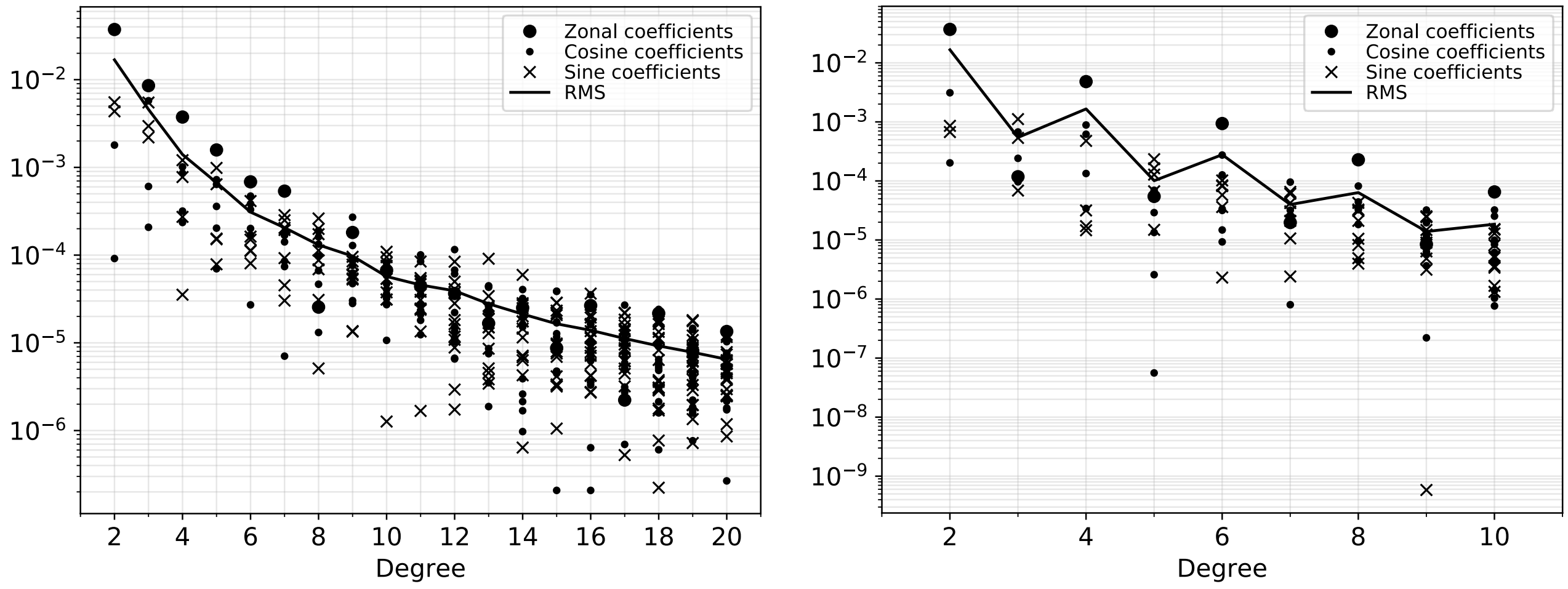

**Figure 6: Gravitational model of Didymos (left) and Dimorphos (right). Crosses and dots: absolute value of spherical harmonics normalized coefficients up to degree 20 and 10, respectively; continuous line: RMS of the coefficients for each degree.**

The rotational models of Didymos and Dimorphos greatly influence the gravitational accelerations acting on Dimorphos and the spacecraft. Therefore, accurately determining the pole orientation parameters is crucial to understanding the mutual interactions between the primary and secondary asteroids. The pole orientation of the bodies, expressed in the Earth Mean Orbit at J2000 (EMO2000) frame, is described by its right ascension α and declination δ values, which are modeled as linear functions of time:

$$\alpha = \alpha_0 + \alpha_1(t - t_0), \tag{12}$$

$$\delta = \delta_0 + \delta_1(t - t_0), \tag{13}$$

where the adopted reference epoch $t_0$ is February 09th 2027 at 16:01:09 TDB.

The orientation of the prime meridian of Didymos with respect to the node is described by the angle w, which is also modeled as a linear function of time (i.e., assuming a uniform rotation):

$$w = w_0 + w_1(t - t_0). \tag{14}$$

Conversely, the prime meridian of Dimorphos is expected to experience small librations, which are modeled using a sinusoid function with angular velocity ω, amplitude $w_a$, and phase ϕ:

$$w = w_0 + w_1(t - t_0) + w_a \sin(\omega(t - t_0) + \varphi). \quad (15)$$

The Didymos rotational model was taken from the Didymos Reference Asteroid (DRA) v5.01. In contrast, the rotational model of Dimorphos was built assuming a synchronous rotation around the primary, with the addition of an assumed libration at the orbital period of amplitude 5 deg. The corresponding numerical values defining the rotational models are collected in Table 2. A drift of less than 15 deg/century in Dimorphos' pole orientation was obtained from the pole dynamical fitting; the drift is caused by the perturbations induced by the Sun and the Earth on the relative orbit of the two asteroids. All the coefficients of the rotational models of Didymos and Dimorphos were estimated during the simulations.

**Table 2: Didymos and Dimorphos rotational models. The base reference frame is Earth Mean Orbit at J2000 (EMO2000). The reference epoch is February 09th, 2027, at 16:01:09 TDB.**

| Parameter | | Nominal value | Comments |
|---|---|---|---|
| **Didymos** | $\alpha_0$ | 311.0 deg | From DRA5.01 |
| | $\alpha_1$ | 0.0 deg/century | Not measured at present. Assumed zero |
| | $\delta_0$ | -79.8 deg | From DRA5.01 |
| | $\delta_1$ | 0.0 deg/century | Not measured at present. Assumed zero |
| | $w_0$ | 0.0 deg | This term defines the prime meridian. Assumed zero |
| | $w_1$ | 159.29 deg/hour | From rotational period [DRA5.01] |
| **Dimorphos** | $\alpha_0$ | -49.07 deg | Fitted to a dynamical synchronous model |
| | $\alpha_1$ | 13.24 deg/century | Fitted to a dynamical synchronous model |
| | $\delta_0$ | -79.79 deg | Fitted to a dynamical synchronous model |
| | $\delta_1$ | -0.98 deg/century | Fitted to a dynamical synchronous model |
| | $w_0$ | 13.13 deg | Fitted to a dynamical synchronous model |
| | $w_1$ | 31.35 deg/hour | Fitted to a dynamical synchronous model |
| | $w_a$ | 5.0 deg | Assumed |
| | $\omega$ | 30.35 deg/hour | Assumed equal to the average orbital period |
| | $\varphi$ | 6.19 deg | Fitted to a dynamical synchronous model |

# 6. Measurement model

Within the Hera RSE, four different measurement types are considered:

- Earth-based two-way range and Doppler via an X-band link (~8.4 GHz);
- Hera optical measurements of surface landmarks on both Didymos and Dimorphos, as well as Dimorphos centroid during ECP;
- Two-way range and Doppler measurements using the S-band Inter-Satellite Link (~2.4 GHz) between Hera and the Juventas and Milani CubeSats;
- Hera PALT LIDAR measurements to Didymos and Dimorphos landmarks and crossovers.

A detailed description of each measurement is provided in the following, and Table 3 summarizes the measurement uncertainties adopted in the simulations.

## 6.1. Radiometric measurements

During the ECP, we assume to collect Earth-Hera two-way range and Doppler measurements at X-band for 4 hours before and 4 hours after each maneuver at the corners in between arcs, corresponding to a hyperbolic trajectory arc of Hera, and for 8 hours around the closest approach (CA) to Didymos. This RSE assumption is conservative since actual operations foresee a 1x 8-hour pass each day to align with the ground operations schedule. We assume no thruster maneuvers, wheel de-saturations, or attitude corrections are performed during this time window. For the DCP, COP, and EXP phases, Earth-based radio tracking is taken only at the beginning and end of each arc. During low-altitude flybys (i.e., below 8 km), the AFC of Hera shall point at the target to allow for autonomous navigation, creating a conflict with the pointing requirements of the high-gain antenna. However, during the actual mission, a few arcs are expected to be fully dedicated to radio science, so the exclusion of radio tracking at CA may represent a conservative assumption.

The Earth-based Doppler noise level adopted in the simulations was computed using empirical models of the primary noise sources [Asmar et al., 2005; Iess et al., 2014a; Lasagni Manghi, 2021, 2023]. The main noise factors, whose value is shown in Figure 7 in terms of the Allan standard deviation, are the solar, interplanetary, and ionospheric plasma, the S/C electronics, the ground station instrumentation, and the Earth's troposphere. The solar plasma contribution, which depends on the Sun-Earth-Probe (SEP) angle, is assessed at the minimum and maximum SEP angles encountered during the Hera mission, corresponding to 20° and 60°, respectively, as indicated in Figure 8.

Furthermore, Figure 8 shows the total noise in terms of Allan standard deviation and range rate at 60 sec integration time. This integration time is usually adopted in gravity science studies as it balances numerical noise, computational burden, and sensitivity to gravity spherical harmonics [Zannoni et al., 2013]. Indeed, the spatial scale of a spherical harmonic of degree is $\Delta l = \frac{\pi}{l} r$ [Milani and Gronchi, 2010], where $r$ is Hera's flyby pericenter radius or Juventas SSTO radius. Then, the theoretical maximum sampling time to correctly reconstruct the gravity field of degree $l$ is $\Delta t = \frac{\Delta l}{v} = \frac{\pi}{l}\frac{r}{v}$ [Zannoni et al., 2020], where $v$ is Hera's flyby relative velocity or Juventas' orbital velocity. The required interaction time associated with the degree 5 (the maximum estimated in the OD filter for this study) is approximately 30 minutes due to a very slow relative dynamic (where $\Delta l$ is roughly 245 m, and assuming a relative velocity of 0.15 m/s, which is an approximate value valid for both Hera's flybys and Juventas orbital velocities). Therefore, a 60-second integration time is adopted, allowing sensitivity to the low-degree gravity field and avoiding numerical noise issues and high computational time. Figure 8 shows that the Doppler noise is dominated by the plasma noise, which increases towards the end of the mission due to lower SEP values. For this study, we conservatively assumed an Allan standard deviation of $1.67 \cdot 10^{-13}$ and applied a safety factor of 2, which results in an Earth-based Doppler uncertainty of 0.1 mm/sec at 60 sec integration time. Regarding the range, the measurements are collected every 300 seconds, with an assumed uncertainty of 1 m and an estimated measurement bias of 10 m for every tracking pass.

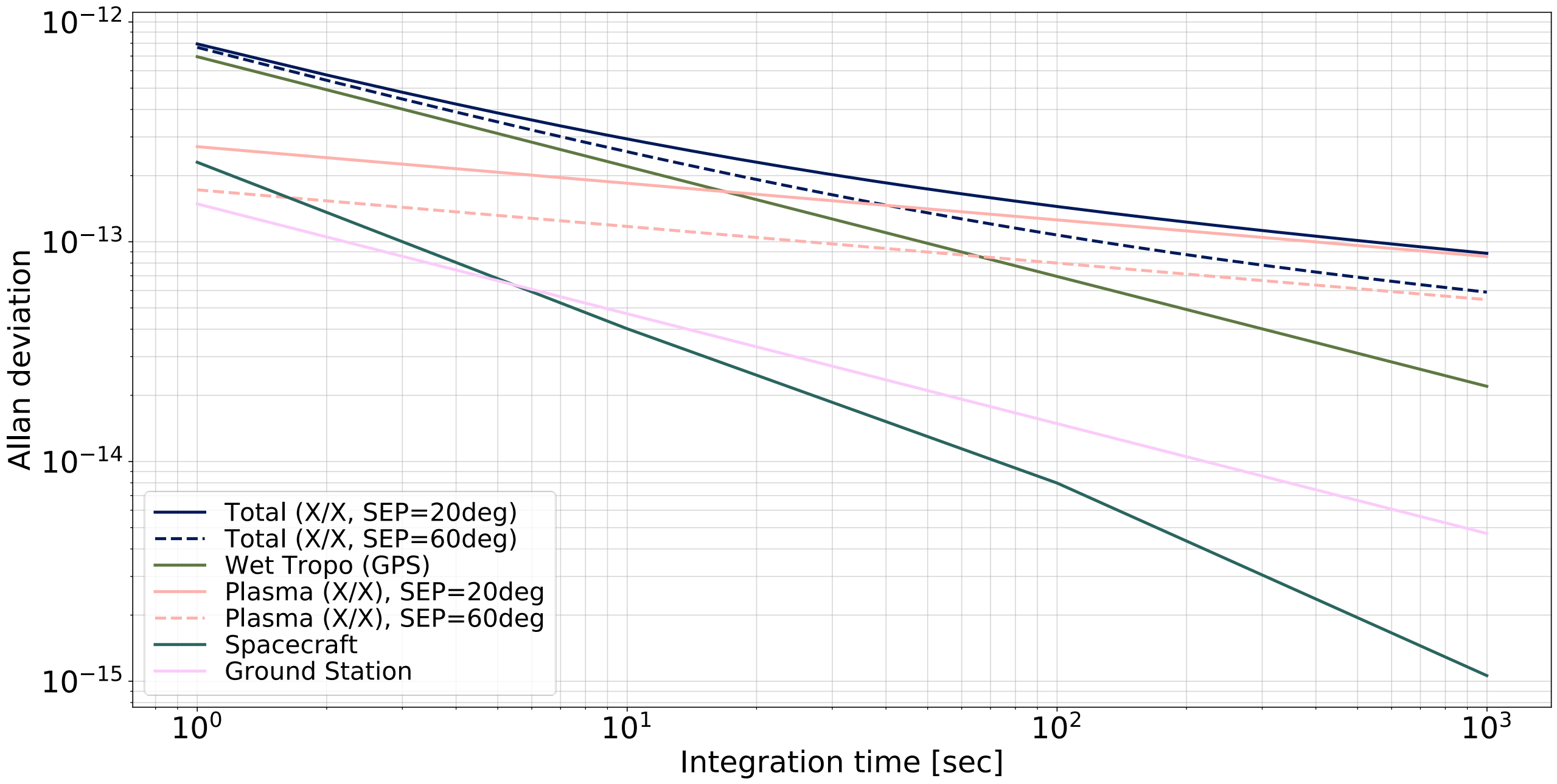

**Figure 7: Allan deviation of the main noise sources for the Hera mission. The noise contributions of the ground stations, the Earth's troposphere, and the plasma are evaluated using analytical models (Asmar et al., 2005; less et al., 2014a). The plasma model incorporates the effects of both the interplanetary plasma, computed at the minimum and maximum Sun-Earth-Probe (SEP) angles encountered by Hera (respectively 20° and 60°, see Figure 8), and the ionospheric plasma. The Hera spacecraft noise reflects the X-DST performance stability tests conducted by TAS-I. The total curves are obtained as the root squared sum of the individual contributions. The primary noise sources are attributed to Earth's troposphere and plasma.**

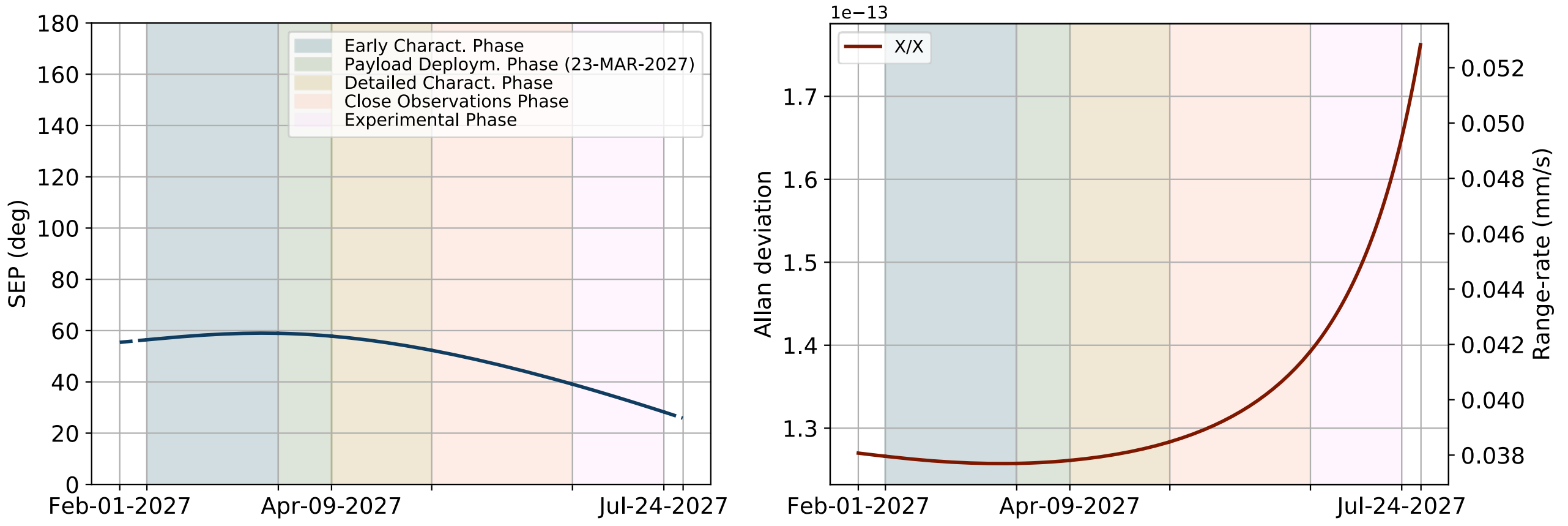

**Figure 8: Left: SEP angle during the Hera mission. Right: expected two-way Doppler noise during Hera mission at 60 sec integration time. The considered noise sources are the plasma, the Earth's atmosphere, the ground station instrumentation, and the S/C electronics. The dominant noise source is the solar plasma.**

Our simulations activate the ISL starting from the CubeSats deployment phase (PDP). Continuous ISL tracking (Doppler with 60 sec integration time and range), with a scheduled 40% duty cycle (DC) (i.e., tracking 2 minutes every 5 minutes of operations), is considered for the nominal results. The ISL observables are taken from Hera-Juventas and Hera-Milani, while the node Juventas-Milani cannot be exploited for mission technological constraints. The uncertainty associated with each measurement is 0.05 mm/s for the Doppler and 50 cm for the ranging. As explained earlier, since the system is still under development, we explored the impact of using different values of the ISL duty cycle and ISL Doppler noise, the results of which are shown in Subsection 8.1.2.

## 6.2. Optical measurements

Optical measurements of the Hera AFC are simulated in MONTE to improve the overall solution and better resolve the asteroids' rotational state, complementing the radiometric data. These measurements consist of sample and line coordinates in the camera images of specific surface features (or landmarks) and Dimorphos centroid to improve the system estimates during ECP, whose positions are defined in the body-fixed frames of the asteroids. For this study, we implemented 258 equally spaced surface landmarks on both Didymos and Dimorphos. The landmark coordinates are solve-for parameters within the OD filter, with *a priori* uncertainties derived from the values observed in the Rosetta mission [Godard et al., 2015]. Specifically, the uncertainty in radial direction is assumed to be $10\ \%$ of the body's mean radius, while the uncertainties in latitude and longitude are equal to $10^{-2}$ rad. When Hera radiometric tracking is inactive, we assume we will acquire one optical image every 2 hours. This represents a conservative assumption with respect to the likely mission operations.

The camera is pointed towards Didymos whenever its apparent diameter is within the camera field of view (FOV) and towards Dimorphos otherwise, as shown in Figure 9. When the camera is pointed at Didymos, we verify if Dimorphos' centroid and its associated landmarks are also in the FOV. Consequently, the targets of each optical image (i.e., the main body and the surface landmarks) are determined based on their presence within the camera's FOV at the time of measurement and on the local Sun Phase Angle (SPA), whose maximum value is set to 90 degrees. Notably, during the ECP and DCP phases, the pointing is always at Didymos, while during the COP, the pointing is switched toward Dimorphos when the flyby altitude is below ~8-10 km. The measurement uncertainty is again obtained by taking the values from the Rosetta mission and applying a safety factor of 2, resulting in a conservative sample and line noise of 2 pixels [Godard et al., 2015]. The S/C attitude uncertainty, while taking images, is considered through the pointing errors per picture, a set of three parameters (scan platform pointing error as a rotation about the M-axis, moving the picture up and down; about the N-axis, moving the picture left and right; about the L-axis, rotating the picture about its center) which are estimated for each picture with uncertainties set to 36 arcsec (0.01 deg).

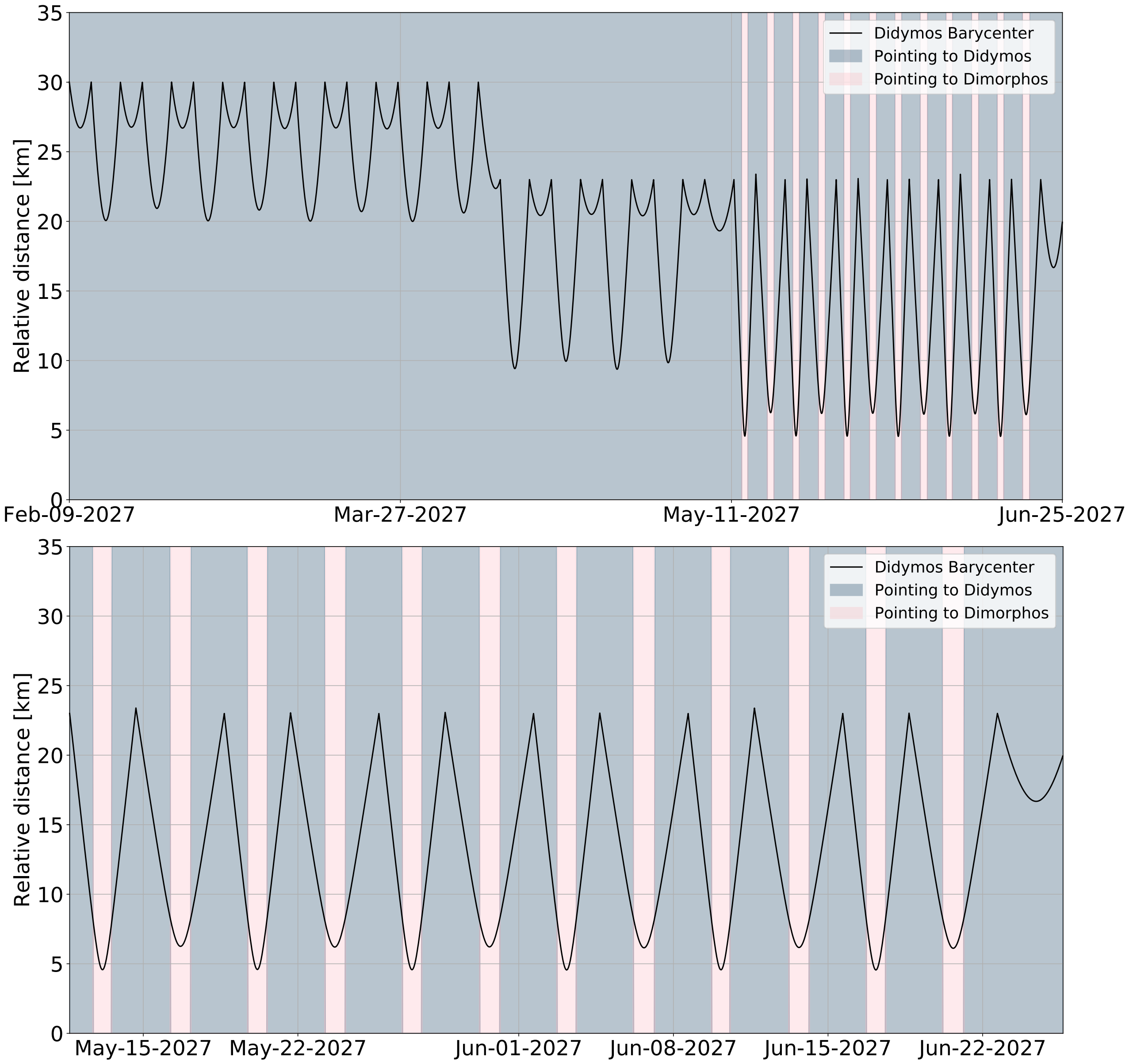

**Figure 9: Hera AFC pointing scheme for the optical measurements. Top: summary of the pointing scheme for all the mission phases; bottom: zoom on the COP.**

## 6.3. LIDAR measurements and crossovers

Various LIDAR instruments have been developed for space missions, including Hayabusa (2003), MESSENGER (2004) [Tsuno et al., 2017], Lunar Reconnaissance Orbiter (LRO) (2009) [Ramos et al., 2005, 2009], Hayabusa2 (2014) [Mizuno et al., 2017], and OSIRIS-REx (2016) [Daly et al., 2017]. These missions have provided invaluable insights and lessons for future endeavors such as the Hera mission. The Hayabusa mission primarily focused on rendezvous and touchdown control, with the LIDAR measuring distances from 50 m to 50 km with meter-level accuracy. Hayabusa2 expanded these capabilities by obtaining the asteroid's albedo in addition to distance measurements. Similarly, the OSIRIS-REx mission aims to return a sample from asteroid 101955 Bennu, utilizing the OSIRIS-REx Laser Altimeter (OLA) to measure the asteroid's shape and provide global maps of slopes. OLA's long- and short-range transmitters enable precise measurements at large and small distances.

Within the OD process, the Hera PALT LIDAR observables can be treated as a standard range measurement, namely:

$$Z_A = |\rho|, \tag{16}$$

where $\rho$ is the vector from the spacecraft to the surface of the illuminated target.

The LIDAR measurements allow a more precise reconstruction of Hera's trajectory. Hence, they can be combined with radiometric and optical observables to enhance the estimation of the gravity field of both bodies and their relative orbits. A key factor for improving the orbit reconstruction is represented by estimating a set of LIDAR crossovers. The crossovers are defined as surface landmarks probed multiple times by different LIDAR swaths and have been widely used in the past to improve the OD solutions provided by radiometric and optical data [Rowlands et al., 1999; Lemoine et al., 2001; Mazarico et al., 2010, 2014, 2015]. Estimating the crossovers within the OD process allows for the constraint of their initial position uncertainty on the surface, which is mainly driven by the asteroids' shape uncertainty. Knowing that the asteroid radius is constant at ground-track intersections (with only minor deformations caused by tides), these measurements serve as valuable constraints for determining and constraining the spacecraft's orbit, resulting in a more accurate reconstruction of Hera's position relative to the asteroid.

The LIDAR crossovers extraction procedure is illustrated in Figure 10. At first, a grid of evenly spaced points with a 2 m separation is defined on the surface of both Didymos and Dimorphos. Then, the Hera ground tracks are evaluated for both bodies according to the latest concept of operations, assuming one LIDAR measurement every 10 seconds and considering the LIDAR footprints. The LIDAR footprints, depicted in Figure 10 as blue circles, have varying radii that depend on the spacecraft's altitude. At an altitude of 14 km, the expected diameter of the LIDAR footprint is approximately 15.4 m, while at 10 km, it reduces to 11 m and further decreases to 5.5 m at 5 km altitude. A safety factor of 50% is incorporated into the LIDAR footprints, meaning the diameters are halved to exclude observations near the edges of the footprints. Finally, to identify the crossovers, denoted by a red cross in Figure 10, each time a LIDAR footprint intersects a specific grid point, the observation's date/time is recorded, and the counter for the illuminated grid point is incremented by one. After completing the process, the crossovers are identified as grid points characterized by a counter greater than one, indicating that they have been illuminated more than once by different LIDAR swaths. Conversely, grid points with a counter value of one are defined as standard landmarks. It is important to note that each selected observation considers only the first illuminated grid point, and we exclude measurements of the same grid point with a revisit time of less than 10 minutes. This avoids measuring different landmarks with the same observation and collecting multiple measurements of the same grid point in close succession.

A trade-off between the desired accuracy and the computational time was required to select the number of retrieved crossovers. As the number of crossovers increases, the computational time for covariance analysis also increases due to the increased number of estimated parameters and partial derivatives to be computed and stored. Therefore, a trade-off between the desired accuracy and the computational time was required. This work presents a scenario with 400 LIDAR landmarks on both bodies, 1068 crossovers on Dimorphos, and 1720 crossovers on Didymos over the nominal mission. The crossovers' locations are shown in Figure 11.

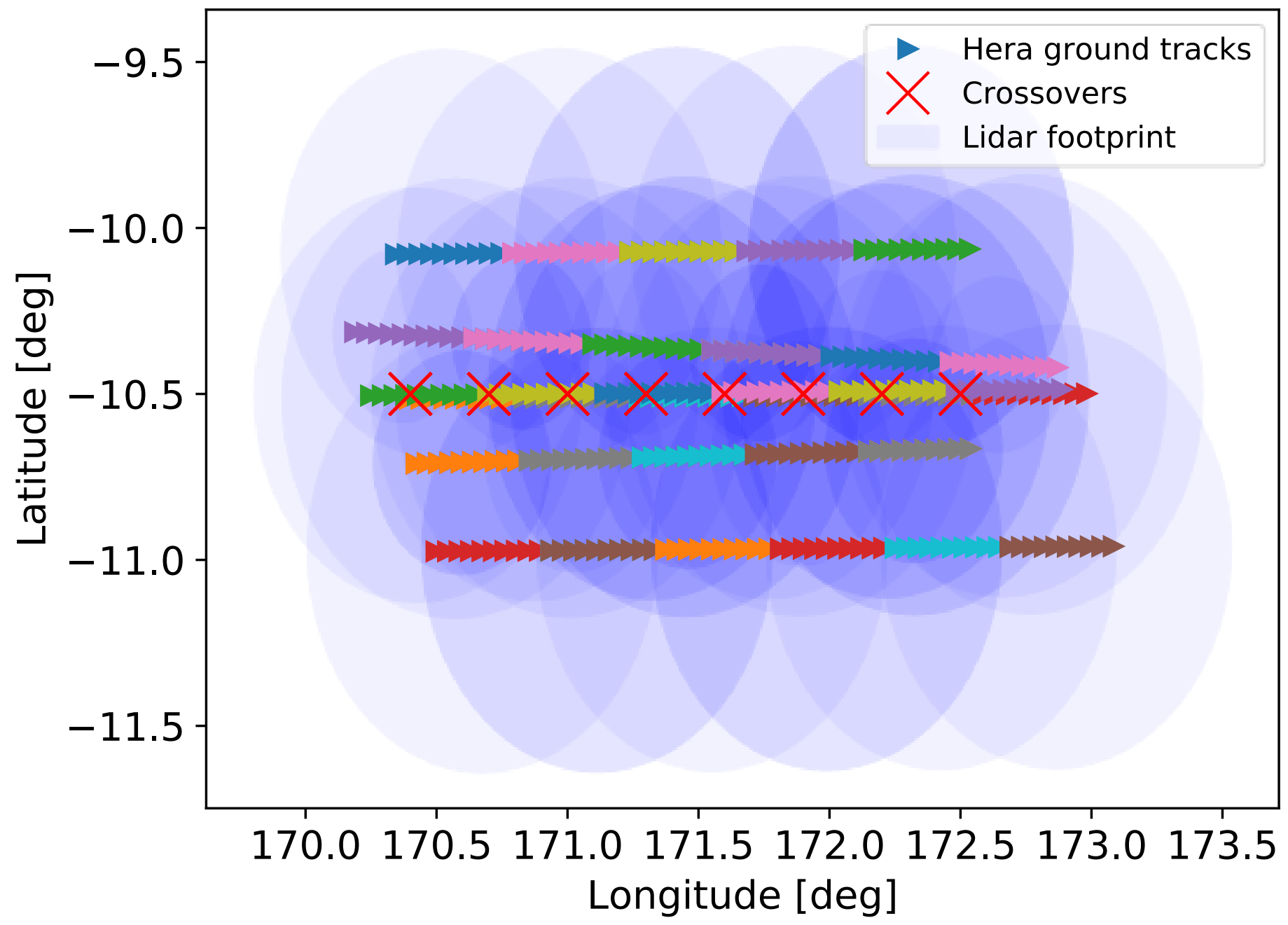

**Figure 10: Graphical representation of the LIDAR crossovers extraction procedure. Each set of colored arrows represents a 10 min subset of Hera's ground track from a specific arc. The blue circles depict the LIDAR footprints, which vary in radius based on the spacecraft's altitude. The red crosses represent our user-defined grid points. When a specific LIDAR footprint intersects a grid point, we record the observation's date/time and increment the counter only of the first illuminated grid point. After the process, crossovers are identified as grid points with a counter greater than 1, indicating they were illuminated multiple times by different LIDAR swaths. Grid points with a counter value of 1 are defined as standard landmarks.**

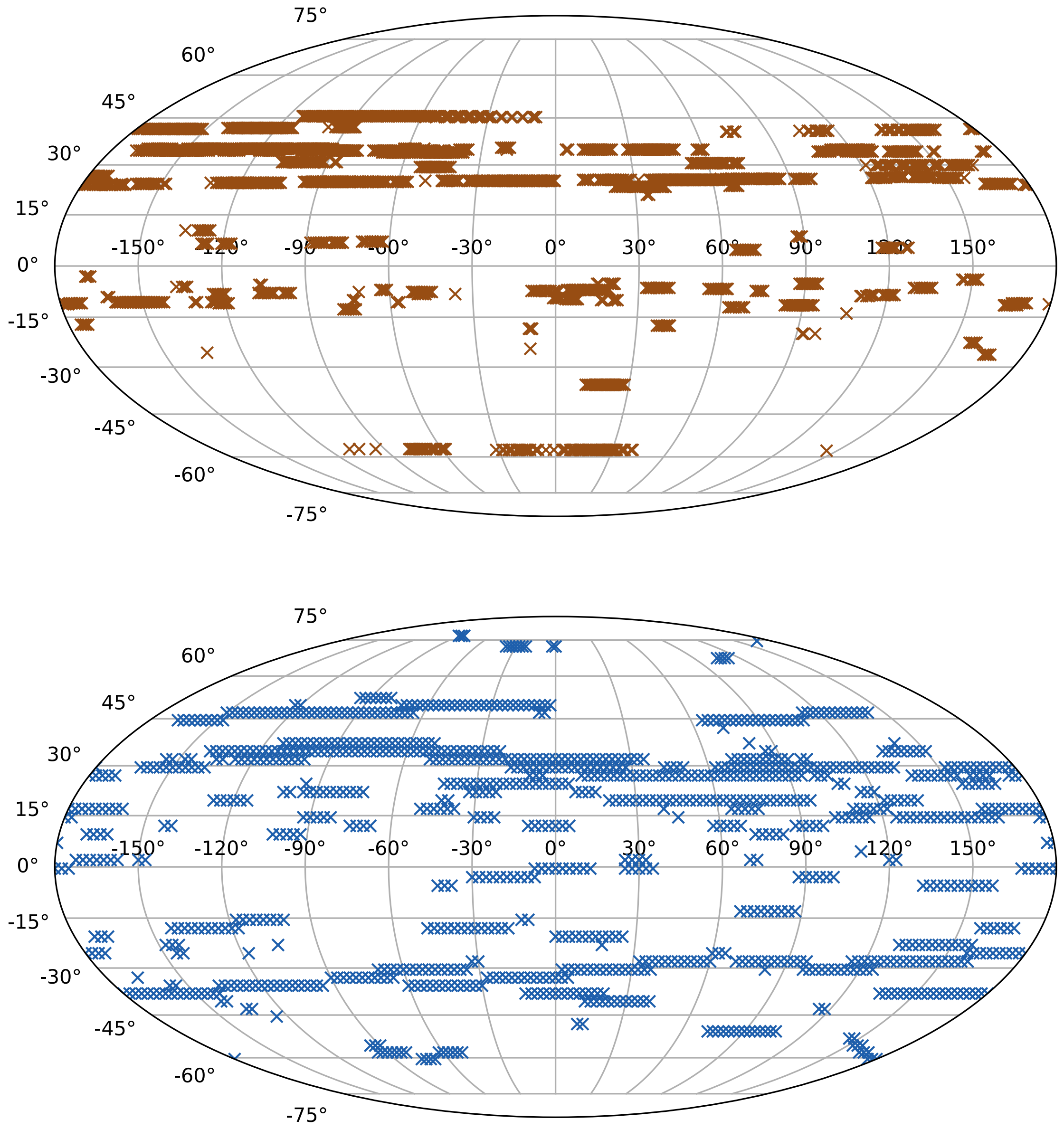

**Figure 11: Ground track of the LIDAR crossovers used for the simulation. Top: Didymos; bottom: Dimorphos. The number of crossovers on Didymos and Dimorphos is 1720 and 1068, respectively, over the entire nominal mission. The a priori uncertainty of the crossover coordinates, estimated within the orbit determination process, is 50 m in the body frame X-Y-Z coordinates.**

The uncertainty of the LIDAR measurements, primarily influenced by the instrument's electronic performance, is expected to improve as the altitude decreases. However, a conservative constant value of 50 cm has been considered [Gramigna et al., 2023a]. A common LIDAR bias is estimated for all measurements, with an *a priori* uncertainty of 200 m (widely open). The initial position uncertainty of the crossovers on the asteroids' surface is set to 50 m in the body-fixed Cartesian coordinates.

**Table 3: Summary of the measurement uncertainties adopted in Hera RSE (1σ).**

| Earth-based tracking | ISL tracking | Optical-AFC camera | Lidar altimeter |
|---|---|---|---|
| Doppler: 0.1 mm/sec at 60 sec integration time | Doppler: 0.05 mm/sec at 60 sec integration time | Sample and line accuracy: 2 pixels (1 pixel from Rosetta mission [Godard et al., 2015] with a safety factor of 2) | Instrument accuracy: 50 cm [Gramigna et al., 2023a] |
| Range: 1.0 m<br>(one measurement every 300 sec)<br>Range Bias: 10 m on each tracking passage (estimated) | Range: 50 cm<br>(one measurement every 60 sec) | Spacecraft attitude: 0.01 deg = 36 arcsec (from Rosetta mission [Godard et al., 2015]) | Bias = 200 m (common to all measurements, estimated) |

## 7. Filter setup

The primary objective of the sensitivity analyses in this study is to estimate the anticipated formal uncertainties associated with critical scientific parameters of interest throughout the Hera mission. These parameters include the gravity field coefficients of Didymos and Dimorphos, their relative orbit and rotational states, and the heliocentric orbit of the system. The spacecraft's state, SRP scale factors, pointing error per picture, and other relevant local solve-for parameters are also estimated. Table 4 provides a comprehensive summary of the estimated parameters within our setup.

**Table 4: Filter setup for the OD simulations of the Hera mission.**

| Parameter | Type | A priori uncertainty (1σ) | Notes |
|---|---|---|---|
| **Spacecrafts state** | | | Hera, Juventas, and Milani states estimated with respect to the Didymos system barycenter. Widely open to account for maneuver errors at the beginning/end of the arcs |
| **Position** | Local | 10 km | |
| **Velocity** | Local | 0.1 m/sec | |
| **Didymos Barycenter state** | | | Didymos barycenter with respect to the Sun. Widely open. |
| **Position** | Global | 100 km | |
| **Velocity** | Global | 1.0 cm/sec | |
| **Dimorphos state** | | | Dimorphos state estimated with respect to Didymos system barycenter. Widely open. |
| **Position** | Global | 0.4 km | |
| **Velocity** | Global | 3 cm/sec | |
| **Didymos Gravity** | | | |
| **GM** | Global | $3.57\cdot10^{-8}$ km$^3$/s$^2$ | Widely open. From the measured uncertainty in the total mass of the Didymos system, scaled by a factor of 10. |
| **$J_2$** | Global | $8.35\cdot10^{-1}$ | $J_2$, $C_{22}$, $S_{22}$, widely open. 1000% of the un-normalized nominal coefficients. |
| **$C_{22}$** | Global | $5.91\cdot10^{-4}$ | |
| **$S_{22}$** | Global | $2.82\cdot10^{-2}$ | |
| **$C_{21}$** | Global | $1.46\cdot10^{-2}$ | $C_{21}$, $S_{21}$, Un-normalized coefficient uncertainty. Computed assuming a maximum misalignment of 10 deg between the body-fixed frame and the principal inertia axes. Widely open. |
| **$S_{21}$** | Global | $1.46\cdot10^{-2}$ | |
| **Dimorphos Gravity** | | | |
| **GM** | Global | $5.88\cdot10^{-10}$ km$^3$/s$^2$ | Widely open. From the measured uncertainty in the total mass of the Didymos system and the diameter |
| | | | ratio between Didymos-Dimorphos, scaled by a factor of 10. |
| **$J_2$** | Global | $8.24 \cdot 10^{-1}$ | $J_2$, $C_{22}$, $S_{22}$, widely open. 1000% of the un-normalized nominal coefficients. |
| **$C_{22}$** | Global | $2.01 \cdot 10^{-2}$ | |
| **$S_{22}$** | Global | $7.01 \cdot 10^{-4}$ | |
| **$C_{21}$** | Global | $1.51 \cdot 10^{-2}$ | $C_{21}$, $S_{21}$, Un-normalized coefficient uncertainty. Computed assuming a maximum misalignment of 10 deg between the body-fixed frame and the principal inertia axes. Widely open. |
| **$S_{21}$** | Global | $1.37 \cdot 10^{-2}$ | |
| **Didymos Frame** | | | |
| ***$\alpha_0$*** | Global | 13.0 deg | From the measured uncertainties, scaled by a factor of 5. Widely open. |
| ***$\delta_0$*** | Global | 2.5 deg | |
| ***$\alpha_1$*** | Global | $3.5 \cdot 10^{-2}$ deg/hour | Equal to w1 uncertainty divided by 10. Typically, it is much smaller than the uncertainty in the rotational period. |
| ***$\delta_1$*** | Global | $3.5 \cdot 10^{-2}$ deg/hour | |
| ***$w_1$*** | Global | $3.5 \cdot 10^{-1}$ deg/hour | From the measured uncertainty in the post-impact rotational period, scaled by a factor of 5. |
| **Dimorphos Frame** | | | |
| ***$\alpha_0$*** | Global | 13.0 deg | From the measured uncertainties, scaled by a factor of 5. Widely open. |
| ***$\delta_0$*** | Global | 2.5 deg | |
| ***$\alpha_1$*** | Global | $4.2 \cdot 10^{-3}$ deg/hour | Equal to w1 uncertainty divided by 10. Typically, it is much smaller than the uncertainty in the rotational period. |
| ***$\delta_1$*** | Global | $4.2 \cdot 10^{-3}$ deg/hour | |
| ***$w_1$*** | Global | $4.2 \cdot 10^{-2}$ deg/hour | From the orbital period uncertainty, scaled by a factor of 5. |
| ***$w_a$*** | Global | 50.0 deg | Widely open |
| ***$\omega$*** | Global | $4.2 \cdot 10^{-2}$ deg/hour | Equal to $w_1$ uncertainty |
| ***φ*** | Global | 50.0 deg | Widely open |
| **Solar radiation pressure** | | | |
| **Scale factor (Hera)** | Local | 1.0 | 100% of the acceleration |
| **Scale factor (Juventas)** | Local | 1.0 | 100% of the acceleration |
| **Scale factor (Milani)** | Local | 1.0 | 100% of the acceleration |
| **Stochastic accelerations** | | | |
| **Hera (X-Y-Z)** | Local | $5 \cdot 10^{-12}$ km/s$^2$ | Set of 3 stochastic accelerations (X-Y-Z) estimated in uncorrelated batches of 18 hours |
| **Juventas (X-Y-Z)** | Local | $5 \cdot 10^{-12}$ km/s$^2$ | Set of 3 stochastic accelerations (X-Y-Z) estimated in uncorrelated batches of 18 hours |
| **Milani (X-Y-Z)** | Local | $5 \cdot 10^{-12}$ km/s$^2$ | Set of 3 stochastic accelerations (X-Y-Z) estimated in uncorrelated batches of 18 hours |
| **Pointing error per picture** | | | |
| **AFC axes rotation** | Local | 0.1 deg (36 arcsec) | From Rosetta [Godard et al., 2015] |
| **Didymos optical landmarks position** | | | |
| **Scale factor** | Global | 0.1 | 10% size scale. Common to all landmarks |
| **Radius** | Global | 39.0 m | 10% of the mean radius in all directions |
| **Lat, Long** | Global | 5.73 deg | |
| **Dimorphos optical landmarks position** | | | |
| **Scale factor** | Global | 0.1 | 10% size scale. Common to all landmarks |
| **Radius** | Global | 8.2 m | 10% of the mean radius in all directions |
| **Lat, Long** | Global | 5.73 deg | |
| **Didymos LIDAR landmarks and crossovers position** | | | |
| **Scale factor** | Global | 0.1 | 10% size scale. Common to all landmarks |
| **Radius** | Global | 50 m | Equivalent to 50 m in X-Y-Z body-fixed |
| **Lat, Long** | Global | 7.5 deg | |
| **Didymos LIDAR landmarks and crossovers position** | | | |
| **Scale factor** | Global | 0.1 | 10% size scale. Common to all landmarks |
| **Radius** | Global | 50 m | Equivalent to 50 m in X-Y-Z body-fixed |
| **Lat, Long** | Global | 37.5 deg | |

Didymos' and Dimorphos' gravity field coefficients are estimated up to degree and order 5 and 3, respectively. For better clarity, Table 4 reports the a priori uncertainties in Didymos' coefficients up to degree and order two only. For higher degrees, the a priori uncertainties are the RMS of each degree scaled by a factor of 5.

# 8. Results

This Section presents the simulation results of the Hera radio science investigations exploiting standard planetary radio science methods. Section 8.1 discusses the results of the nominal mission scenario, which incorporates Earth-based and satellite-to-satellite radiometric measurements and Hera's optical observables. Within this Section, we conducted a parametric analysis to assess the influence of various factors on the scientific parameters of interest. These factors include the mission phases (see Subsection 8.1.1, where we also incorporated the experimental phase), ISL duty cycle and ISL Doppler noise (Subsection 8.1.2), and spacecraft stochastic accelerations (Subsection 8.1.3). Finally, Section 8.2 examines the effects of LIDAR measurements and crossovers' estimation in the OD process within Hera's nominal mission.

## 8.1. Nominal mission scenario

The results for Hera's nominal mission, comprising the ECP, DCP, and COP phases (see Figure 1), are presented herein. These results incorporate the Earth-based Doppler and range measurements, the satellite-to-satellite Doppler and range measurements, AFC optical images to surface landmarks of both asteroids, and centroid data to Dimorphos in ECP.

Figure 12 illustrates the uncertainty in Didymos's recovered gravity field at the end of Hera's nominal mission under different simulation scenarios. The plot provides a quick understanding of the expected performance in retrieving the spherical harmonic coefficients. The black curve represents the magnitude of the simulated gravity field for each degree. At the same time, the colored lines indicate the accuracy of each test case, measured as the RMS of the estimated formal covariances at a given degree. The corresponding curve must lie below the simulated field curve to observe a specific degree of Didymos' gravity field.

Figure 12 shows that using only Earth-based radiometric measurements, the extended gravity field of Didymos is not observable. However, the addition of the AFC optical measurements (*Hera-only* curve) allows to better constrain the relative position of Hera within the Didymos system, enabling the observability of the degree two, as well as the bodies' rotational states and Dimorphos librations. Under this scenario, the overall formal uncertainty for degree two is approximately 50%. The inclusion of ISL measurements with the CubeSats (*Nominal* curve), particularly Juventas with its lower altitude orbits, plays a crucial role in reducing the uncertainty of degree two to about 0.5% and expanding the gravity estimation to degree three and potentially degree four, depending on the selected ISL duty cycle (see Subsection 8.1.2).

The same results are also displayed in Table 5 and Table 6, which show the estimated GMs and spherical harmonics coefficients of Didymos and Dimorphos for the Hera-only and Nominal scenarios of Figure 12, respectively. Table 5 shows that, within the *Hera-only* scenario, the GM of Didymos and Dimorphos can be recovered with a formal uncertainty of 0.068% and 0.93%, respectively. At the same time, the degree 2 is not observable. Furthermore, adding the ISL improves Didymos' and Dimorphos' GMs of about one order of magnitude, with a formal uncertainty of 0.004% and 0.079%, respectively, see Table 6. Dimorphos' mass accuracy provided by the Hera RSE thus meets the requirement of 10% relative error and the goal of 1%. For their $J_2$ coefficients, the corresponding 1σ uncertainties from the *Nominal* scenario are 0.084% and 6.6%, respectively.

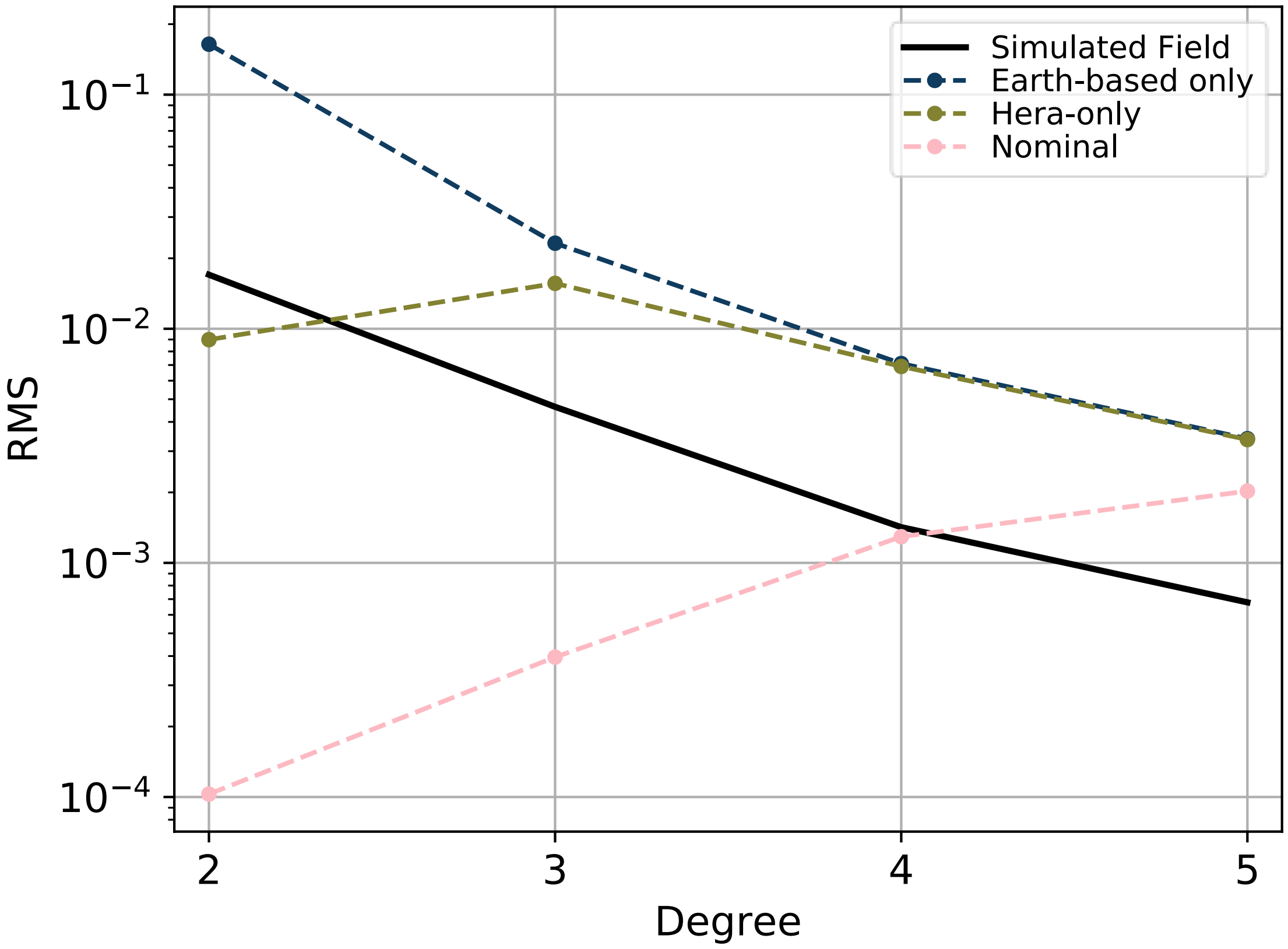

**Figure 12: Power spectra of the extended gravity field of Didymos. Continuous line: simulated field; dashed lines: formal uncertainty of the estimated field at the end of the nominal mission under different scenarios. Blue: Earth-based Doppler and range; green: same as blue with the addition of Hera optical navigation images; pink: same as green with the addition of ISL Doppler and range.**

**Table 5: Summary of the estimated formal 1σ uncertainties for GMs and un-normalized spherical harmonics coefficients of the *Hera-only* mission scenario (without ISL measurements from the CubeSats) of Figure 12.**

| Coefficient | Nominal value | Formal uncertainty (1σ) | Relative uncertainty |
|---|---|---|---|
| **Didymos** | | | |
| **GM** ($km^3/s^2$) | $4.0071\cdot10^{-8}$ | $2.7368\cdot10^{-11}$ | 0.068 % |
| **$J_2$** | $8.35\cdot10^{-2}$ | $3.92\cdot10^{-2}$ | 46.95 % |
| **$C_{21}$** | $2.32\cdot10^{-3}$ | $5.12\cdot10^{-3}$ | not observable |
| **$S_{21}$** | $-7.14\cdot10^{-3}$ | $5.05\cdot10^{-3}$ | 70.73 % |
| **$C_{22}$** | $-5.91\cdot10^{-5}$ | $5.87\cdot10^{-4}$ | not observable |
| **$S_{22}$** | $-2.82\cdot10^{-3}$ | $5.17\cdot10^{-3}$ | not observable |
| **Dimorphos** | | | |
| **GM** ($km^3/s^2$) | $3.4522\cdot10^{-10}$ | $3.2033\cdot10^{-12}$ | 0.93 % |
| **$J_2$** | $8.24\cdot10^{-2}$ | $6.71\cdot10^{-1}$ | not observable |
| **$C_{21}$** | $2.61\cdot10^{-4}$ | $3.02\cdot10^{-3}$ | not observable |
| **$S_{21}$** | $8.67\cdot10^{-4}$ | $3.80\cdot10^{-3}$ | not observable |
| **$C_{22}$** | $2.01\cdot10^{-3}$ | $3.32\cdot10^{-5}$ | 1.65 % |
| **$S_{22}$** | $-5.53\cdot10^{-4}$ | $2.39\cdot10^{-5}$ | 4.32 % |

**Table 6: Summary of the estimated formal 1σ uncertainties for GMs and un-normalized spherical harmonics coefficients of the *Nominal* mission scenario of Figure 12. This table highlights the improvement provided by the ISL in the GM and spherical harmonics coefficients.**

| Coefficient | Nominal value | Formal uncertainty (1σ) | Relative uncertainty |
|---|---|---|---|
| **Didymos** | | | |
| **GM** ($km^3/s^2$) | $4.0071\cdot10^{-8}$ | $1.6125\cdot10^{-12}$ | 0.004 % |
| **$J_2$** | $8.35\cdot10^{-2}$ | $7.02\cdot10^{-5}$ | 0.084 % |
| **$C_{21}$** | $2.32\cdot10^{-3}$ | $8.14\cdot10^{-5}$ | 3.51 % |
| **$S_{21}$** | $-7.14\cdot10^{-3}$ | $6.99\cdot10^{-5}$ | 0.98 % |
| **$C_{22}$** | $-5.91\cdot10^{-5}$ | $9.93\cdot10^{-5}$ | not observable |
| **$S_{22}$** | $-2.82\cdot10^{-3}$ | $9.44\cdot10^{-5}$ | 3.35 % |
| **$J_3$** | $-2.27\cdot10^{-2}$ | $2.37\cdot10^{-4}$ | 1.04 % |
| **$C_{31}$** | $-6.22\cdot10^{-3}$ | $1.40\cdot10^{-4}$ | 2.25 % |
| **$S_{31}$** | $-5.94\cdot10^{-3}$ | $1.39\cdot10^{-4}$ | 2.34 % |
| **$C_{32}$** | $-7.10\cdot10^{-5}$ | $1.21\cdot10^{-4}$ | not observable |
| **$S_{32}$** | $1.01\cdot10^{-3}$ | $1.20\cdot10^{-4}$ | 11.9 % |
| **$C_{33}$** | $8.47\cdot10^{-5}$ | $8.83\cdot10^{-5}$ | not observable |
| **$S_{33}$** | $-3.07 \cdot 10^{-4}$ | $8.90 \cdot 10^{-5}$ | 29.0 % |
| **Dimorphos** | | | |
| **GM** ($km^3/s^2$) | $3.4522 \cdot 10^{-10}$ | $2.7321 \cdot 10^{-13}$ | 0.079 % |
| **$J_2$** | $8.24 \cdot 10^{-2}$ | $5.44 \cdot 10^{-3}$ | 6.6 % |
| **$C_{21}$** | $2.61 \cdot 10^{-4}$ | $2.22 \cdot 10^{-3}$ | not observable |
| **$S_{21}$** | $8.67 \cdot 10^{-4}$ | $2.45 \cdot 10^{-3}$ | not observable |
| **$C_{22}$** | $2.01 \cdot 10^{-3}$ | $2.53 \cdot 10^{-5}$ | 1.26 % |
| **$S_{22}$** | $-5.53 \cdot 10^{-4}$ | $2.27 \cdot 10^{-5}$ | 4.11 % |

Table 7 shows the formal 1σ uncertainties achievable for Didymos and Dimorphos pole parameters. The nominal covariance analysis conducted in this study reveals that, based on the RSE data, we anticipate fitting Didymos' spin pole right ascension and declination to approximately 1 degree (in line with the requirements), with a spin rate accuracy in the order of $4 \cdot 10^{-6}$ deg/hour. The prime meridian $w_0$ of Didymos can be determined from the estimation of the optical landmarks group offset, a scale factor characterized by an uncertainty of $6.15 \cdot 10^{-3}$, while the prime meridian rate uncertainty is $1.5 \cdot 10^{-6}$ deg/hour. For Dimorphos, the right ascension and declination accuracies are 0.33 deg and 0.15 deg (which also satisfy the requirements), with rate accuracies of $1.24 \cdot 10^{-4}$ deg/hour and $5.41 \cdot 10^{-5}$ deg/hour, respectively. The uncertainties for Dimorphos' prime meridian and prime meridian rate are $6.32 \cdot 10^{-3}$ deg and $1.21 \cdot 10^{-4}$ deg/hour, respectively. Regarding Dimorphos' libration, the expected 1σ uncertainties are in the order of $2.28 \cdot 10^{-2}$ deg for the libration amplitude, $7.63 \cdot 10^{-4}$ deg/hour for the angular velocity (equivalent to 1.1 seconds for the libration's period), and 2.06 degrees for the phase.

Furthermore, the estimation of optical landmarks reveals that the height uncertainty for Didymos landmarks is approximately 2.4 m. For Dimorphos, the landmarks at latitudes below 65 deg N are recovered with uncertainties in the order of 0.5 m, while uncertainties of roughly 8 m characterize the ones in the northern region.

The accurate estimation of Didymos and Dimorphos spin states, facilitated by the RSE, holds significant scientific importance for three main reasons among the others: firstly, it provides insights into the rotational state of the bodies after the DART impact; secondly, it allows for the determination of non-gravitational accelerations, such as YORP, acting on the bodies; thirdly, it enables the estimation of wobbling and librations in Dimorphos, which in turn allow the estimate of the moments of inertia of the body (see Section 9).

**Table 7: Pole parameters: summary of formal 1σ uncertainties achievable for the Hera nominal mission scenario of Figure 12.**

| Coefficient | Nominal value | A posteriori 1σ |
|---|---|---|
| **Didymos** | | |
| $\boldsymbol{\alpha_0}$ (deg) | 311.00 | 1.00 |
| $\boldsymbol{\delta_0}$ (deg) | -79.80 | 0.82 |
| $\boldsymbol{\alpha_1}$ (deg/hour) | 0.00 | $4.15 \cdot 10^{-6}$ |
| $\boldsymbol{\delta_1}$(deg/hour) | 0.00 | $3.40 \cdot 10^{-6}$ |
| $\boldsymbol{w_1}$ (deg/hour) | 159.29 | $1.47 \cdot 10^{-6}$ |
| **Dimorphos** | | |
| $\boldsymbol{\alpha_0}$ (deg) | -49.07 | 0.33 |
| $\boldsymbol{\delta_0}$ (deg) | -79.80 | 0.15 |
| $\boldsymbol{\alpha_1}$ (deg/hour) | $1.51 \cdot 10^{-5}$ | $1.24 \cdot 10^{-4}$ |
| $\boldsymbol{\delta_1}$(deg/hour) | $-1.11 \cdot 10^{-6}$ | $5.41 \cdot 10^{-5}$ |
| $\boldsymbol{w_1}$ (deg/hour) | 31.35 | $1.21 \cdot 10^{-4}$ |
| $\boldsymbol{w_a}$ (deg) | 5.00 | $2.28 \cdot 10^{-2}$ |
| $\boldsymbol{\omega}$ (deg/hour) | 30.35 | $7.63 \cdot 10^{-4}$ |
| $\boldsymbol{\varphi}$ (deg) | 6.19 | 2.06 |

Furthermore, Figure 13 shows the position uncertainties of Hera, measured in the Radial Tangential Normal (RTN) frame relative to Didymos. During the ECP phase, Hera exhibits position uncertainties in the order of meters. Following the deployment of the CubeSats, Hera's position uncertainty decreases to the sub-meter level thanks to the ISL tracking before returning to a meter-level uncertainty when the CubeSats conclude their mission. Similarly, Figure 14 shows Juventas position uncertainty in the same RTN frame with respect to Didymos. During the initial 3.3 km SSTO orbit, the uncertainty is in the order of a few meters, while as the CubeSat transitions towards its lower 2.0 km SSTO orbit, the position uncertainty gradually decreases, reaching the sub-meter level in all three components. On the other hand, due to its higher altitude flybys, Milani is characterized by higher position uncertainties between the meter and kilometer levels (see Figure 15).

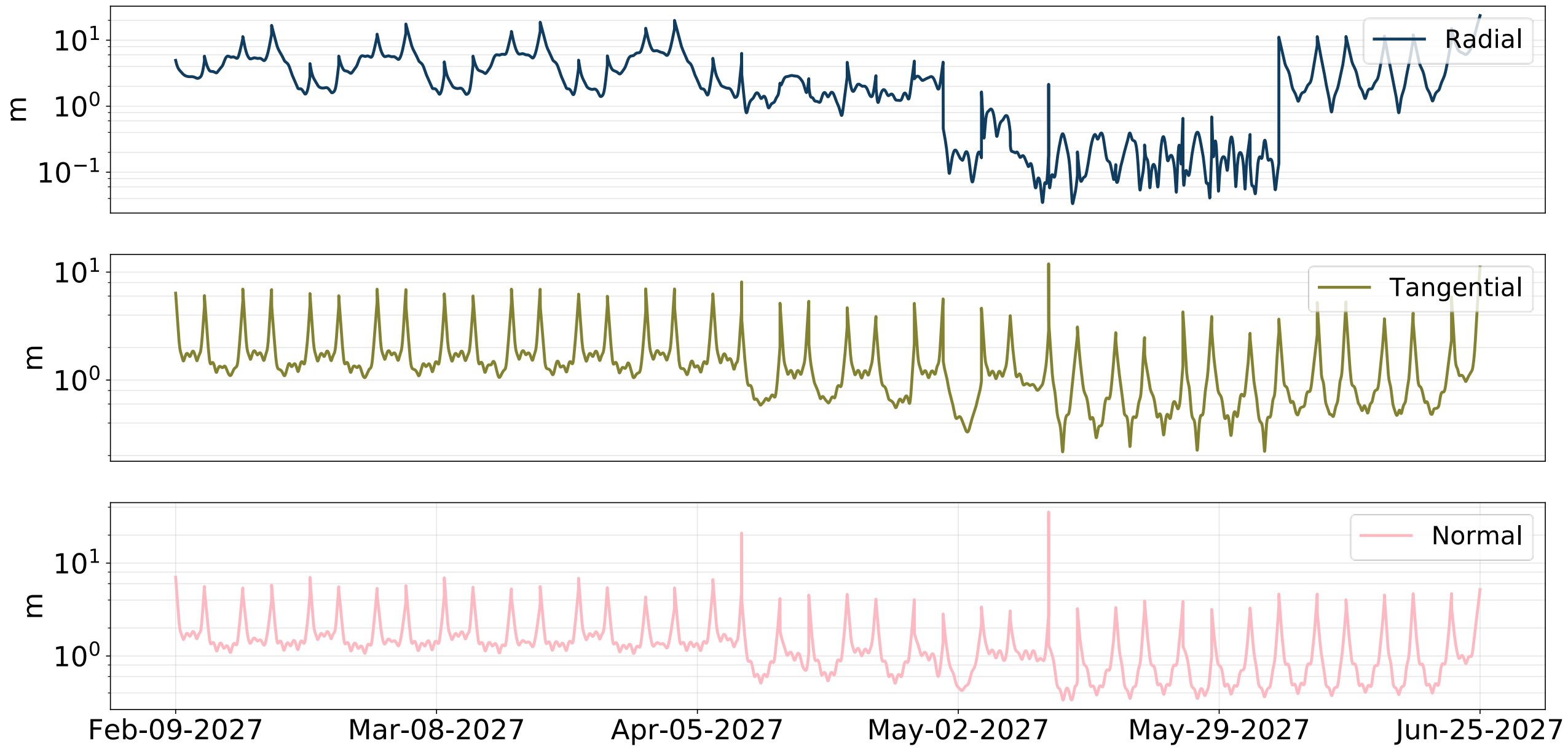

**Figure 13: Expected accuracy of Hera's reconstructed orbits. 1σ position uncertainty with respect to Didymos in the radial, along-track, and normal directions.**

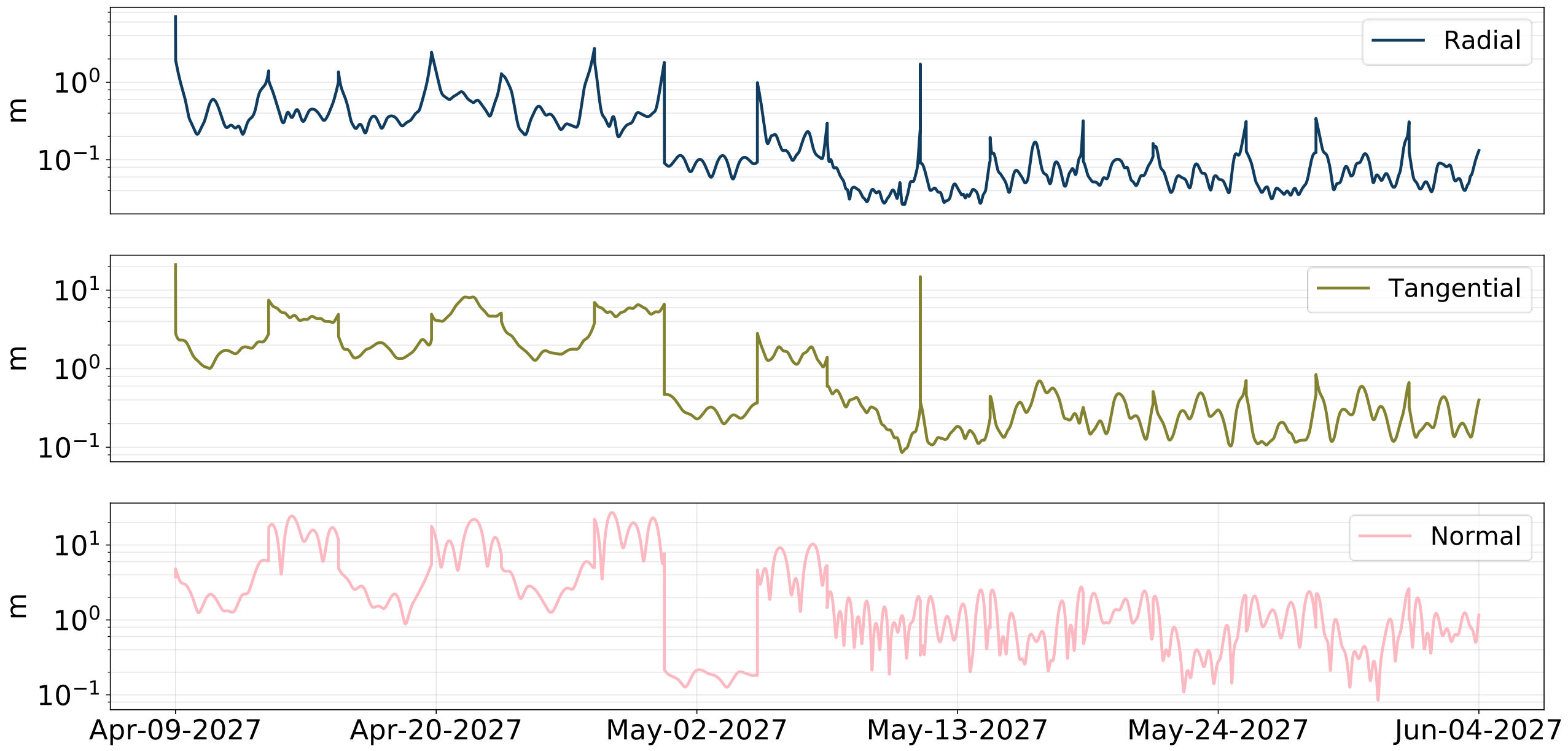

**Figure 14: Expected accuracy of Juventas reconstructed orbit. 1σ position uncertainty with respect to Didymos in the radial, along-track, and normal directions.**

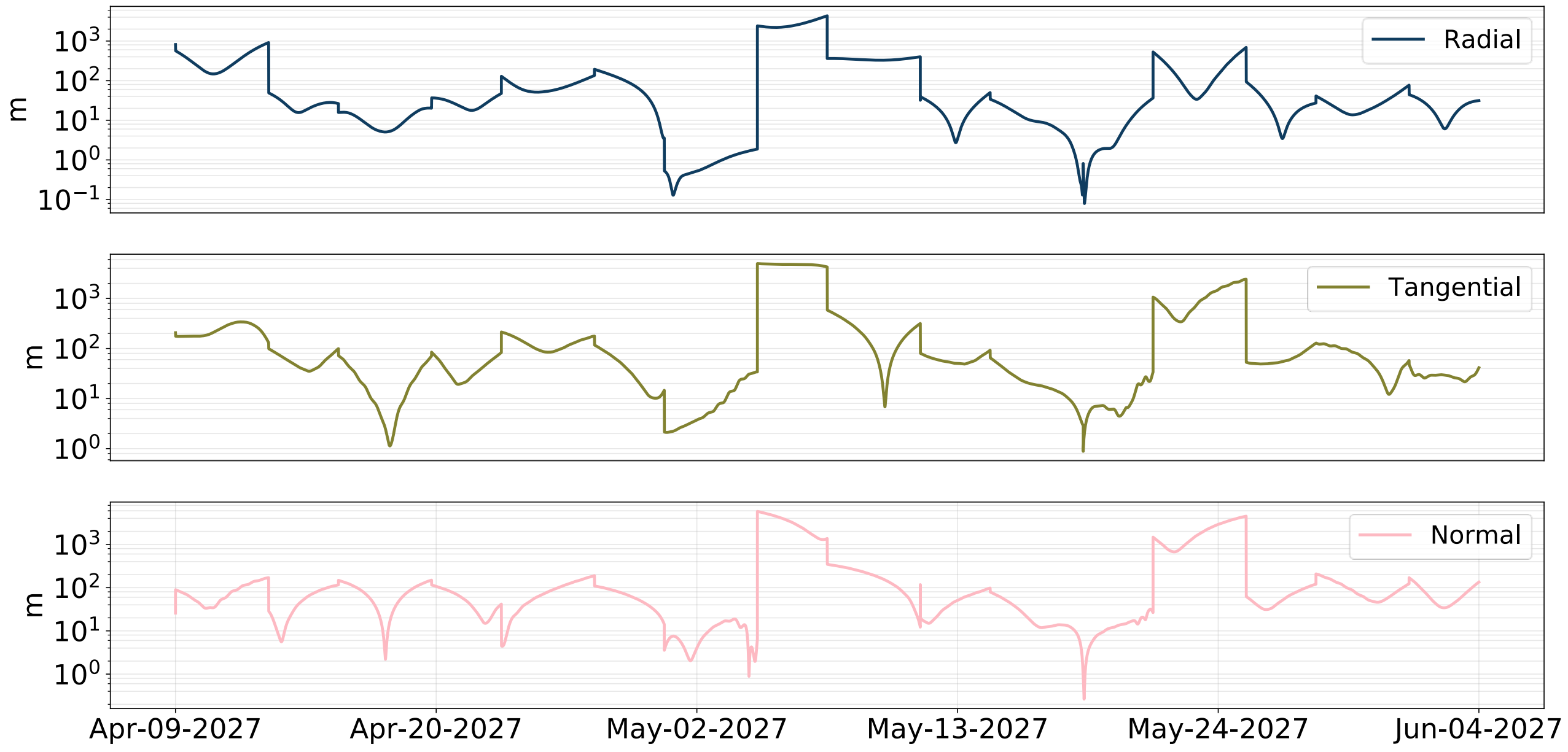

**Figure 15: Expected accuracy of Milani reconstructed orbit. 1σ position uncertainty with respect to Didymos in the radial, along-track, and normal directions.**

Figure 16 and Figure 17 illustrate the formal uncertainty of Dimorphos' position and velocity in the RTN frame with respect to Didymos. The estimated position uncertainties are at the sub-meter level throughout the mission, with approximately 20 cm for the radial and normal components and 30-100 cm for the along-track component. In contrast, the velocity uncertainties are in the order of $10^{-1}$-$10^{-2}$ mm/sec. The corresponding semimajor-axis and eccentricity uncertainties are in the order of $10^{-1}$ m and $10^{-4}$, respectively, and the implications of these results will be explored in more detail in Section 9 (see Figure 23). Overall, the accuracies in the reconstructed relative orbit satisfy the RSE requirement outlined in Section 3. In this context, the CubeSats' optical images could further improve Dimorphos' relative orbit.

Lastly, Figure 18 shows the formal 1σ position uncertainty of Didymos with respect to the Sun in the RTN components. The expected ephemerides uncertainty provided by Hera data is estimated to vary between 10-100 m for the radial and tangential components and 1-2 km for the normal component. This results in eccentricity and semimajor axis uncertainties of $10^{-10}$ and 240 m, respectively. The ephemerides uncertainties of this work are obtained by estimating Didymos barycenter, Dimorphos, and the spacecraft, using all the observables (including the Earth-based range), as explained earlier. To get a more accurate estimation of the heliocentric orbit, a standard and classical approach is the one discussed in Konopliv et al., 2006, 2018 and Park et al., 2017, which will be followed during the operational Hera mission to reconstruct Didymos ephemerides precisely. First, we will estimate the orbit of Hera with respect to Didymos' center of mass (COM), including Didymos' and Dimorphos' gravity fields, and without using Hera's Earth-based range data. With the estimated orbit for Hera, we will adjust the Earth-based range measurements to Didymos COM, obtaining the so-called pseudorange data. Subsequently, we will process the pseudorange data along with other Didymos observations, such as radar astrometry data, to estimate the heliocentric orbit of Didymos. To evaluate the expected accuracy of the Hera pseudorange data, we first performed a simulation without estimating Didymos barycenter and without including Earth-based range data (we only use Earth-based Doppler, Hera AFC images, and ISL Doppler/range); second, using this simulation, we evaluated the Line of Sight (LOS) uncertainty of Hera with respect to Didymos as seen from the Earth; third, we computed the pseudorange accuracy as the root sum square (RSS) of the expected Earth-based range uncertainty (1-2 m) and the Hera-Didymos LOS position uncertainty, which from our results is expected to be about 1 m. Considering a safety factor of 2, this procedure results in an expected Hera pseudorange data accuracy of ~ 5 m.

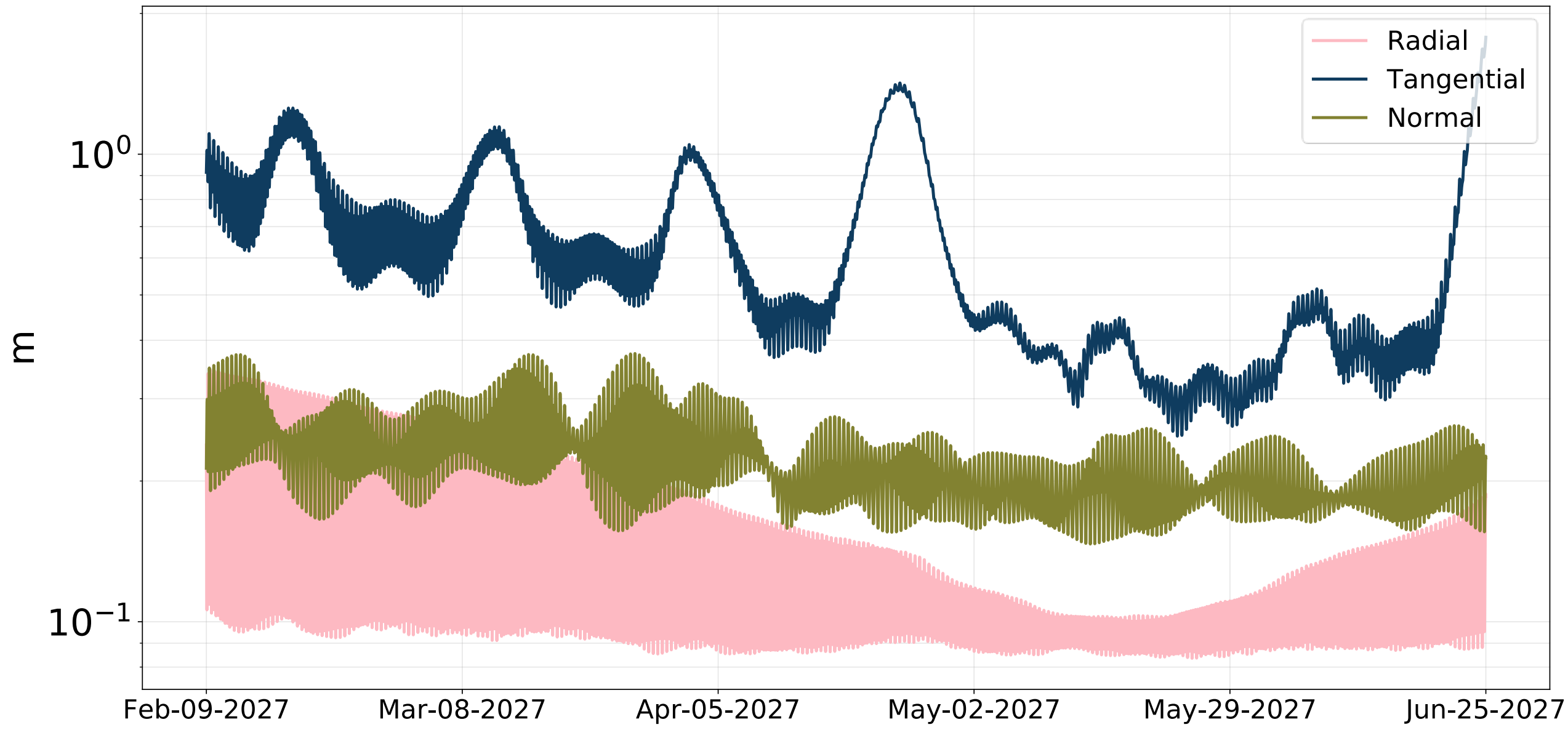

**Figure 16: Expected accuracy of Dimorphos' orbit. 1σ position uncertainty with respect to Didymos in the radial, along-track, and normal directions.**

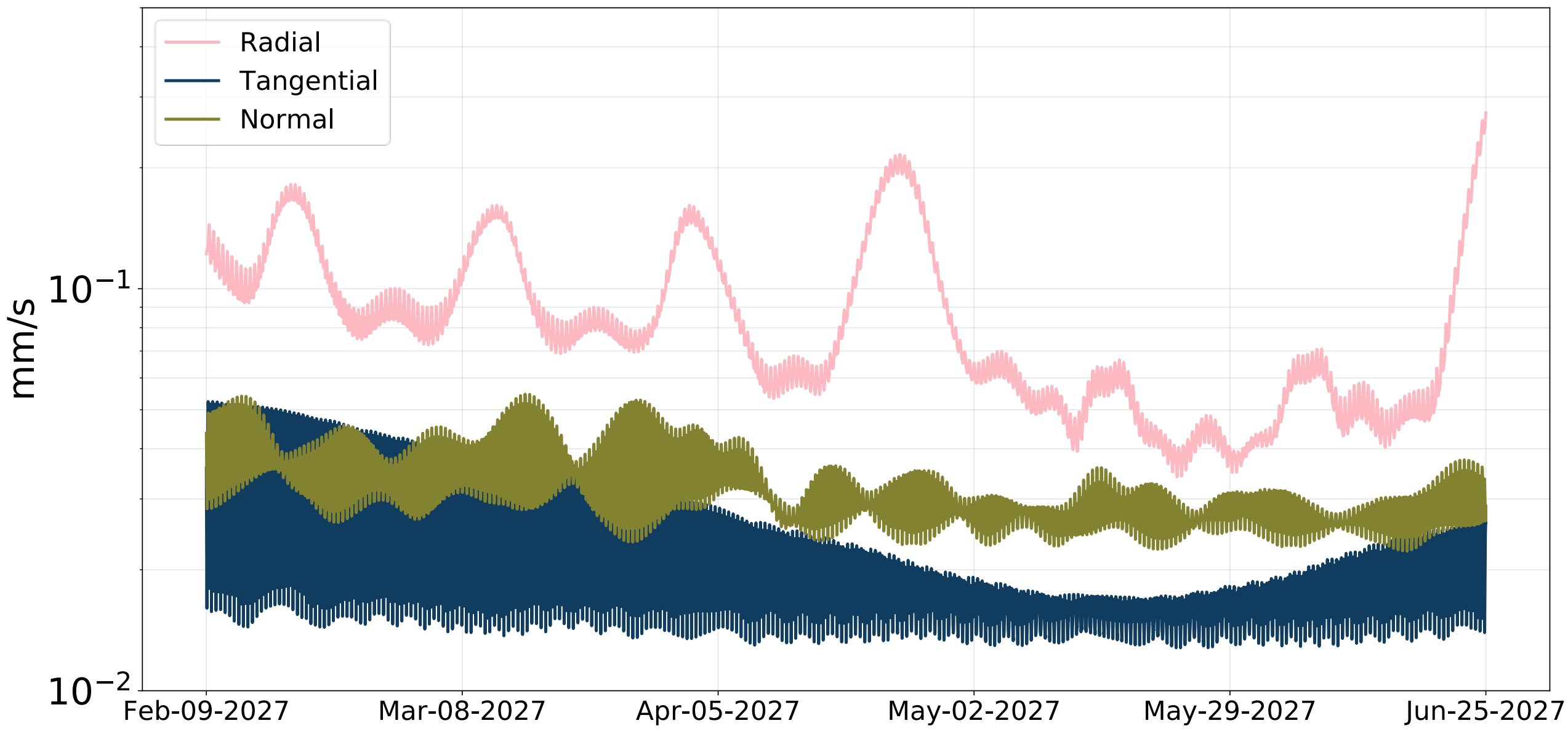

**Figure 17: Formal uncertainty (1σ) of Dimorphos velocity with respect to Didymos in the radial, along-track, and normal directions.**

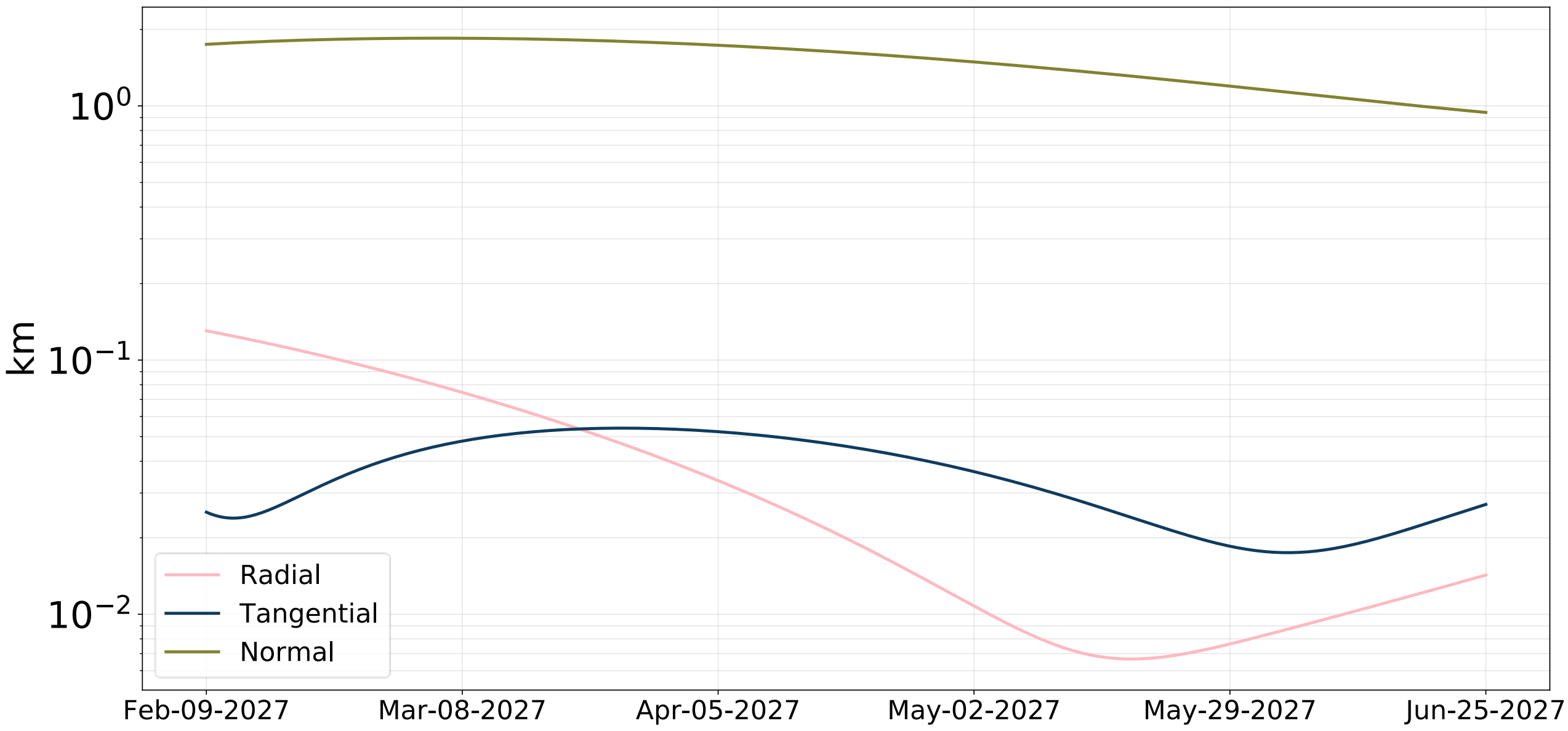

**Figure 18: Expected accuracy of Didymos heliocentric orbit. 1σ position uncertainty with respect to the Sun in the radial, along-track, and normal directions.**

### *8.1.1* Mission phases

Simulations were conducted to assess the influence of each mission phase on the estimation accuracies as the mission progresses. The results are shown in Figure 19 and Table 8. During the ECP, Hera's considerable distance from the Didymos system only allows for estimating Didymos' GM and $J_2$ with a formal uncertainty of 0.11% and 57.91%, respectively. For Dimorphos, the corresponding value for the GM is 2.18%, while the $J_2$ is not observable. Transitioning to the DCP, the deployment of CubeSats and utilization of ISL observables result in a one-two order of magnitude improvement in gravity reconstruction. Formal uncertainties of Didymos' GM and $J_2$ at the end of the DCP are 0.015% and 0.20%, respectively. Including the odd-order terms, the recovery of Didymos' degree two harmonics exhibits an overall uncertainty of roughly 1%. Within this phase, the $J_2$ of Dimorphos becomes observable with a formal uncertainty of 16.6%. The COP phase, thanks to Juventas' 2 km altitude SSTO orbit, further improves the mass and gravity field recovery.

It should be noted that the experimental phase does not significantly improve the accuracy of the gravity field obtained in the Nominal mission at the end of the COP. Although a couple of EXP flybys occur at lower altitudes, compared to the COP flybys, the distance is still too high to enhance the information gathered in the previous phases significantly. In this context, the ISL Doppler measurements from Juventas play a crucial role, particularly during its 2 km altitude orbit around Didymos.

Nevertheless, the additional Hera low-altitude flybys enable higher-resolution optical measurements of the asteroids and their landmarks, resulting in an improved understanding of their rotational dynamics during the EXP phase, see Table 8. Specifically, most of the pole parameters for Didymos and Dimorphos improve by a factor between 1.2-2.1 compared to the estimations obtained at the end of the COP. For example, the recovered Didymos right ascension and declination uncertainties are 0.79 degrees and 0.60 degrees, respectively, with rates of $3.3{\cdot}10^{-6}$ and $2.5{\cdot}10^{-6}$ deg/hour deg/hour. In the case of Dimorphos, the uncertainties for the same parameters are 0.27 degrees, 0.07 degrees, $9.4 \cdot 10^{-5}$ deg/hour, and $2.4 \cdot 10^{-5}$ deg/hour, with the 1σ prime meridian rate at $9.1 \cdot 10^{-5}$ deg/hour.

The estimation of Dimorphos' libration also improves by a factor between 1.3-1.9 at the end of the EXP: the libration magnitude uncertainty drops to $1.8 \cdot 10^{-2}$ degrees, the phase uncertainty is 1.2 degrees, and the libration angular velocity is $4.0 \cdot 10^{-4}$ deg/hour (equivalent to an uncertainty of 0.5 seconds in the libration period).

To conclude, Table 8 also shows the evolution of the uncertainties in Dimorphos' orbital period and position. The orbital period is characterized by an uncertainty of 157 seconds at the end of the ECP and drops to 11.2, 4.4, and 4.3 seconds at the end of the DCP, COP, and EXP, respectively. Regarding the position uncertainty (provided in RTN components with respect to Didymos), the improvement factor at the end of the COP, with respect to the ECP, is between 2.2 and 4.9 for all the components. The final RTN uncertainties at the end of the EXP are in the order of 0.2-0.5 m.

Hera flybys (or Juventas orbit) at altitudes lower than 1.5 km would be required to increase the gravity field accuracies further. Additionally, the imaging and tracking of the CubeSats during their landing phase and the monitoring of ejecta particles, if present (see Chesley et al. 2020 for the methodology), have the potential to increase our understanding of the higher-order gravity fields of Didymos and Dimorphos. Moreover, tracking the CubeSats after their landing on the surface of the asteroids may be crucial to improving the recovery of their rotational state.

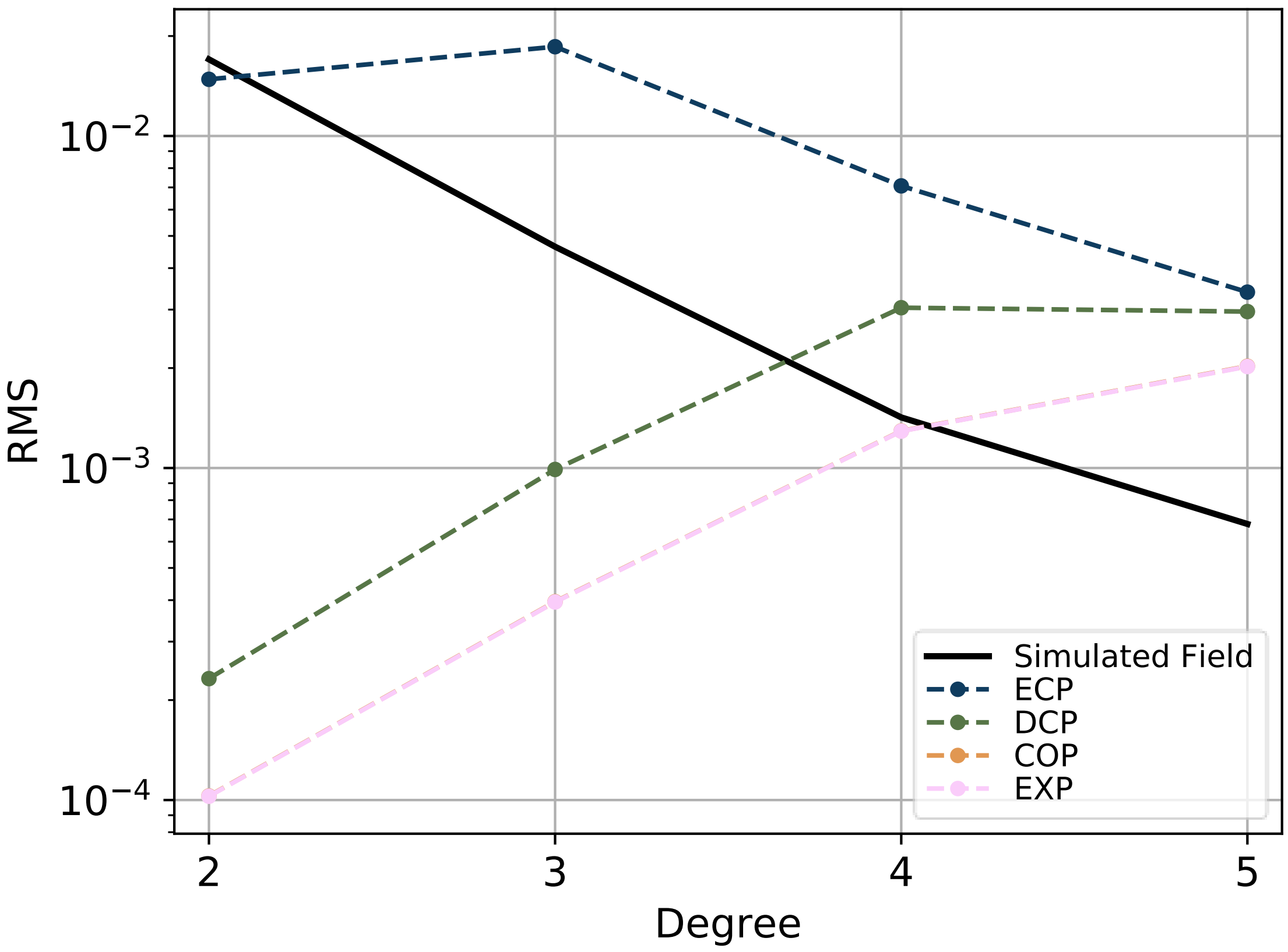

**Figure 19: Power spectra of the simulated Didymos gravity field (black) and of the recovered gravity field uncertainty at the end of each mission phase, considering Hera range-Doppler-opnav and ISL range-Doppler. Early characterization phase (blue); ECP + detailed characterization phase (green); ECP+DCP + close observation phase (orange); ECP+DCP+COP + experimental phase (pink). The EXP curve overlaps with the COP one.**

**Table 8: Estimated formal uncertainties (1σ) of Didymos and Dimorphos GM and $J_2$ at the end of each mission phase. Between squared brackets, the relative uncertainty is displayed.**

| Coefficient | ECP | DCP | COP | EXP |
|---|---|---|---|---|
| **Didymos** | | | | |
| **GM** ($km^3/s^2$) | $4.7049 \cdot 10^{-11}$ [0.11 %] | $6.0967 \cdot 10^{-12}$ [0.015 %] | $1.6125 \cdot 10^{-12}$ [0.004 %] | $1.5996 \cdot 10^{-12}$ [0.004 %] |
| **$J_2$** | $4.84 \cdot 10^{-2}$ [57.91 %] | $1.68 \cdot 10^{-4}$ [0.20 %] | $7.02 \cdot 10^{-5}$ [0.084 %] | $6.98 \cdot 10^{-5}$ [0.084 %] |
| $\alpha_0$ (deg) | 3.94 | 1.90 | 1.00 | 0.79 |
| $\boldsymbol{\delta_0}$ (deg) | 2.06 | 1.34 | 0.82 | 0.60 |
| $\boldsymbol{\alpha_1}$ (deg/hour) | $1.65\cdot10^{-5}$ | $7.98\cdot10^{-6}$ | $4.15\cdot10^{-6}$ | $3.31\cdot10^{-6}$ |
| $\boldsymbol{\delta_1}$ (deg/hour) | $8.65\cdot10^{-6}$ | $5.61\cdot10^{-6}$ | $3.40\cdot10^{-6}$ | $2.50\cdot10^{-6}$ |
| $\boldsymbol{w_1}$ (deg/hour) | $1.49\cdot10^{-6}$ | $1.48\cdot10^{-6}$ | $1.47\cdot10^{-6}$ | $1.46\cdot10^{-6}$ |
| **Dimorphos** | | | | |
| **GM** ($km^3/s^2$) | $7.5264\cdot10^{-12}$ [2.18 %] | $8.3736\cdot10^{-13}$ [0.24 %] | $2.7321\cdot10^{-13}$ [0.079 %] | $2.6951\cdot10^{-13}$ [0.078 %] |
| $\mathbf{J_2}$ | $8.19\cdot10^{-1}$ [not observable] | $5.45\cdot10^{-3}$ [6.61 %] | $5.44\cdot10^{-3}$ [6.60 %] | $5.40\cdot10^{-3}$ [6.55 %] |
| $\boldsymbol{\alpha_0}$ (deg) | 3.26 | 2.90 | 0.33 | 0.27 |
| $\boldsymbol{\delta_0}$ (deg) | 0.85 | 0.67 | 0.15 | $7.44\cdot10^{-2}$ |
| $\boldsymbol{\alpha_1}$ (deg/hour) | $3.3\cdot10^{-3}$ | $2.2\cdot10^{-3}$ | $1.24\cdot10^{-4}$ | $9.40\cdot10^{-5}$ |
| $\boldsymbol{\delta_1}$ (deg/hour) | $1.0\cdot10^{-3}$ | $5.8\cdot10^{-4}$ | $5.41\cdot10^{-5}$ | $2.41\cdot10^{-5}$ |
| $\boldsymbol{w_1}$ (deg/hour) | $3.3\cdot10^{-3}$ | $2.1\cdot10^{-3}$ | $1.21\cdot10^{-4}$ | $9.14\cdot10^{-5}$ |
| $\boldsymbol{w_a}$ (deg) | 0.81 | 0.33 | $2.28\cdot10^{-2}$ | $1.82\cdot10^{-2}$ |
| $\boldsymbol{\omega}$ (deg/hour) | $2.7\cdot10^{-2}$ | $5.1\cdot10^{-3}$ | $7.63\cdot10^{-4}$ | $4.06\cdot10^{-4}$ |
| $\boldsymbol{\varphi}$ (deg) | 19.18 | 8.93 | 2.06 | 1.20 |
| **Position uncertainty** (m) (average of R,T,N at the end of each phase) | R: 0.69 T: 1.40 N: 0.99 | R: 0.29 T: 1.17 N: 0.73 | R: 0.14 T: 0.62 N: 0.22 | R: 0.12 T: 0.53 N: 0.18 |
| **Orbital period** (sec) | 156.8 | 11.2 | 4.4 | 4.3 |

### 8.1.2 ISL parametric analyses

Parametric analyses were also conducted to evaluate the impact of the ISL duty cycle and ISL Doppler noise on the estimation of the scientific parameters of interest, since the ISL Doppler channel is still under development. The left plot in Figure 20 presents the Didymos gravity spectra for different ISL duty cycles of 20%, 40%, and 80% (i.e., 1-2-4 minutes of ISL tracking every 5 minutes of operations). The right plot demonstrates the effect of varying ISL Doppler noise levels at 60 s integration time, comparing a worst-case scenario of 0.07 mm/s to the nominal case of 0.05 mm/s and a best-case scenario of 0.02 mm/s. Notably, in the best-case ISL noise scenario, the observability of degree 4 in the gravity field becomes possible, with uncertainties in the order of 40%. Furthermore, the best-case ISL Doppler noise improves the estimation of Didymos' GM and $J_2$ by a factor of 1.4 and 2.1, respectively, compared to the nominal case. Dimorphos' improvement factors are 1.8 for the GM and 2.2 for the $J_2$. As expected, a higher ISL duty cycle and lower Doppler noise correspond to reduced uncertainties in the recovered gravity fields.

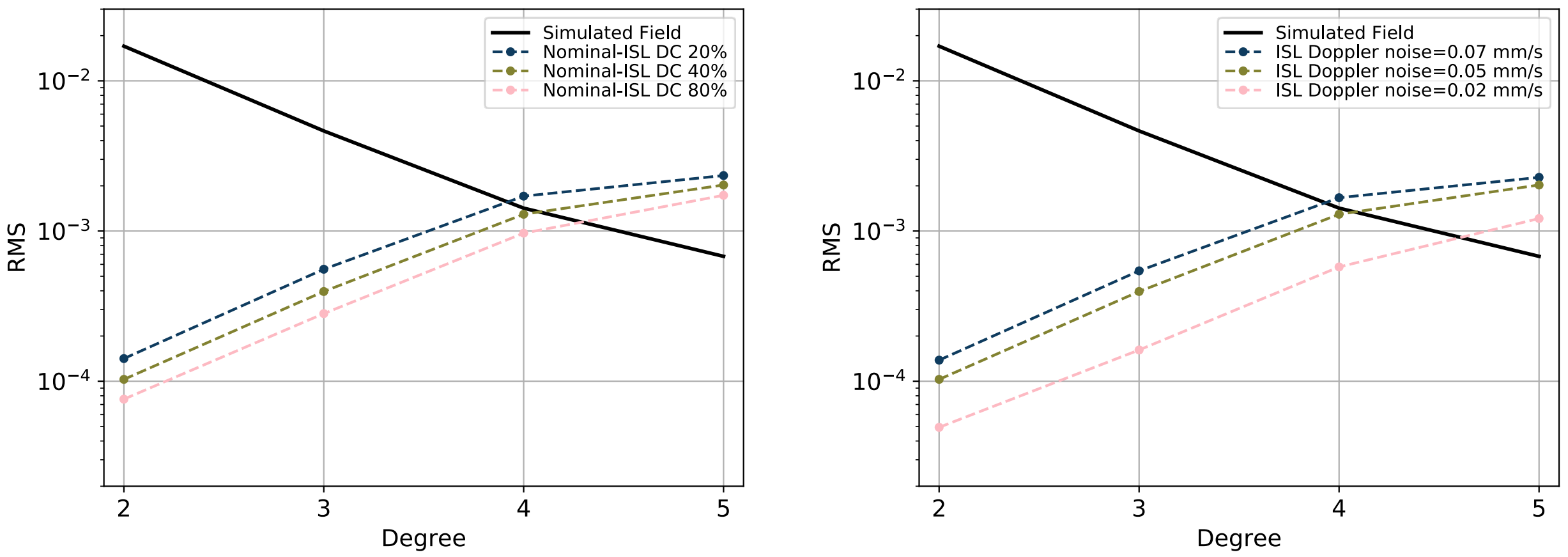

**Figure 20: Power spectra of the simulated Didymos gravity field (black) and of the recovered gravity field uncertainty at the end of the nominal mission for different ISL duty cycles (left plot) and ISL Doppler noises (right plot).**

### 8.1.3 Spacecraft stochastic accelerations

An analysis was also conducted to investigate the impact of different spacecraft stochastic accelerations on the recovered parameters. Specifically, we compared the nominal case (i.e., stochastic accelerations of 5.0·10-12 km/s$^2$ on all three spacecraft) with a worst-case scenario of $1.0 \cdot 10^{-11}$ km/s$^2$ and a best-case scenario of $1.0 \cdot 10^{-12}$ km/s$^2$. The results of these comparisons are summarized in Table 9. The Didymos gravity spectra and Dimorphos' position uncertainty do not exhibit significant differences, and the primary affected gravity coefficients are the GM and $J_2$ of the asteroids, with Didymos GM being the most affected. The ratio between the worst and best-case scenarios is 2.1 for Didymos GM and 1.2 for the $J_2$, while for Dimorphos, the ratios are 1.4 and 1.1, respectively.

**Table 9: Estimated formal uncertainties of Didymos and Dimorphos GM and $J_2$ for different magnitudes of the stochastic accelerations. Between squared brackets, the relative uncertainty is displayed.**

| **Formal uncertainty 1σ** | **Stochs $1.0 \cdot 10^{-11}$ (km/s$^2$)** | **Stochs $5.0 \cdot 10^{-12}$ (km/s$^2$)** | **Stochs $1.0 \cdot 10^{-12}$ (km/s$^2$)** |
|---|---|---|---|
| **Didymos** | | | |
| **GM** (km$^3$/s$^2$) | $1.8958 \cdot 10^{-12}$ [0.005 %] | $1.6125 \cdot 10^{-12}$ [0.004 %] | $9.1270 \cdot 10^{-13}$ [0.002 %] |
| **$J_2$** | $7.33 \cdot 10^{-5}$ [0.088 %] | $7.02 \cdot 10^{-5}$ [0.084 %] | $6.19 \cdot 10^{-5}$ [0.074 %] |
| **Dimorphos** | | | |
| **GM** (km$^3$/s$^2$) | $2.8997 \cdot 10^{-13}$ [0.084 %] | $2.7321 \cdot 10^{-13}$ [0.079 %] | $2.1129 \cdot 10^{-13}$ [0.061 %] |
| **$J_2$** | $5.55 \cdot 10^{-3}$ [6.74 %] | $5.44 \cdot 10^{-3}$ [6.61 %] | $5.06 \cdot 10^{-3}$ [6.14 %] |

## 8.2. LIDAR altimetry data and crossovers

As discussed in Section 6.3, a LIDAR scenario was analyzed within Hera's nominal mission, characterized by 400 LIDAR landmarks on both bodies and 1720 and 1068 LIDAR crossovers on Didymos and Dimorphos, respectively. As anticipated, due to the nature of altimetric measurements, the radial component of the crossovers' position is the one best determined, being primarily constrained by the uncertainty in Hera's radial position, which is in the order of m-level (see Figure 13).

The LIDAR measurements provide critical redundancy to the orbit reconstruction obtained from the radiometric and optical observables and allow the reconstruction of the shape models. Furthermore, their inclusion in the nominal solution enables the constraint of Hera's orbit during the flybys, resulting in a decreased uncertainty in Hera's position. Consequently, this improvement in Hera's position uncertainty leads to a higher accuracy in reconstructing the relative orbits of the asteroids and the Didymos system's gravity field.

Figure 21 shows the relative improvement in the position of Hera with respect to Didymos (in the RTN frame) when adding the LIDAR observables to the nominal case presented in Section 8.1. The relative improvement provided by the LIDAR and the crossovers estimation is up to 100% for the radial component and around 30% and 10% for the tangential and normal components. During the ECP (09-Feb to 09-Apr 2027), we do not observe dramatic improvements in the estimation. This is explained by the local nature of the spacecraft state estimation and by the absence of the LIDAR observables during this phase due to operation constraints (i.e., PALT can only operate at Hera-target distances lower than 14 km). For clarity, only the radial component is reported in Figure 21 since it is the most constrained by the LIDAR altimetry data. In contrast, the tangential and normal ones are shown in Appendix A.

Similarly, Figure 22 shows the relative improvement in the reconstructed orbit of Dimorphos, which is roughly 60% for the radial and tangential components and 40% for the normal component. This time, a more consistent reduction in uncertainty is also observed in the early mission phases due to the global nature of the Dimorphos ephemeris estimation.

The uncertainty of the gravity field coefficients for the LIDAR scenario is reported in Table 12 in Appendix A. Among them, the most affected parameters are the GMs of Didymos and Dimorphos, which show improvement factors of 1.2 in the LIDAR scenario with respect to the nominal case without altimetry data and crossovers. Conversely, the knowledge of Didymos higher-order harmonics is not significantly improved. Regarding Dimorphos, the most improved parameter is the $C_{22}$, which has an improvement of 1.83. The rotational state of both asteroids is also better determined, with factors in the order of 1.02-1.10 depending on the studied parameter. Dimorphos librations phase and angular velocity are better recovered by a factor of 1.3.

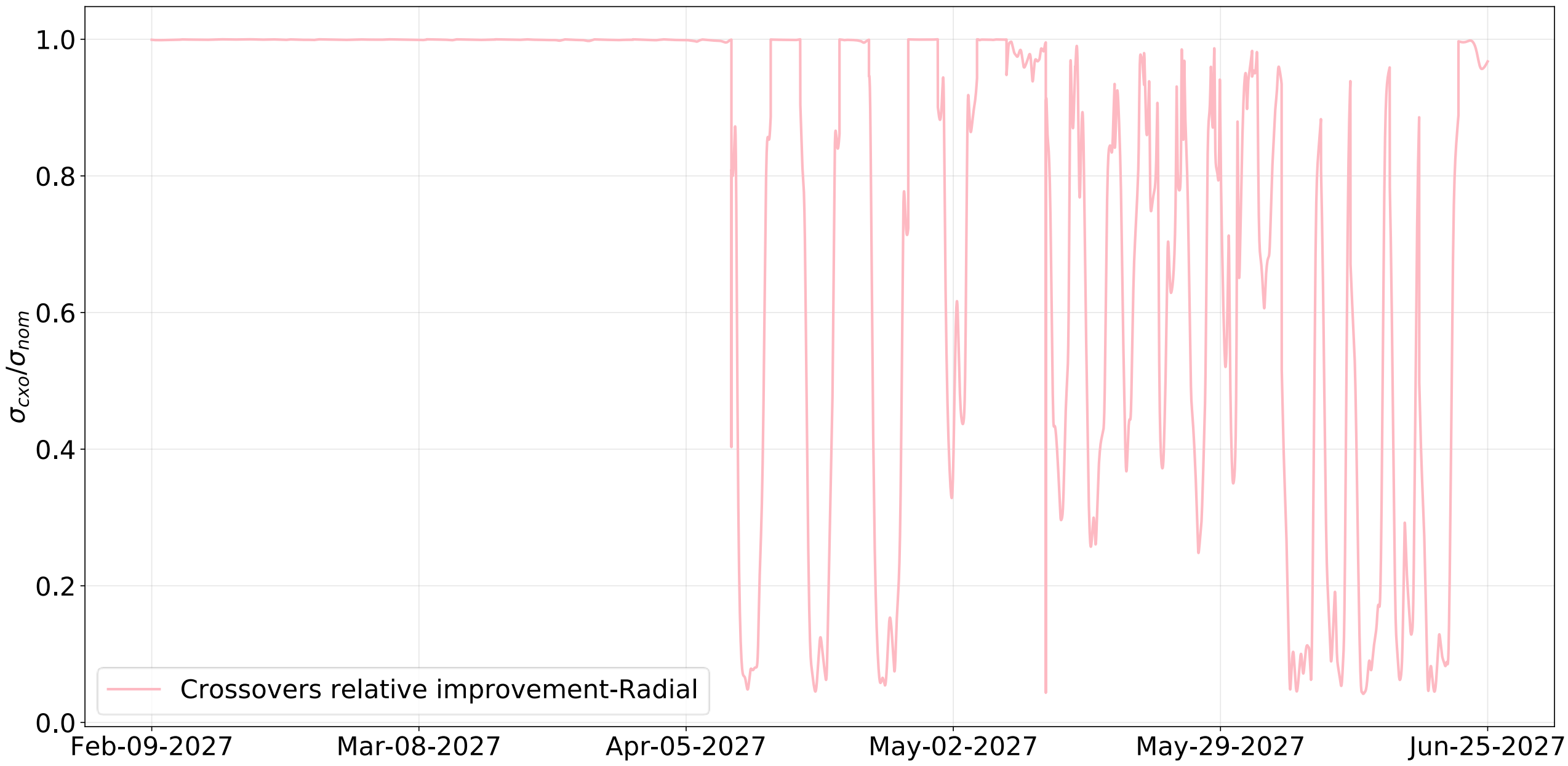

**Figure 21: Relative improvement in Hera's radial position uncertainty provided by including LIDAR observables and crossovers estimation. For clarity, only the radial component of the RTN frame with respect to Didymos is reported here since it is the one constrained by the LIDAR observables. The other two components are reported in Appendix A, see Figure 29.**

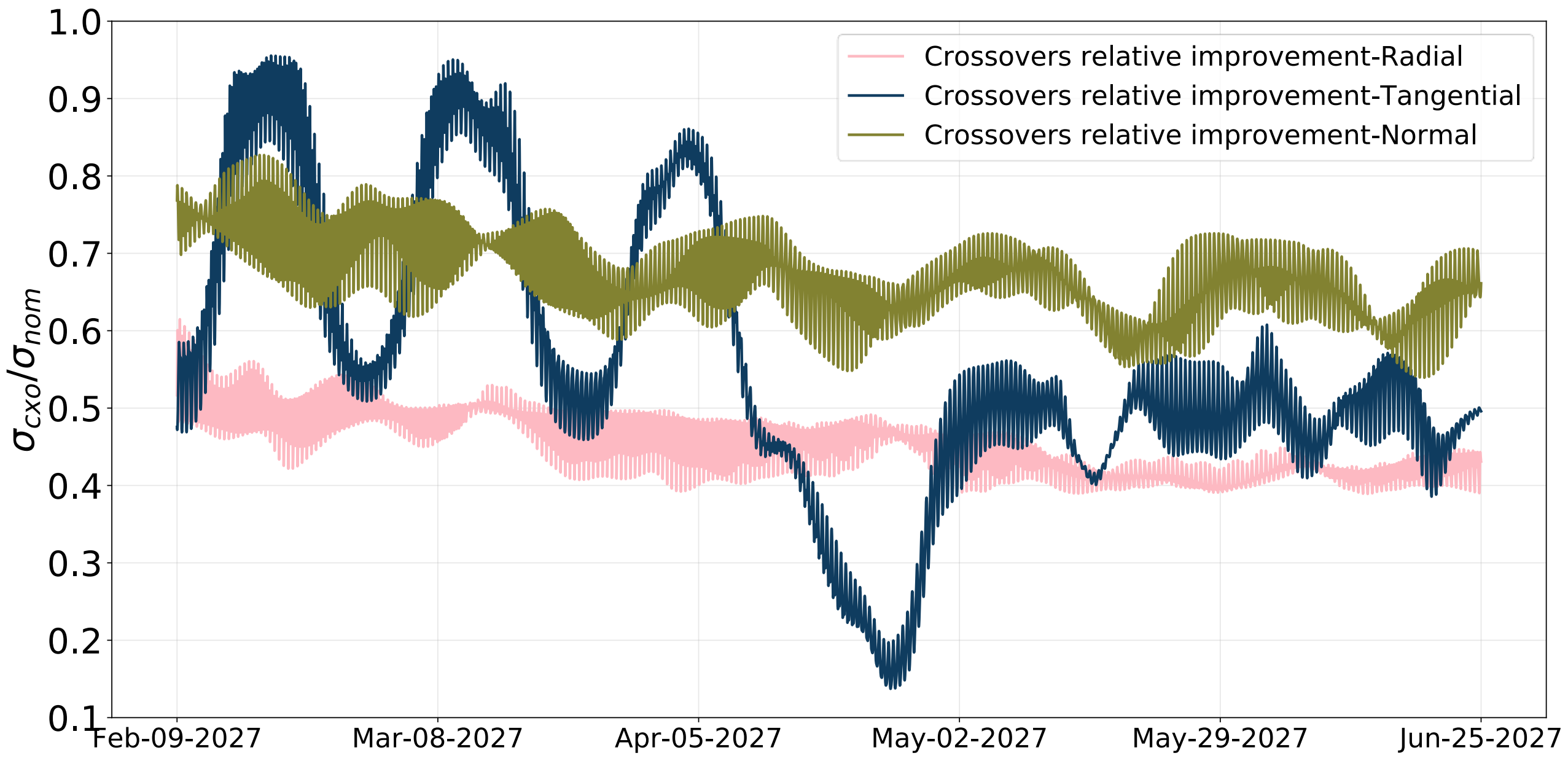

**Figure 22: Relative improvement provided by including LIDAR observables and crossovers estimation in formal uncertainty of Dimorphos' position, given in RTN components with respect to Didymos. The relative improvement offered by the LIDAR scenario is evaluated with respect to the nominal mission results presented in Section 8.1, which considers Hera Doppler-range-optical + CubeSats ISL Doppler-range measurements.**

The covariance analysis conducted in this Section employed a standard planetary radio science analysis with simplified dynamical models, which omitted the translational and rotational coupling of the asteroids known as the Full Two-Body Problem (F2BP). Despite the simplicity of these models, the results remain highly valuable in the context of small bodies investigations, as they provide:

1. Sensitivity and parametric analyses. Regardless of the adopted dynamical model, these analyses offer indications of the expected relative contributions from different mission segments, measurements, and instrument performances.
2. Insights into the disparities between covariance analysis outcomes using simplified and higher-fidelity dynamical models (e.g., F2BP).

Regarding point 2), from a broad perspective, both models exhibit the same general translational motion, showcasing non-uniform (or non-sinusoidal) orbit deviations induced by the F2BP interaction. Consequently, given the covariance analysis context, we anticipate comparable accuracy estimates for the main asteroid parameters (GMs, $J_2$, orbits) when adopting the F2BP model. However, the F2BP may introduce correlations or help de-correlate the parameters, potentially influencing their uncertainty. This aspect is explored in Section 10.

# 9. Implications for the DART impact assessment

The Hera RSE, through the estimation of the Didymos system's gravity field, provides valuable insights into the geophysical characteristics of the bodies. The gravity field coefficients of Didymos and Dimorphos are intricately linked to the internal mass distribution within these asteroids. Their estimation allows certain constraints to be placed on the mass distribution (e.g., Le Maistre et al. (2019), Caldiero et al. (2022)). However, the spherical harmonics coefficients, integrals over the mass distribution, do not uniquely determine the local mass distribution. Specifically, the un-normalized second-degree gravity coefficients are related to the moments of inertia (MOI) of the bodies according to (e.g., Scheeres (2012)):

$$\begin{aligned} C_{20} &= \frac{-2I_{zz} + I_{xx} + I_{yy}}{2MR^2} = -J_2, \\ C_{22} &= \frac{I_{yy} - I_{xx}}{4MR^2}, \\ S_{22} &= \frac{-I_{xy}}{2MR^2}, \\ C_{21} &= \frac{-I_{xz}}{MR^2}, \\ S_{21} &= \frac{-I_{yz}}{MR^2}. \end{aligned} \tag{17}$$

Degree-2 gravity coefficients cannot be used alone to separately retrieve the principal MOI ($I_{xx} < I_{yy} < I_{zz}$). However, when combined with the libration amplitude, they can fully determine the MOI. Indeed, assuming non-chaotic rotation, the forced libration amplitude also depends on the moments of inertia according to the following expression [Willner et al., 2010]:

$$w_a = \frac{2e}{1 - \frac{I_{zz}}{3(I_{yy} - I_{xx})}}, \tag{18}$$

where *e* is the orbital eccentricity. Therefore, by combining Equations (17) and (18), one can show that:

$$\frac{I_{zz}}{MR^2} = 12C_{22}\left(1 - \frac{2e}{w_a}\right),$$

$$\frac{I_{xx}}{MR^2} = 12C_{22}\left(1 - \frac{2e}{w_a}\right) + C_{20} - 2C_{22}, \tag{19}$$

$$\frac{I_{yy}}{MR^2} = 12C_{22}\left(1 - \frac{2e}{w_a}\right) + C_{20} + 2C_{22}.$$

The principal MOI are now independently determined, constituting a valuable source of information for constraining the bodies' interior structure since they characterize their radial internal mass distribution. However, if the body is in a very tumbling dynamical state, free libration modes can be excited. Their amplitudes mixing with the forced libration ones would then prevent us from deducing the MOI as described above. In such a case, we could nevertheless still constrain the internal mass distribution from the determination of the natural frequencies themselves, and will be addressed in a future work. Finally, by measuring the mass and shapes of both bodies, we can also constrain their bulk densities.

Furthermore, the orbital parameters of Dimorphos and its librations are directly linked to the energy dissipation of the system [Meyer et al., 2023]. The phase of the libration can also be helpful to constrain the dissipation since we have $Q^{-1} = \sin(\Delta\varphi)$, where $\Delta\varphi$ is the phase lag (or constant misalignment of the tidal bulge, for more details see e.g. Section 12 of Le Maistre et al. (2013)). Therefore, it is interesting to examine the anticipated uncertainties provided by Hera for these parameters, as they can be valuable for subsequent studies. Figure 23 reports the expected formal uncertainty in the eccentricity and semi-major axis of Dimorphos' relative orbit, showing that Hera can achieve accuracies of roughly $10^{-4}$ for the eccentricity and $10^{-1}$ m for the semimajor axis.

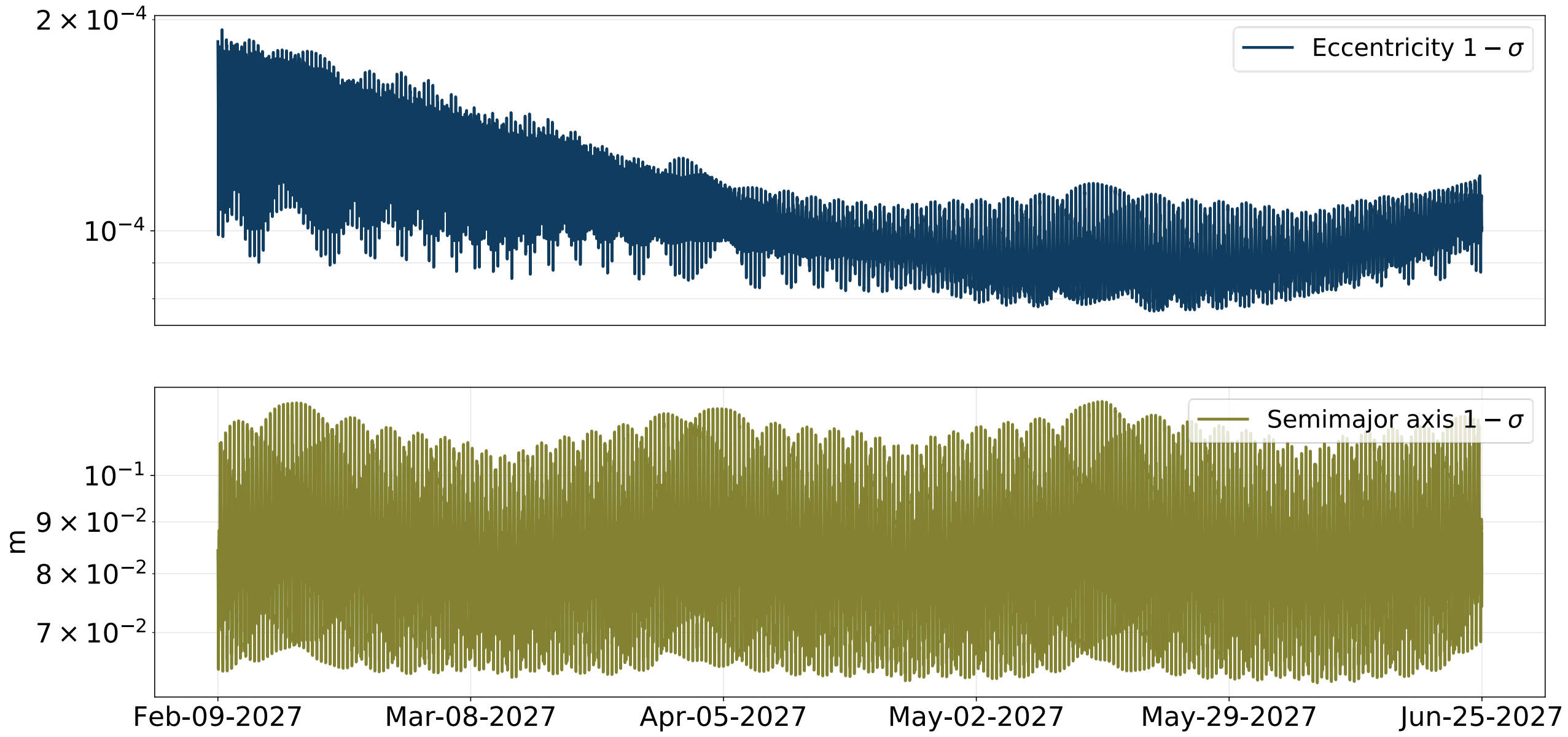

**Figure 23: Expected 1σ accuracy of Dimorphos relative orbit during the nominal mission: eccentricity (upper panel) and semimajor axis (lower panel).**

Table 7 shows the expected accuracy in estimating Dimorphos' libration, which was modeled as a single sinusoid term having a period equal to the orbital period. Specifically, we observe an amplitude accuracy of 0.02 degrees, an angular velocity accuracy of $7.6 \cdot 10^{-4}$ deg/hour (equivalent to 1.1 s in libration period), and a libration phase accuracy of 2.0 degrees. The accuracy of the libration amplitude corresponds to a spatial resolution on the surface of Dimorphos of approximately 3 cm, which aligns with the performance capabilities of the AFC camera. This high level of accuracy is primarily due to the precise reconstruction of the landmarks' positions in the body-fixed frames of the asteroids, which was facilitated by optical images.

Based on the analysis conducted by Meyer et al. (2023) and their findings presented in Figures 18 and 20, Hera is expected to provide sufficient resolution in eccentricity, libration, semimajor axis, and orbital parameters to characterize the short-term implications following the DART impact. The observed variation in libration amplitude, considering different elongations between the DART impact and Hera's arrival, ranges in the order of degrees. In contrast, the change in eccentricity ranges from $10^{-2}$ to $10^{-4}$ (as shown in Figure 18-20 of Meyer et al. (2023)). These values are consistent with the expected accuracies shown in Figure 23 and Table 7. As a result, Hera's accurate predictions on these parameters will enhance our understanding of the energy dissipation processes occurring in the Didymos system.

## 9.1. Beta uncertainty evaluation

An accurate estimation of the mass of Dimorphos allows us to determine the transfer efficiency from the DART impact, which is defined through the momentum enhancement factor β. This factor is defined as the ratio of the actual system momentum change to the momentum change caused by an inelastic impact [Meyer et al., 2023; Cheng et al., 2023; Feldhacker et al., 2017; Richardson et al., 2023], namely:

$$\overline{\Delta v} = \frac{M_{DART}}{M_B}(\bar{u} + (\beta - 1)(\hat{u} \cdot \bar{u})\hat{n}), \tag{20}$$

where $\overline{\Delta v}$ is the impact-induced change in Dimorphos' orbital velocity, $M_{DART}$ is the DART spacecraft mass, $\bar{u}$ is the impact velocity vector, and $\hat{n}$ is the outward surface normal at the impact site.

The estimation of Dimorphos' mass and the constraints on its surface and interior structure, enabled by the Hera RSE, will allow us to fully characterize the momentum transfer efficiency, currently estimated between 2 and 5 based solely on DART data [Cheng et al., 2023], as well as potentially help in understanding the scaling of the momentum transfer efficiency to different asteroids.

We are interested in obtaining a first-order estimation of the expected uncertainty in β that we could retrieve with the sensitivity provided by Hera. To achieve this, we leverage the Monte Carlo analysis conducted by Cheng et al. (2023). Their analysis is highly valuable as it accounts for the most relevant uncertainties, including DART's impact velocity, direction, and $\overline{\Delta v}$ on Dimorphos. The primary remaining uncertainty is related to Dimorphos' density. The expected uncertainty in Dimorphos' mass is the one obtained through our covariance analysis, while the uncertainty in Dimorphos' volume is the one supplied by the DART team. Starting from the Equation presented in Figure 3 of Cheng et al. (2023):

$$\beta = (A \pm \sigma_A)\frac{\rho_B}{2400\, kg/m^3} - C \pm \sigma_C \tag{21}$$

where $\rho_B = \frac{M_B}{V_B}$ is Dimorphos' density, $A = 3.61$ and $C = 0.03$ are constants derived by a linear fit of the Monte Carlo results with their corresponding (1σ) confidence intervals $\sigma_A = 0.2$ and $\sigma_C = 0.02$. Assuming these terms are uncorrelated, the uncertainty in β can be written as:

$$\sigma_\beta = \sqrt{\left(\frac{\rho_B}{2400\, kg/m^3}\sigma_A\right)^2 + \left(\frac{A}{2400\, kg/m^3}\sigma_{\rho_B}\right)^2 + \sigma_C^2} \tag{22}$$

Additionally, we note that Cheng et al. (2023) did not consider the rotational state of Dimorphos, as explained in their Methods section; this aspect is also not addressed within the scope of this work.

As a first-order approximation, $\sigma_{\rho_B}$ can be quantified as:

$$\sigma_{\rho_B} = \sqrt{\left(\frac{1}{V_B}\sigma_{M_B}\right)^2 + \left(\frac{M_B}{V_B^2}\sigma_{V_B}\right)^2} \cong 151.98\frac{kg}{m^3} \tag{23}$$

In this expression, we have used $M_B = \frac{3.4522\cdot 10^{-10}(km^3/s^2)}{G}$ kg and $\sigma_{M_B} = \frac{2.7321\cdot 10^{-13}(km^3/s^2)}{G}$ ($G$ is the gravitational constant) from our Hera RSE nominal covariance analysis; conversely, $V_B = 1760000\, m^3$ and $\sigma_{V_B} = 91000\, m^3$ were taken from DART data.

Rewriting Equation (22) and using the retrieved $\sigma_{\rho_B}$, the expected uncertainty in β is:

$$\sigma_{\beta} \cong 0.34\ (1\sigma). \tag{24}$$

It should be pointed out that this value might be affected by post-impact variations in Dimorphos' volume and mass. Schereich et al. (2024) indicate a reduction of Dimorphos' volume by (9±3) % (1σ) as a result of DART's impact. Conversely, the most significant reduction in Dimorphos' mass due to ejecta, corresponding to $5.5 \cdot 10^{7}$ kg, is the one estimated by Ferrari et al. (2023). Assuming a decrement in Dimorphos' volume and mass after DART's impact using the values above and keeping a 5% volume uncertainty for the new value (as provided by DART for the pre-impact scenario) does not significantly affect $\sigma_{\beta}$ (still $\cong 0.34\ (1\sigma)$). At the same time, we can expect Hera's estimates to enhance Dimorphos' volume accuracy by roughly a factor of 10 with respect to DART's data, consistently with Bennu's volume uncertainty obtained by OSIRIS-REX [Barnouin et al., 2019]. In this scenario, the uncertainty could be reduced up to $\sigma_{\beta} \cong 0.25\ (1\sigma)$.

Similarly, the Hera pseudorange are expected to improve, and further constrain, the reconstructed ephemeris of the Didymos system, allowing us to estimate the heliocentric momentum enhancement factor $\beta_{\odot}$, which models Didymos' heliocentric orbit changes as a result of the DART's impact.

## 10. The Full Two-Body Problem

The F2BP characterizes the dynamic interplay between two distributions of mass, leading to interconnected translational and rotational motion driven by the gravitational potential shared by irregular mass distributions. Theoretical models addressing the dynamics of the F2BP, its stability, and the observability of its physical parameter are presented, among others, in Maciejewski (1995), Ashenberg (2007), Tricarico (2008), Bellerose and Scheeres (2008), Scheeres (2009), McMahon and Scheeres (2013), Hou et al. (2017), Agrusa et al. (2020) and Richardson et al. (2022).

The Didymos binary system is a perfect illustration of the F2BP, where the integrated rotational and translational dynamics stem from the irregular shapes of the objects and their close spatial proximity [Agrusa et al., 2021]. Its dynamic evolution is mainly influenced by the asteroids' shape, initial position, and orientation. Therefore, higher-fidelity models (with respect to the ones employed in Section 8), such as the F2BP, may be required to comprehensively describe the system's dynamics in preparation for the actual data analysis. This represents one of the most significant challenges of the Hera mission.

This Section briefly describes the implementation of the F2BP and the preliminary results of the covariance analyses exploiting this higher-fidelity model. A more detailed description of the theoretical background for the F2BP and its implementation within the OD filter will be the subject of a forthcoming dedicated manuscript, which will also explore a variety of dynamic scenarios in the aftermath of the DART impact.

The F2BP model adopted within this research is a full 3D model based on the Ashenberg (2007). In this formulation, which relies on inertial Cartesian coordinates, accelerations, and torques acting on the bodies are functions of the mutual gravitational potential, which is a function of the inertia integrals associated with the principal reference frames of each body and the relative rotation matrix. Within this work, the mutual gravitational potential and torque of two bodies of arbitrary shape are expanded to the second order. Moments of inertia of the bodies are obtained from the generalized inertia tensor $N_A^{(l,m,n)}$ associated to the polyhedron, which is computed assuming uniform density and using the formulation of Hou et al. (2017).

Our MONTE F2BP model has been tested and validated against the Aristotle University of Thessaloniki F2BP propagator, which was previously validated against the General Use Binary

Asteroid Simulator (GUBAS) [Davis and Scheeres, 2020]. The same tool was also used to estimate the DART impact's momentum enhancement factor in Cheng et al. (2023).

The filter setup used for this analysis is consistent with the one presented in Section 8, with only a couple of differences:

- In the F2BP implementation, the rotational states of the asteroids are numerically integrated using a dedicated torque class, establishing a full coupling between the orbital and rotational states. This stands as the primary distinction from the covariance analyses conducted in Section 8, where the rotational states of the bodies were defined through analytical functions for the pole and prime meridian.
- The degree-two gravity terms are evaluated using a custom F2BP force model, incorporating the complete degree-two figure-to-figure effects. In contrast, the higher-order harmonics accelerations are estimated using the standard gravity functions of MONTE (up to degree 5 for Didymos, consistent with the approach outlined in the previous Sections). In the case of Dimorphos, the gravity is estimated up to degree 2 using the F2BP model.

Utilizing the F2BP model enables us to derive estimates for the moments of inertia of the bodies directly. However, for ease of analysis, it is common to examine the spherical harmonics coefficients. Therefore, we introduced filter constraints between the moments of inertia and the degree-2 coefficients to facilitate a more straightforward visualization of the solution. The constraints between these quantities are explicitly defined in Equations (17). Notably, these filter constraints do not alter the solution; instead, they provide an alternative basis or representation for visualization purposes.

## 10.1. F2BP covariance analyses

The simulation parameters, instrument performances, ISL duty cycle, measurement noises/models, and spacecraft stochastic accelerations remain identical to the nominal case in Section 8.1. This consistency is maintained to facilitate a direct comparison of the dynamical models.

Two scenarios are analyzed, namely:

- A pre-impact relaxed case where Dimorphos is assumed to be tidally locked and in a circular orbit around Didymos. The initial state and attitude corresponding to this relaxed condition are obtained using an approach similar to that described by Agrusa et al. (2021).
- An excited post-impact state (non-chaotic). For this case, we use the same initial conditions of the pre-impact case and introduce the DART impact in the dynamical model, treating it as an impulsive variation in Dimorphos' orbital velocity. The values for this velocity variation are obtained from the JPL solution *Dimorphos_s527* [NASA JPL SSD].

Figure 24 shows the Euler angles of the propagated body frame of Dimorphos in the pre-impact scenario. Notably, we observe approximately 1 degree of physical librations in yaw and about 0.01 degrees of librations for both pitch and roll. The orbit eccentricity is about 0.016, and the mean anomaly is approximately 0, indicative of Dimorphos being trapped at periapsis. This is not a fully relaxed case, as we did not fit the rotation initial conditions, which accounts for observing very small librations in pitch and roll, as well as a free libration in yaw. However, this does not pose a problem, as the covariance analysis enables us to assess sensitivity to angular velocities irrespective of the particular values chosen.

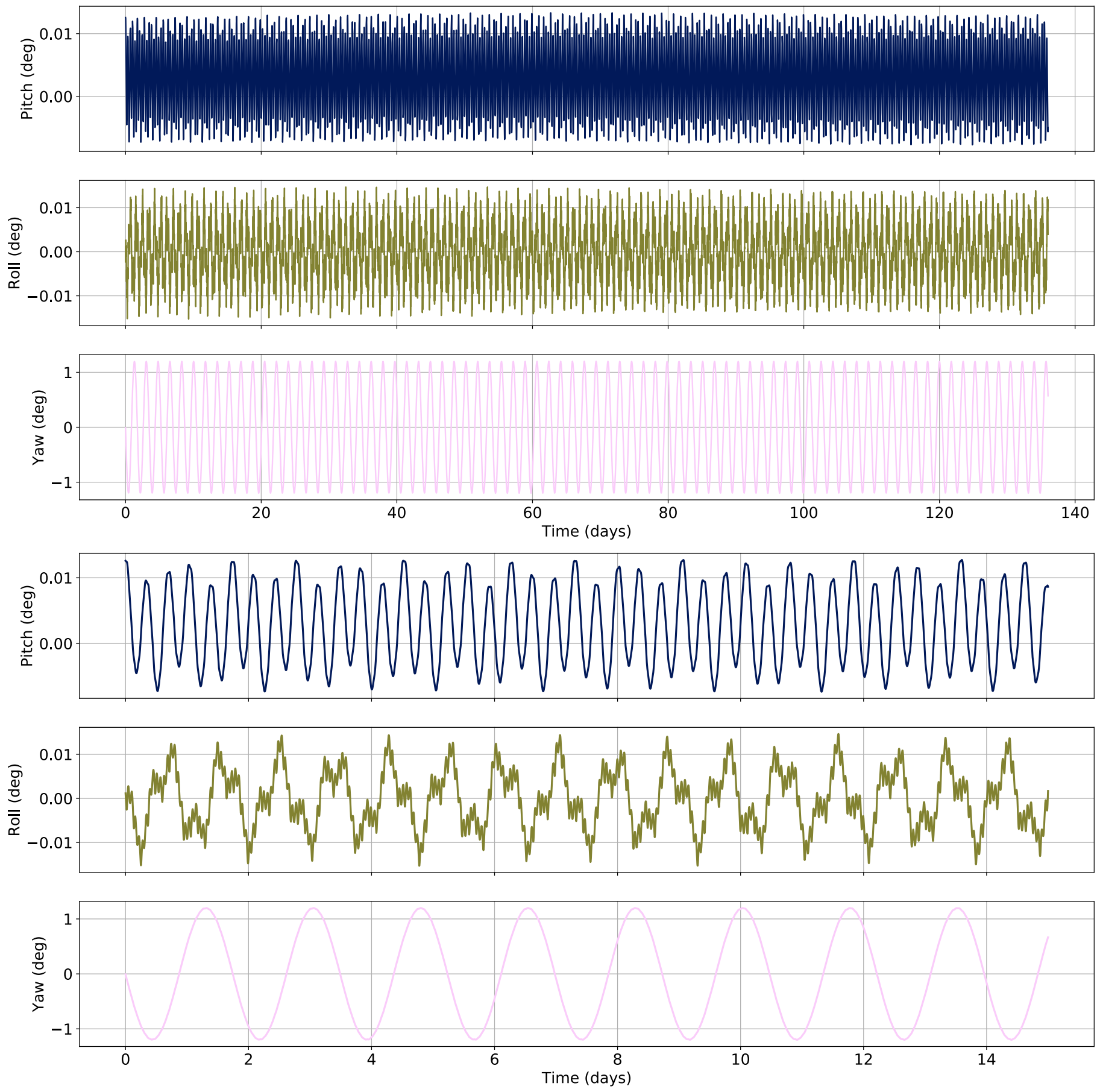

**Figure 24: Pre-impact propagated attitude (Euler 3-2-1 angles) of Dimorphos. The bottom panel shows the same results for a shorter time window, highlighting the short time-scale frequencies.**

Furthermore, Figure 25 presents the Dimorphos' simulation parameters for the post-impact case. Notably, we observe approximately 10 degrees of librations in yaw, which include both a physical and free libration contribution, while both pitch and roll exhibit librations of less than 1 degree at short (less than one day) and long (approximately one year) period frequencies. The average eccentricity for this case is roughly 0.03, accompanied by a relative distance oscillation of about 80 meters. The mean anomaly maintains a clocking rate in line with the orbital rate, while the argument of periapsis undergoes a slow precession.

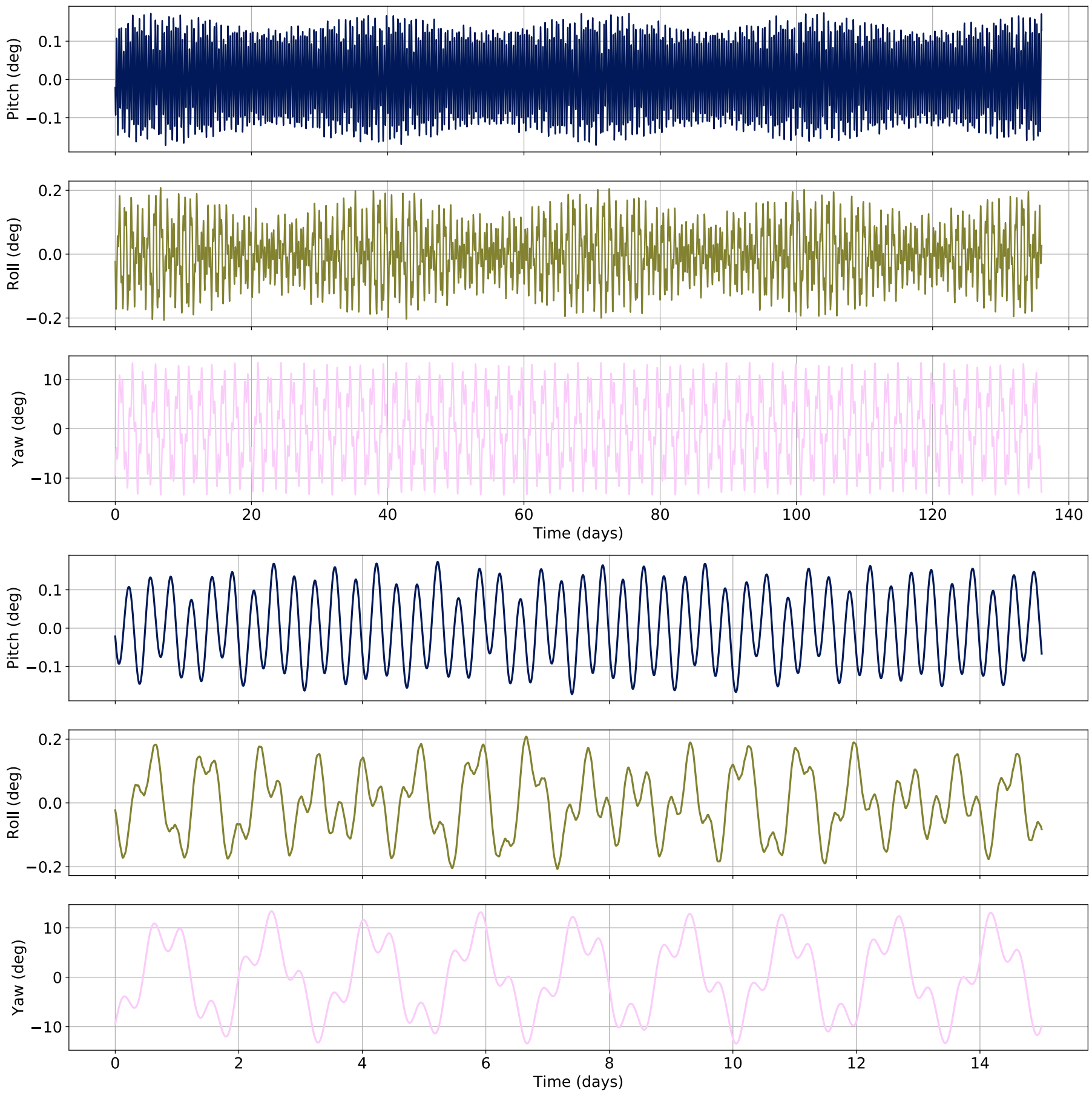

**Figure 25: Post-impact Dimorphos F2BP propagated attitude (Euler 3-2-1 angles). The bottom panel shows the same results for a shorter time window, highlighting the short time-scale frequencies.**

Concerning the librational state of Dimorphos, a considerable body of work has already been presented in the literature (refer to, for example, Naidu and Margot (2015), and Agrusa et al., (2020, 2021)), and it stands as another intriguing topic that Hera will investigate. In the pre-impact scenario, it is not anticipated for Dimorphos to exhibit free librations, as these motions should naturally diminish over time due to tidal forces, and the forced libration is not observable due to its nearly circular orbit. However, Dimorphos demonstrates free libration in our tested case, with a frequency described by Equation (25), as we did not fit the initial rotation conditions. As previously mentioned, this discrepancy does not present an issue within the covariance analysis. Moreover, it provides a scenario closely resembling the one analyzed in Section 8, facilitating comparing results under similar conditions but with different models. Additionally, in the post-impact scenario, we posit that the DART impact induces an additional free libration. Furthermore, the forced libration becomes observable due to the non-circular orbit caused by the impact, with a frequency approximately equivalent to the orbital period and an amplitude described by Equation (18). The free libration is anticipated to persist upon Hera's arrival, as recent models for tidal dissipation in binary asteroids suggest that any free libration would dissipate over 100-year timescales [Meyer et al., 2023]. The amplitude and phase of the free libration depend on the impact's geometry and the beta factor, while its frequency is contingent on the internal structure. In the decoupled spin-orbit motion, the frequency of free libration for a synchronous satellite on a circular orbit is namely:

$$\omega_0 = \mathrm{n}\left(3\frac{I_{yy} - I_{xx}}{I_{zz}}\right)^{\frac{1}{2}}. \tag{25}$$

It is worth mentioning that the impact of DART may have excited Dimorphos' spin, leading to attitude instability and even possibly chaotic tumbling, as detailed in Agrusa et al. (2021). Notably, the analysis of chaotic scenarios demands deeper investigations, including selecting suitable durations for the gravity science arcs, numerical propagation tolerances, and evaluating predictability and computability horizons [Spoto and Milani, 2016; Serra et al., 2018]. This specific problem is not addressed within this work, as it warrants a dedicated study that will be the focus of a forthcoming paper. We want to emphasize also that the rotation pole of Didymos may undergo rapid precession of tens of degrees per day if the orbit is found to have some inclination [Fahnestock and Scheeres, 2008], which will be investigated in the subsequent work. In conclusion, the gravity field parameters of Dimorphos can also be determined by observing its dynamical motion, as outlined by Davis and Scheeres (2020). This method is independent of radiometric tracking, adding valuable confirmation and consolidation to the obtained RSE results.

## 10.2. Results

Table 10 provides a summary of the uncertainties for the GM, $J_2$, and pole orientation parameters for the analyzed cases. Overall, we observe that the formal uncertainties for the GMs are similar to the ones presented in Section 8.1. This consistency is expected, given the similarity in orbits. Significant improvements are observed in the $J_2$ and pole uncertainties of Didymos when leveraging the F2BP, with a factor of approximately 3. Conversely, estimation uncertainties in the parameters of Dimorphos remain consistent with those from previous simulations, albeit with a slight degradation by a factor ~ 2 in the pole parameters.

**Table 10: Formal uncertainties (1σ) in the main parameters of interest for the Nominal simulation of Section 8.1 and the F2BP scenarios.**

| Coefficient | Nominal | F2BP pre-impact | F2BP post-impact |
|---|---|---|---|
| **Didymos** | | | |
| **GM** ($km^3/s^2$) | $1.6125 \cdot 10^{-12}$ | $1.5444 \cdot 10^{-12}$ | $1.4509 \cdot 10^{-12}$ |
| $J_2$ | $7.02 \cdot 10^{-5}$ | $2.32 \cdot 10^{-5}$ | $2.16 \cdot 10^{-5}$ |
| $\alpha_0$ (deg) | 1.00 | 0.37 | 0.37 |
| $\delta_0$ (deg) | 0.82 | 0.21 | 0.21 |
| **Dimorphos** | | | |
| **GM** ($km^3/s^2$) | $2.7321 \cdot 10^{-13}$ | $3.4847 \cdot 10^{-13}$ | $3.3467 \cdot 10^{-13}$ |
| $J_2$ | $5.44 \cdot 10^{-3}$ | $5.64 \cdot 10^{-3}$ | $4.21 \cdot 10^{-3}$ |
| $\alpha_0$ (deg) | 0.33 | 0.76 | 0.72 |
| $\delta_0$ (deg) | 0.15 | 0.31 | 0.32 |

Furthermore, Table 11 presents our preliminary uncertainties in the moments of inertia of Didymos and Dimorphos. As expected, due to the uniform rotation of the pre-impact scenario, the observability of Didymos' principal inertia moments is rather limited. However, the augmented librational state of Dimorphos in the post-impact scenario significantly enhances the observability of the inertia moments of both the asteroids, resulting in relative uncertainties of approximately 13-18% for Didymos and 3-5% for Dimorphos. Further analysis on the observability of the inertia moments will be conducted in a dedicated F2BP study.

**Table 11: Formal uncertainties (1σ) in the moments of inertia given by the F2BP scenarios. The values are rounded to 4 decimal places and provided in kg·km², while the apriori covariances are unconstrained.**

| Coefficient | Value | F2BP pre-impact | F2BP post-impact |
|---|---|---|---|
| **Didymos** | | | |
| $I_{XX}$ | $2.9221 \cdot 10^{10}$ | $5.4572 \cdot 10^{10}$ | $5.1878 \cdot 10^{9}$ |
| $I_{YY}$ | $3.1394 \cdot 10^{10}$ | $5.4572 \cdot 10^{10}$ | $5.1878 \cdot 10^{9}$ |
| $I_{ZZ}$ | $3.9236 \cdot 10^{10}$ | $5.4572 \cdot 10^{10}$ | $5.1879 \cdot 10^{9}$ |
| $I_{XY}$ | 0.00 | $1.4205 \cdot 10^{7}$ | $1.4000 \cdot 10^{7}$ |
| $I_{XZ}$ | 0.00 | $4.0777 \cdot 10^{7}$ | $4.0777 \cdot 10^{7}$ |
| $I_{YZ}$ | 0.00 | $3.3463 \cdot 10^{6}$ | $3.3642 \cdot 10^{6}$ |
| **Dimorphos** | | | |
| $I_{XX}$ | $1.0781 \cdot 10^{7}$ | $7.7777 \cdot 10^{5}$ | $5.3560 \cdot 10^{5}$ |
| $I_{YY}$ | $1.1215 \cdot 10^{7}$ | $8.0784 \cdot 10^{5}$ | $5.5731 \cdot 10^{5}$ |
| $I_{ZZ}$ | $1.5383 \cdot 10^{7}$ | $1.0647 \cdot 10^{6}$ | $7.6387 \cdot 10^{5}$ |
| $I_{XY}$ | 0.00 | $3.2838 \cdot 10^{3}$ | $3.1180 \cdot 10^{3}$ |
| $I_{XZ}$ | 0.00 | $2.6994 \cdot 10^{4}$ | $2.6995 \cdot 10^{4}$ |
| $I_{YZ}$ | 0.00 | $1.1637 \cdot 10^{4}$ | $1.1233 \cdot 10^{4}$ |

Figure 26 illustrates the Didymos gravity spectra. Overall, the three simulations exhibit consistency, with the F2BP scenarios displaying slightly higher uncertainties in the gravity coefficients. Nevertheless, these uncertainties are deemed negligible in a broader context.

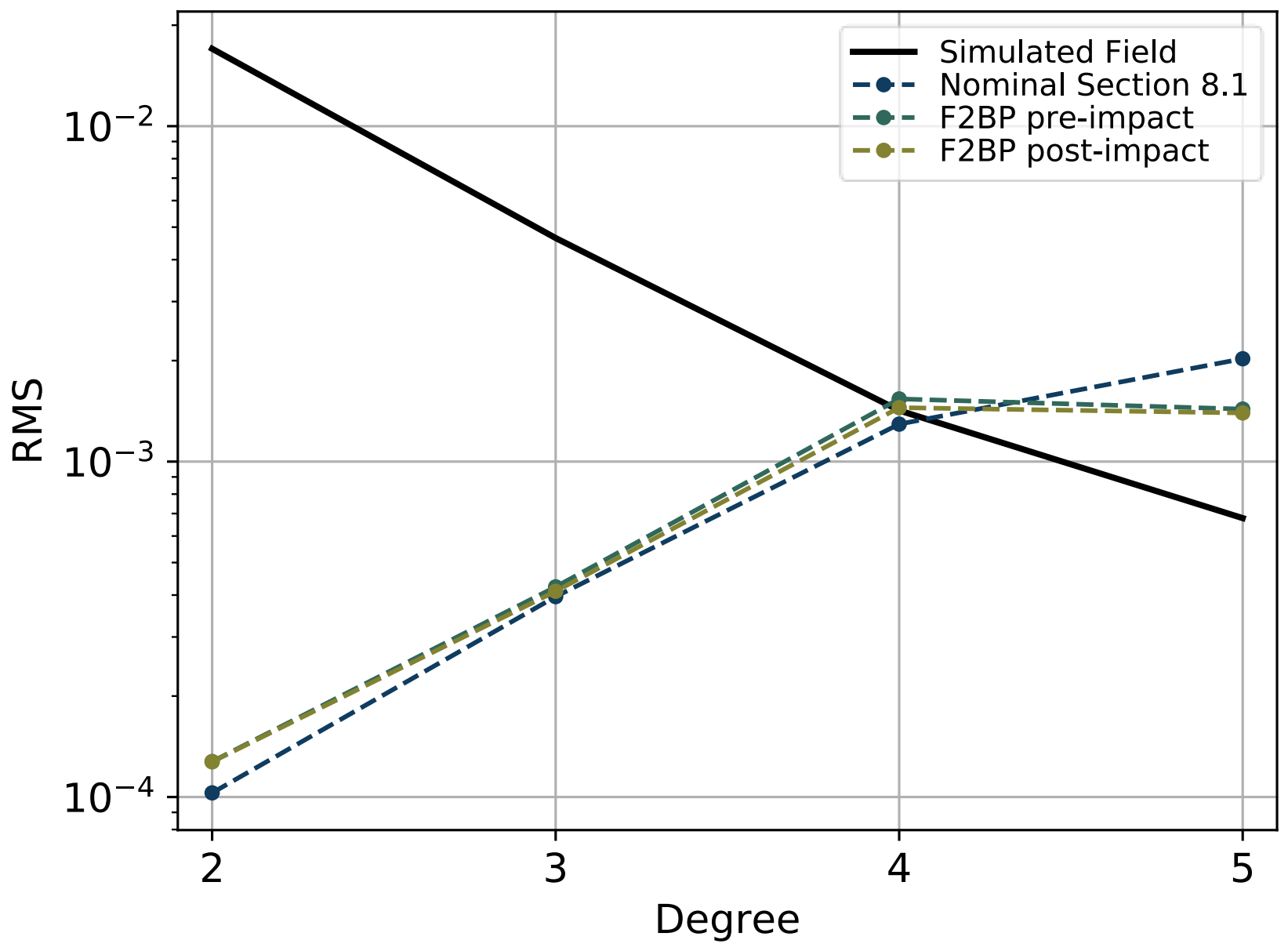

**Figure 26: Power spectra of the extended gravity field of Didymos. Continuous line: simulated field; dashed lines: formal uncertainty of the estimated field at the end of the nominal mission under different scenarios. Blue: Nominal scenario of Section 8.1; green: F2BP pre-impact scenario; olive: post-impact scenario.**

Figure 27 depicts the recovered 1σ uncertainties for Dimorphos' body-fixed angular velocities using the F2BP model for both pre- and post-impact cases. Overall, the two scenarios exhibit similar uncertainties in the angular velocities, with values in the order of $10^{-5}$ deg/sec. In the case of Didymos, not reported here for clarity, the uncertainties for $\omega_x$ and $\omega_y$ remain consistent between the two scenarios, with values in the order of $10^{-4}$ deg/s and $10^{-5}$ deg/s, respectively. The Z component shows better recovery for both asteroids in the post-impact scenario, with an improvement of up to one order of magnitude for Didymos (see Figure 30 in Appendix B). This improvement is likely attributed to the lower correlations between the Didymos angular velocities in the post-impact case.

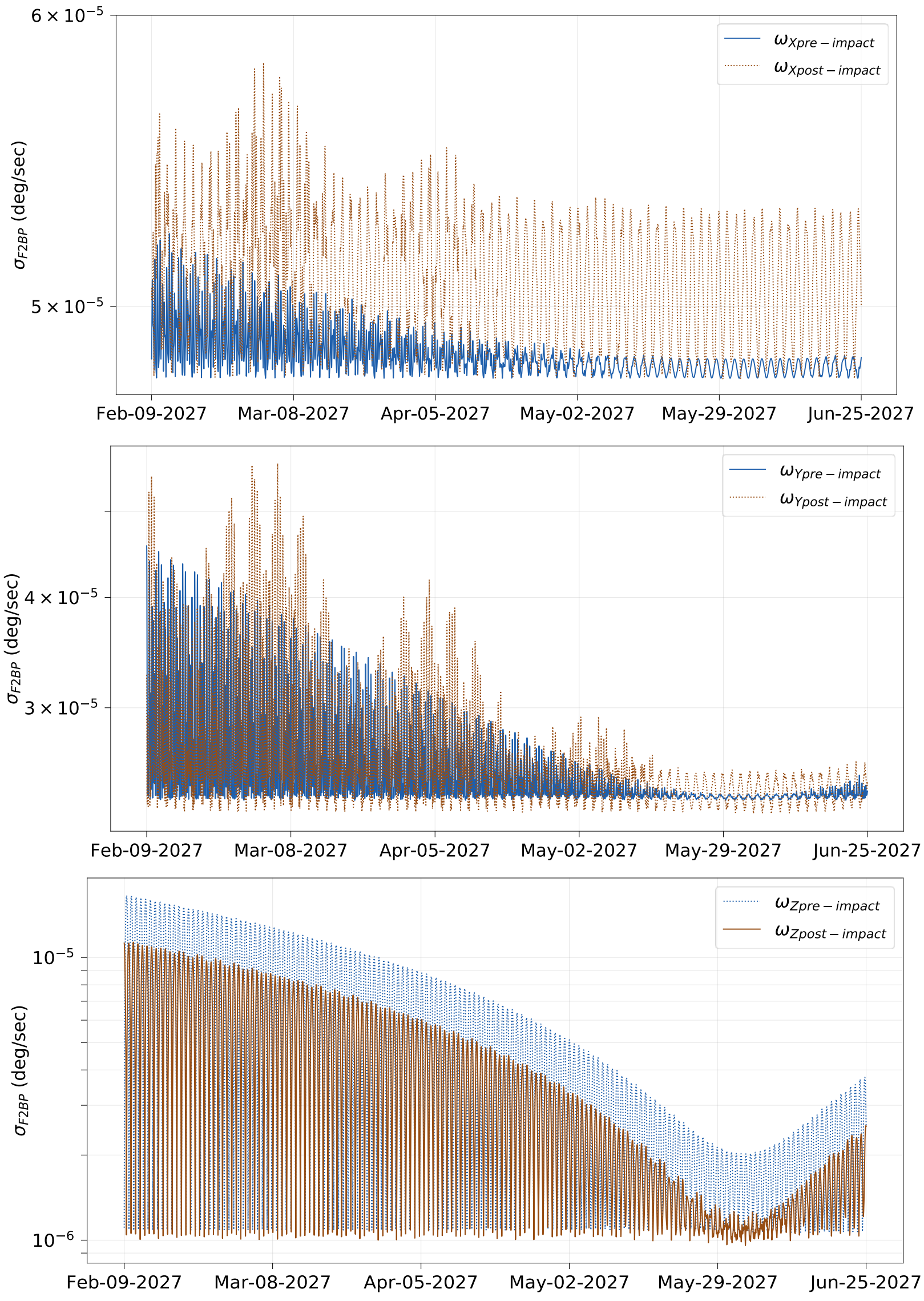

**Figure 27: Formal uncertainties (1σ) for Dimorphos' X-Y-Z angular velocities in the pre- and post-impact scenarios using the F2BP.**

Figure 28 provides a comparison of Dimorphos' reconstructed position uncertainties between the F2BP scenarios and the nominal case. During the early stages of the mission, particularly the ECP, observability of Dimorphos is limited and primarily obtained through centroiding. Consequently, the post-impact scenario exhibits higher uncertainties in the early stages than the nominal scenario, mainly attributed to the increased complexity in determining the intricate rotational state during these early phases. As Hera moves closer to the Didymos system, the observability of the system, including its rotational and librational state, increases, facilitated by landmark observations as well. Thus, in the later stages, the F2BP enables higher accuracies in the relative orbits.

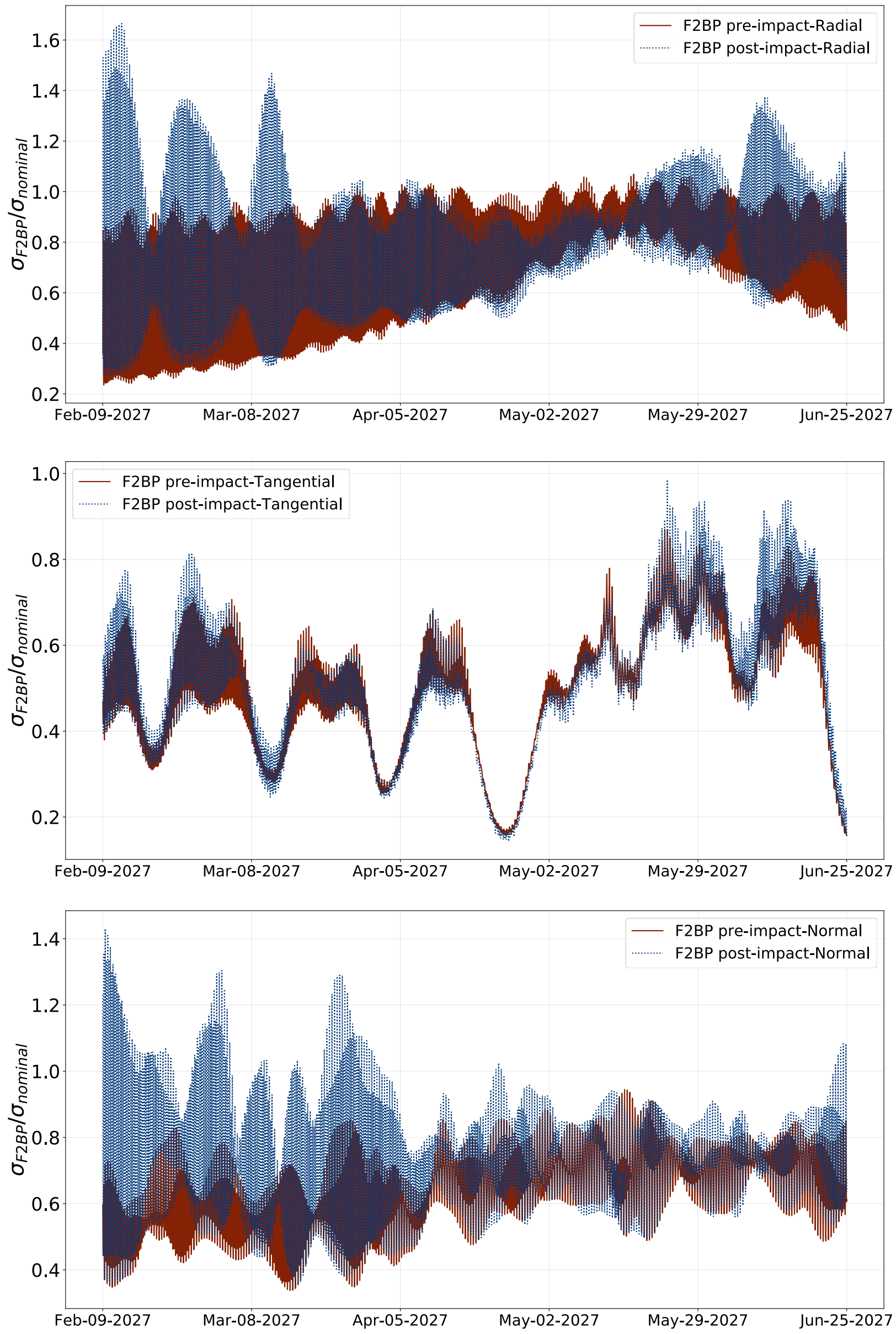

**Figure 28: Ratio between the 1σ uncertainties of Dimorphos' position, given in RTN components with respect to Didymos, for the F2BP scenarios and the nominal simulation of Section 8.1.**

In Appendix B, we report the correlation matrices for the nominal scenario of Section 8.1 and the F2BP pre- and post-impact simulations (i.e., Figure 31, Figure 32, and Figure 33). Examining the correlation matrices, the observed improvement in Dimorphos' orbit within the F2BP simulations, compared to the nominal scenario presented in Section 8.1, can be attributed to the reduced correlations between Dimorphos' state and the asteroid body-fixed frames. This reduction is facilitated by the coupled attitude and orbit integration in the F2BP simulations.

In conclusion, assuming a non-chaotic post-impact scenario, the primary scientific parameters of interest in these preliminary findings exhibit uncertainties comparable to those presented in Section 8.1. This behavior was anticipated, considering the similar orbits and corresponding translational motion in both analyses, demonstrating non-uniform or non-sinusoidal orbit deviations caused by the F2BP interaction. If Hera discovers the Didymos system in a non-chaotic librating state, these methods should yield similar estimations for the main parameters of interest, particularly Dimorphos' mass, the primary objective of the Hera mission.

The decision to employ the F2BP or the standard approach depends on selecting the model that best fits the data during the operational mission. Conversely, in the event that a chaotic motion is triggered by DART's impact (e.g., due to shape or angular momentum variations on Dimorphos), the anticipated uncertainties in this study may not be entirely applicable and a potential degradation in the results could be expected. Considering the intricate nature of chaotic systems, it is difficult to anticipate how the results will vary. Therefore, a dedicated study should be undertaken, forming the focal point of future investigations. In this context, we also plan to enhance our F2BP MONTE model by extending it to higher orders and degrees.

## 11. Conclusions and Future Works

The Hera radio science experiment will significantly contribute to completing the planetary defense post-impact survey of DART, and it will advance asteroid science as the first RSE investigation of a binary asteroid system, contributing to understanding their formation scenario.

By combining radio tracking data, optical measurements, Inter-Satellite Link observables, and LIDAR altimetry data, we can accurately estimate the mass, mass distribution, orbits, and rotational states of Didymos and its moon Dimorphos, which are crucial to validate the kinetic impactor technique for future asteroid deflection efforts.

We conducted a multi-arc covariance analysis of the Hera RSE during the expected asteroid phase to assess the attainable accuracy of the scientific parameters of interest. Within the nominal mission, we found that the Hera RSE successfully achieves all the experiment objectives and required accuracies in retrieving the asteroids' masses, relative orbits (e.g., semimajor axis and eccentricity), rotational states, mass distribution, and Dimorphos librations. Notably, the ISL data enables the estimation of the extended gravity fields of Didymos up to degree 3 (potentially also degree 4, depending on the ISL performances) and Dimorphos up to degree 2. The masses of Didymos and Dimorphos can be retrieved with relative accuracies of better than 0.01% and 0.1%, respectively. Additionally, their $J_2$ formal uncertainties are better than 0.1% and 10%, respectively. Regarding their rotational state, the absolute spin pole orientations of the bodies can be recovered to better than 1 degree, and the Dimorphos spin rate to better than $10^{-3}$ %. Furthermore, Hera can achieve an uncertainty level of approximately $10^{-4}$ for Dimorphos eccentricity, $10^{-1}$ m for semimajor axis, sub-m level for Dimorphos reconstructed orbit, and $10^{-2}$ degrees for its libration amplitude. The retrieved values satisfy the Hera RSE requirements and goals. As a result, Hera's accurate predictions on these parameters will enhance our understanding of the energy dissipation processes occurring in the Didymos system.

Through an accurate estimation of the mass of Dimorphos, the Hera RSE will also determine the momentum transfer efficiency β from DART's impact. A first-order estimation of the expected uncertainty obtainable with Hera results in a value of $\sigma_\beta \simeq 0.25$, which represents a significant improvement with respect to current estimates.

By incorporating LIDAR altimetry data and crossovers estimation into the orbit determination process, we observe improvements with respect to the nominal scenario. Hera's position is better determined when the LIDAR is activated, with relative enhancements up to 100%, 30%, and 10% for the RTN components. Similarly, Dimorphos' reconstructed orbit shows lower uncertainties, in the order of 60% for the radial and tangential components and 40% for the normal one, while Didymos and Dimorphos GM improve by a factor of 1.2. In this context, Didymos' extended gravity field does not show sensible improvements with respect to the nominal case.

Within this work we demonstrated the crucial role represented by the Inter-Satellite Link data, thanks to the vicinity of the CubeSats to the system. For this reason, we performed parametric and sensitivity analysis to evaluate the impact of the ISL duty cycle and ISL Doppler noise on the estimation of the scientific parameters of interest. As expected, a higher ISL duty cycle and lower Doppler noise correspond to reduced uncertainties in the recovered gravity fields. Notably, the best case scenario for ISL Doppler noise and duty cycle allows us to start observing Didymos degree 4 with accuracies of 40-67%. In contrast, the worst cases do not dramatically degrade the results with respect to the nominal one.

In addition, we conducted a parametric analysis to assess the influence of various factors on the scientific parameters of interest, such as the spacecraft measurements, mission phases, ISL duty cycle, ISL Doppler noise, and spacecraft stochastic accelerations.

Sensitivity analyses were also performed to assess the influence of each mission phase on the estimation accuracies as the mission progresses. At the end of ECP, Dimorphos' mass formal uncertainty already satisfies the Hera RSE requirement. However, Didymos' extended gravity field can be retrieved only after the DCP, which also allows us to constrain Dimorphos' orbital period by approximately one order of magnitude with respect to the ECP. The COP and EXP further improve the accuracies of the estimated parameters. In particular, the EXP improves Dimorphos' rotational state and its librations by a factor of 1.2-2.0 with respect to the COP, while Dimorphos' orbit estimation improves by a factor of 2.2-4.5 (depending on the selected component).

Another important sensitivity analysis considered the impact of the spacecraft stochastic acceleration models on the recovered parameters. The results are not exceptionally sensible to the magnitude of the stochastic accelerations, with the GM of Didymos, the most impacted parameter, varying by a factor of 2.1 between the best-case and worst-case scenarios.

The second part of the study focused on developing a comprehensive F2BP model, aiming for a higher-fidelity representation of the coupled motion between rotational and orbital dynamics within the Didymos system. Such models may be crucial for Hera's operational data analysis and pose one of the most significant challenges for the mission. To address this issue, we implemented the F2BP model within MONTE and conducted preliminary covariance analyses on both pre- and post-impact scenarios. Using the currently available shape models of the asteroids and assuming no mass loss and no shape variations from DART's impact results in a non-chaotic excited state, which simplifies the analysis and allows for a direct comparison with the standard radio science model described above. Initial findings show comparable uncertainties in the main scientific parameters of interest, as expected due to the similarities in orbits and translational motion in both analyses.

In the event that a chaotic motion is onset by DART's impact, the anticipated uncertainties in this study may not be fully applicable, suggesting a potential degradation in the results. Due to the intricate nature of chaotic systems, predicting the extent of this degradation is challenging and a dedicated and thorough study should be undertaken. This upcoming study will expand the F2BP representation to higher order and degree and will delve into a more detailed discussion on the aftermath of DART's impact.

Future works will also consider implementing more realistic non-gravitational accelerations and torques induced by the Yarkoswki, YORP and BYORP effects. Furthermore, opportunity measurements such as the CubeSats' optical images, Juventas' gravimetry data, and tracking of in situ ejecta particles will be considered.

The simulations carried out in this work are based on conservative assumptions, serving as a robust baseline for the expected results that the Hera Mission will provide if the system is determined to be in an excited but non-chaotic (or tumbling) state.

# Appendix A - LIDAR

In this appendix, we report all the results of the LIDAR scenario studied in Section 8.2.

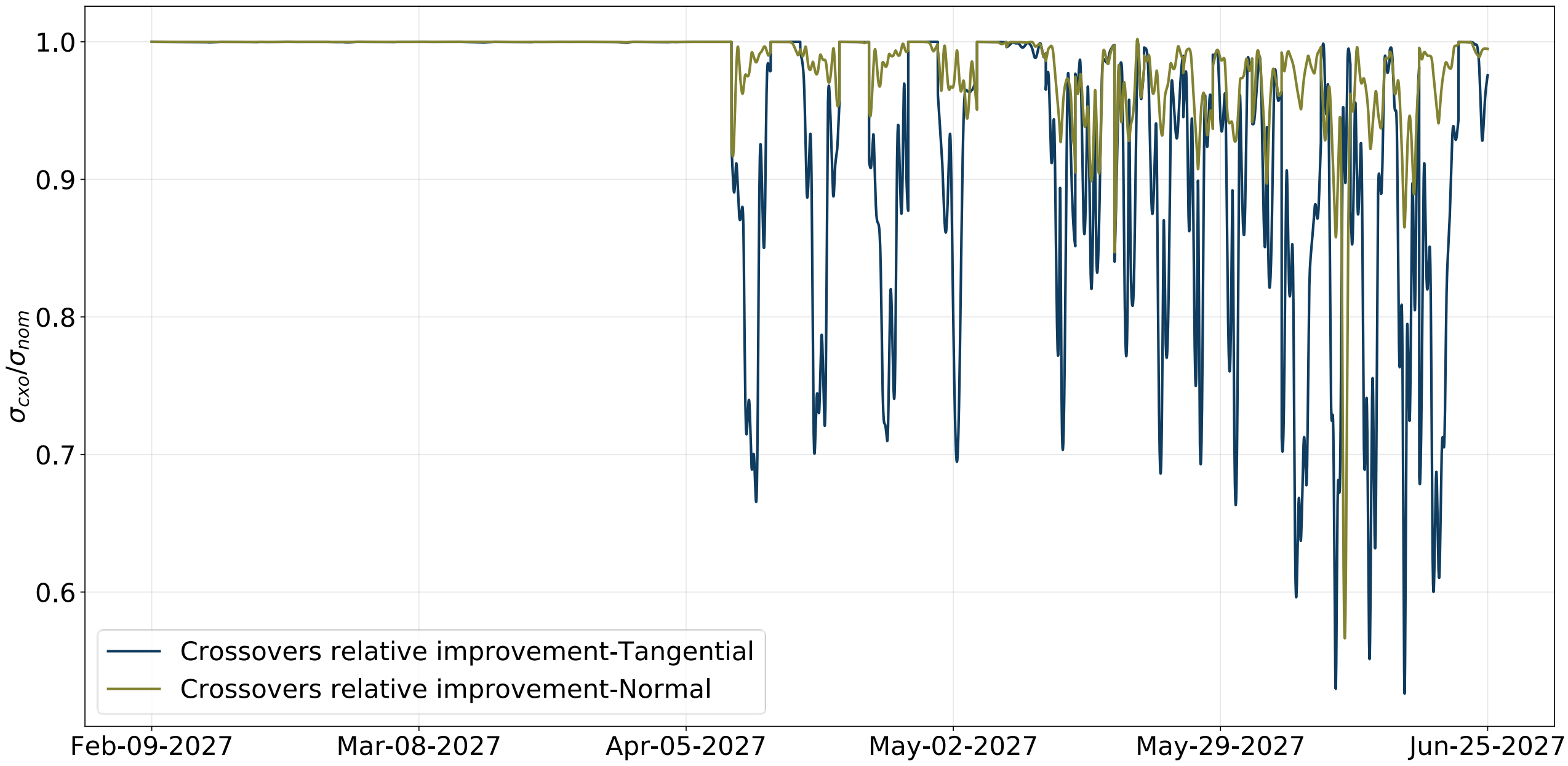

**Figure 29: Relative improvement in Hera's tangential and normal position uncertainty provided by including LIDAR observables and crossovers estimation. The position uncertainty is provided in the RTN frame with respect to Didymos.**

**Table 12: Summary of the estimated formal 1σ uncertainties for GMs, un-normalized spherical harmonics coefficients, pole parameters, and librations of the LIDAR scenario. The improvement factor of each parameter with respect to the nominal scenario of Section 8.1 is provided in the last column.**

| **Coefficient** | **Nominal value** | **Formal uncertainty (1σ)** | **Improvement factor wrt Nominal scenario** |
|---|---|---|---|
| | | **Didymos** | |
| **GM** ($km^3/s^2$) | $4.0071 \cdot 10^{-8}$ | $1.3333 \cdot 10^{-12}$ | 1.21 |
| $\mathbf{J_2}$ | $8.35 \cdot 10^{-2}$ | $6.62 \cdot 10^{-5}$ | 1.06 |
| $\mathbf{C_{21}}$ | $2.32 \cdot 10^{-3}$ | $7.86 \cdot 10^{-5}$ | 1.04 |
| $\mathbf{S_{21}}$ | $-7.14 \cdot 10^{-3}$ | $6.91 \cdot 10^{-5}$ | 1.01 |
| $\mathbf{C_{22}}$ | $-5.91 \cdot 10^{-5}$ | $9.81 \cdot 10^{-5}$ | 1.01 (still not observable) |
| $\mathbf{S_{22}}$ | $-2.82 \cdot 10^{-3}$ | $9.42 \cdot 10^{-5}$ | 1.00 |
| **$J_3$** | $-2.27 \cdot 10^{-2}$ | $2.30 \cdot 10^{-4}$ | 1.03 |
| **$C_{31}$** | $-6.22 \cdot 10^{-3}$ | $1.35 \cdot 10^{-4}$ | 1.04 |
| **$S_{31}$** | $-5.94 \cdot 10^{-3}$ | $1.35 \cdot 10^{-4}$ | 1.03 |
| **$C_{32}$** | $-7.10 \cdot 10^{-5}$ | $1.20 \cdot 10^{-4}$ | 1.00 (still not observable) |
| **$S_{32}$** | $1.01 \cdot 10^{-3}$ | $1.20 \cdot 10^{-4}$ | 1.00 |
| **$C_{33}$** | $8.47 \cdot 10^{-5}$ | $8.81 \cdot 10^{-5}$ | 1.00 (still not observable) |
| **$S_{33}$** | $-3.07 \cdot 10^{-4}$ | $8.88 \cdot 10^{-5}$ | 1.00 |
| $\boldsymbol{\alpha_0}$ (deg) | 311.00 | 0.97 | 1.02 |
| $\boldsymbol{\delta_0}$ (deg) | -79.80 | 0.74 | 1.10 |
| $\boldsymbol{\alpha_1}$ (deg/hour) | 0.00 | $4.06 \cdot 10^{-6}$ | 1.02 |
| $\boldsymbol{\delta_1}$(deg/hour) | 0.00 | $3.09 \cdot 10^{-6}$ | 1.10 |
| $\boldsymbol{w_1}$ (deg/hour) | 159.29 | $1.33 \cdot 10^{-6}$ | 1.10 |
| **Dimorphos** | | | |
| **GM** ($km^3/s^2$) | $3.4522 \cdot 10^{-10}$ | $2.3337 \cdot 10^{-13}$ | 1.17 |
| **$J_2$** | $8.24 \cdot 10^{-2}$ | $5.03 \cdot 10^{-3}$ | 1.08 |
| **$C_{21}$** | $2.61 \cdot 10^{-4}$ | $1.53 \cdot 10^{-3}$ | 1.45 (still not observable) |
| **$S_{21}$** | $8.67 \cdot 10^{-4}$ | $1.61 \cdot 10^{-3}$ | 1.52 (still not observable) |
| **$C_{22}$** | $2.01 \cdot 10^{-3}$ | $1.38 \cdot 10^{-5}$ | 1.83 |
| **$S_{22}$** | $-5.53 \cdot 10^{-4}$ | $2.18 \cdot 10^{-5}$ | 1.02 |
| $\boldsymbol{\alpha_0}$ (deg) | -49.07 | 0.32 | 1.03 |
| $\boldsymbol{\delta_0}$ (deg) | -79.80 | 0.13 | 1.13 |
| $\boldsymbol{\alpha_1}$ (deg/hour) | $1.51 \cdot 10^{-5}$ | $1.21 \cdot 10^{-4}$ | 1.03 |
| $\boldsymbol{\delta_1}$(deg/hour) | $-1.11 \cdot 10^{-6}$ | $4.83 \cdot 10^{-5}$ | 1.12 |
| $\boldsymbol{w_1}$ (deg/hour) | 31.35 | $1.19 \cdot 10^{-4}$ | 1.02 |
| $\boldsymbol{w_a}$ (deg) | 5.00 | $2.23 \cdot 10^{-2}$ | 1.02 |
| $\boldsymbol{\omega}$ (deg/hour) | 30.35 | $5.92 \cdot 10^{-4}$ | 1.29 |
| $\boldsymbol{\varphi}$ (deg) | 6.19 | 1.56 | 1.32 |

# Appendix B - F2BP

In this appendix, we report additional results and figures for the full two-body problem studied in Section 10.

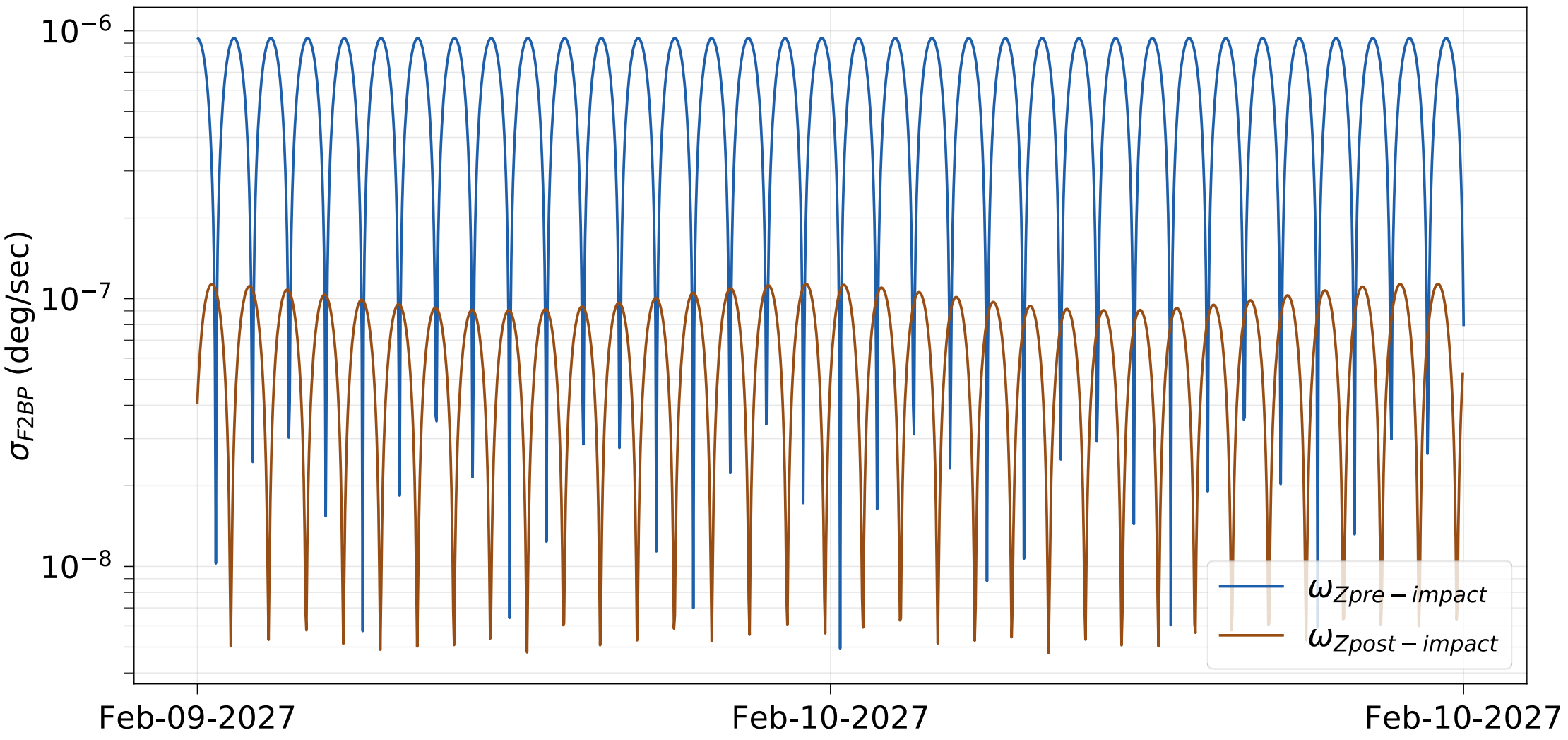

**Figure 30: Formal uncertainty (1σ) of Didymos' body-fixed frame angular velocity. Blue: pre-impact scenario; brown: post-impact scenario. The uncertainty pattern, shown for a single day, is consistent throughout the mission. Only the Z-component is reported for clarity. The X- and Y-components do not show sensible differences between the two cases, as expected.**

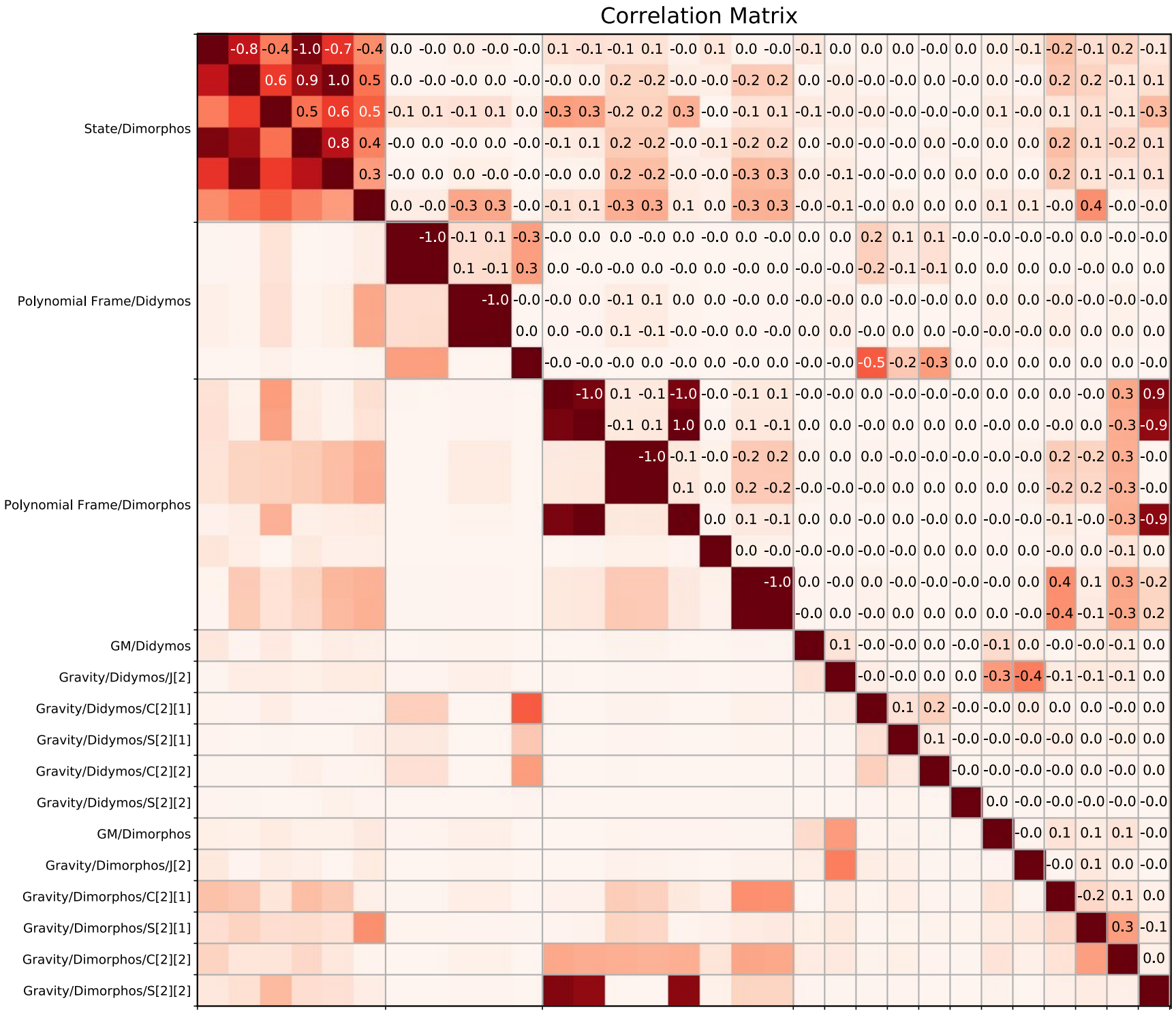

**Figure 31: Correlation matrix of the nominal simulation of Section 8.1 for the main parameters of interest. For clarity, some parameters in the correlation matrices are grouped: the state of Dimorphos parameters includes [X, Y, Z, DX, DY, DZ], while the polynomial frame of Section 8.1 includes [$\alpha_0, \alpha_1, \delta_0, \delta_1, w_1$], with the addition of [$w_a, \omega, \varphi$] parameters to model Dimorphos' librations amplitude, phase, and angular velocity, respectively.**

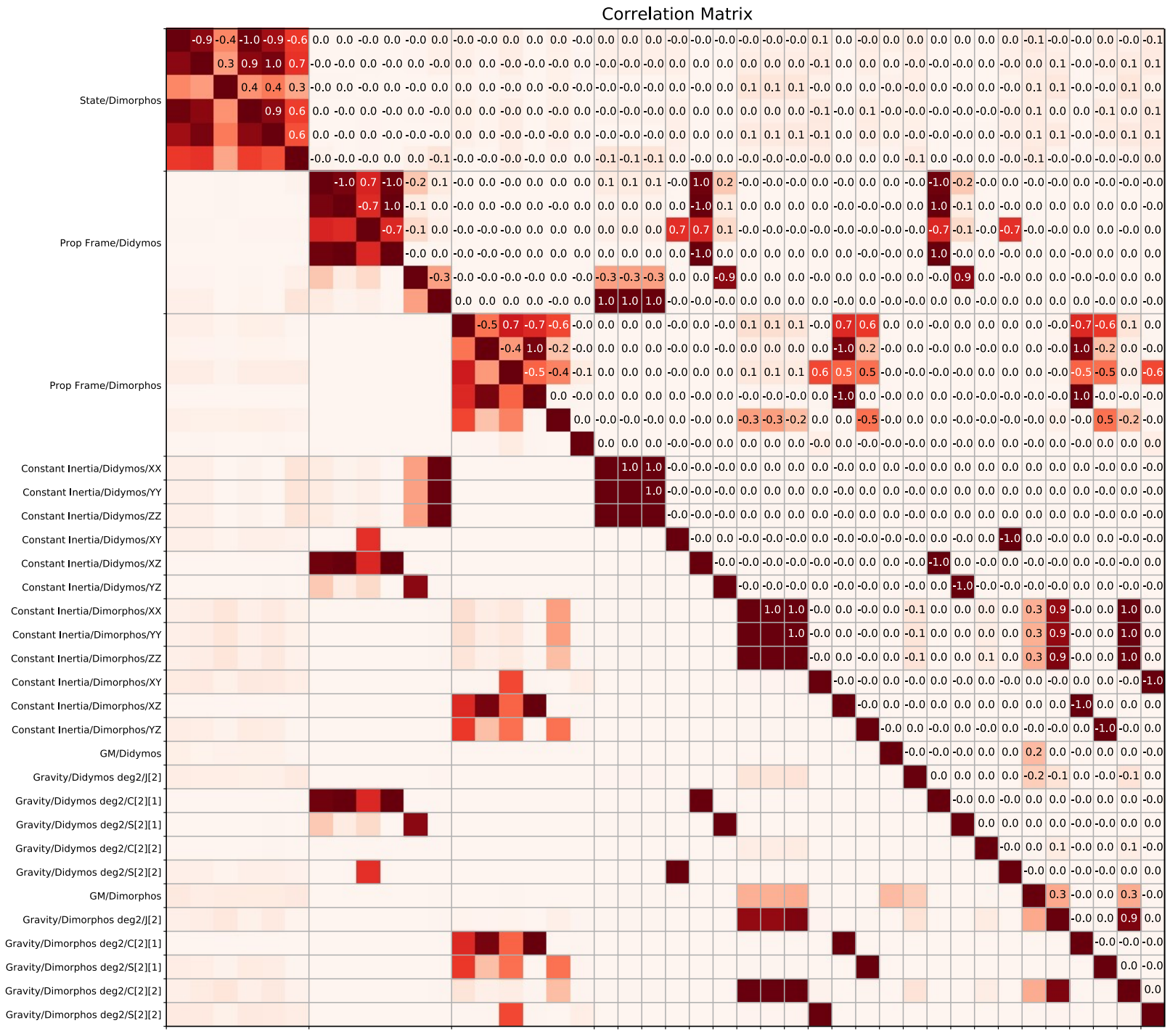

**Figure 32: Correlation matrix of the F2BP pre-impact relaxed simulation for the main parameters of interest. For clarity, some parameters in the correlation matrices are grouped: the state of Dimorphos parameters includes [X, Y, Z, DX, DY, DZ], the F2BP propagated frame comprises [RA, DEC, W, $\omega_X, \omega_Y, \omega_Z$].**

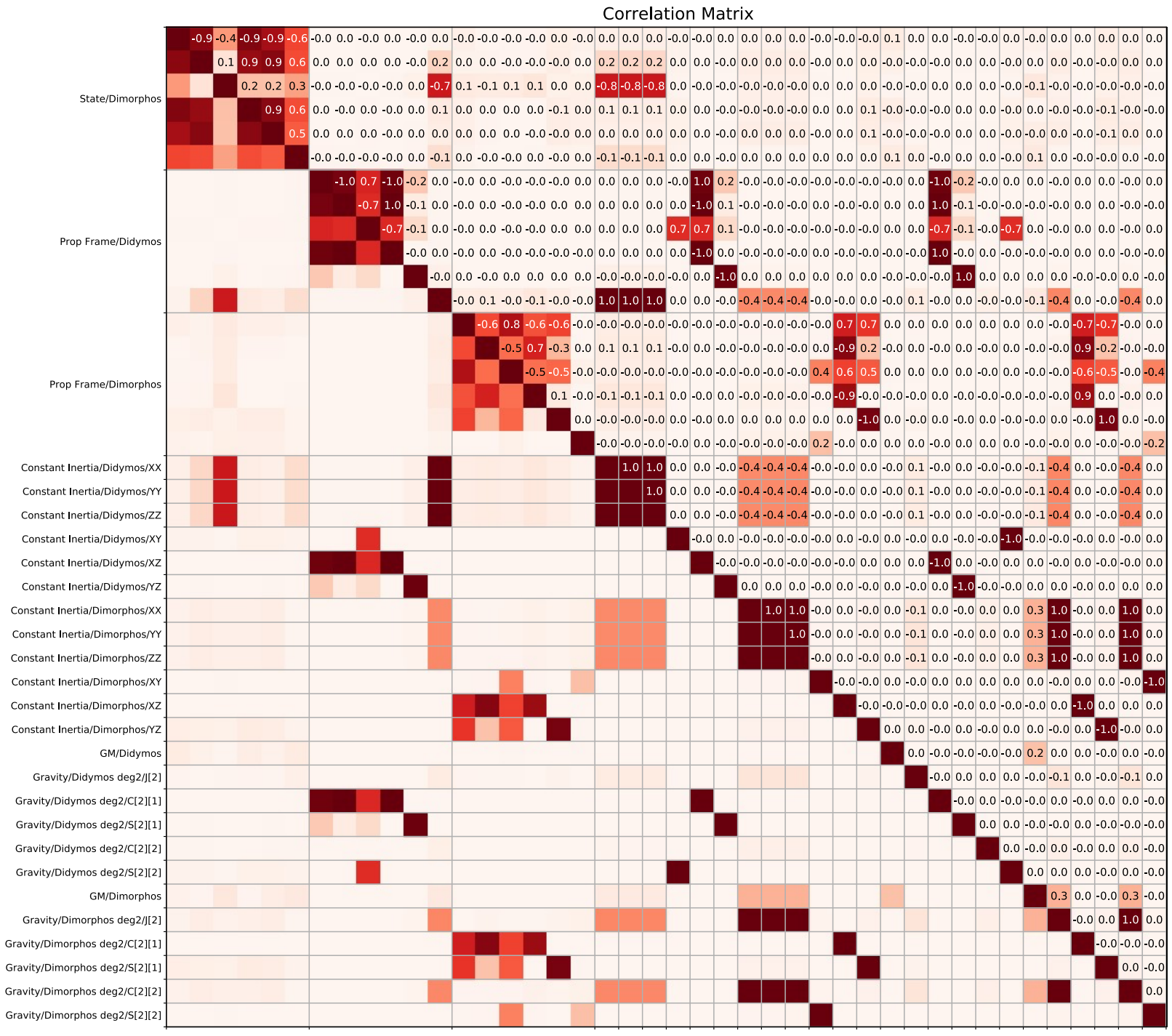

**Figure 33: Correlation matrix of the F2BP post-impact relaxed simulation for the main parameters of interest. For clarity, some parameters in the correlation matrices are grouped: the state of Dimorphos parameters includes [X, Y, Z, DX, DY, DZ], the F2BP propagated frame comprises [RA, DEC, W, $\omega_X, \omega_Y, \omega_Z$].**

## Acknowledgments

This work was partially carried out at the Jet Propulsion Laboratory, California Institute of Technology, under a contract with the National Aeronautics and Space Administration. Government sponsorship acknowledged. All rights reserved.

EG, RLM, MZ, and PT wish to acknowledge Caltech and the NASA Jet Propulsion Laboratory for granting the University of Bologna a license to an executable version of MONTE Project Edition S/W. EG, RLM, MZ, PT, and GT are grateful to the Italian Space Agency (ASI) for financial support through Agreement No. 2022-8-HH.0 in the context of ESA's Hera mission. EG is grateful to Fondazione Cassa dei Risparmi di Forlì for the financial support of his PhD fellowship. This project has received funding from the European Union's Horizon 2020 research and innovation programme under grant agreement No 870377 (project NEO-MAPP).

Special thanks are extended to the anonymous reviewers whose insightful comments and suggestions significantly contributed to the improvement of this paper.

The Scientific colour map *batlow* and *roma* (Crameri, 2018) are used in this study to prevent visual distortion of the data and exclusion of readers with colour-vision deficiencies (Crameri et al., 2020).